\documentclass[10pt]{article}

\usepackage{hyperref}

\usepackage{amsmath}
\usepackage{amssymb}

\usepackage{algpseudocode}
\usepackage{algorithm}
\usepackage{array}
\usepackage{breakurl}
\usepackage{multirow}
\usepackage{rotating}
\usepackage{indentfirst}
\usepackage{mdwlist}
\usepackage{setspace}

\usepackage{booktabs}

\usepackage{graphicx}

\usepackage{subcaption}

\newcommand{\pmf}{\operatorname{pmf}}

\usepackage[T1]{fontenc}
\usepackage{mathtools}

\begin{document}

\sloppy

\title{The $p$-convolution forest: a method for solving graphical models
  with additive probabilistic equations}

\author{Oliver Serang\\
  University of Montana\\
  32 Campus Drive\\
  Social Science 408\\
  Missoula, MT\\
}

\maketitle

\section*{Abstract}
\noindent Convolution trees, loopy belief propagation, and fast
numerical $p$-convolution are combined for the first time to
efficiently solve networks with several additive constraints between
random variables. An implementation of this ``convolution forest''
approach is constructed from scratch, including an improved trimmed
convolution tree algorithm and engineering details that permit fast
inference in practice, and improve the ability of scientists to
prototype models with additive relationships between discrete
variables. The utility of this approach is demonstrated using several
examples: these include illustrations on special cases of some classic
NP-complete problems (subset sum and knapsack), identification of
GC-rich genomic regions with a large hidden Markov model, inference of
molecular composition from summary statistics of the intact molecule,
and estimation of elemental abundance in the presence of overlapping
isotope peaks.

\section*{Introduction}
The idea of tree decomposition and the junction tree algorithm made a
profound impact on graphical models, providing a formal approach for
identifying the graphs on which dynamic programming algorithms could
be applied to a problem much faster than the worst-case complexity for
the problem in general. These discoveries included implications on
NP-hard problems and cases under which they may be solved efficiently
in practice \cite{robertson:graph}. The junction tree algorithm has
been used for RNA secondary structure with
pseudoknots \cite{zhao:rapid}, genotyping on pedigrees with
loops \cite{totir:efficient}, and protein inference in mass
spectrometry \cite{serang:faster}.

\subsection*{Symmetry-based probabilistic message passing algorithms}
Recently, symmetry has been exploited to achieve much faster
performance in graphical models where vertices include several
directed edges in (which necessarily implies high treewidth, due to a
large clique formed in the moral graph). When a vertex has $n$
directed edges in, symmetry permits the induced $n$-dimensional table
from the clique in the moral graph to be replaced by dynamic
programming \cite{heckerman:causal}. Fast practical instances of this
approach for use on applied problems have been derived from
scratch \cite{kim:spectral, howbert:computing}. If each of the $n$
incoming edges carries a discrete distribution with $k$ distinct
support values, the cost of dynamic programming will be $O(n^2
k^2)$. Although this is much faster than the $O(k^n)$ required by the
$n$-dimensional table, the cost of dynamic programming is still
prohibitive for even moderately sized problems, and so its successes
in practice have been when neither $n$ nor $k$ are very large.

\subsection*{Additive models and probabilistic convolution trees}
The additive case, where the relationship induced by the vertex is of
the form $Y = X_1 + X_2 + \cdots + X_n$ (or equivalently, $Y = X_1
\cdot X_2 \cdot ~\cdots~ \cdot X_n$ in an exponentiated space), is
quite common in practice (indeed, both  \cite{kim:spectral,
  howbert:computing} rely on additive symmetry). Tarlow \emph{et al.}
created the first subquadratic approach for the additive case under
the condition $k=2$ and when the dimension of all distributions is
1. Their approach decreases the $O(n^2 k^2) = O(n^2)$ runtime to $O(n
\log(n) \log(n))$ \cite{tarlow2012fast}. That approach was
independently discovered in a more general form as the ``probabilistic
convolution tree'', where $k$ and the dimension of the distributions
can take any value, and where the runtime in practice was decreased by
narrowing the support of distributions on internal nodes during the
backward pass \cite{serang:probabilistic}.

The essential idea behind the probabilistic convolution tree uses the
fact that addition of random variables corresponds to convolution of
the probability mass functions (PMFs), and these convolutions can be
performed in subquadratic time via the fast Fourier transform
(FFT) \cite{cooley:algorithm}. Rather than combine variables
left-to-right in a chain $Y = (~(~(X_1 + X_2) ~+X_3) ~+X_4) ~+\cdots$,
they can be combined in a balanced binary tree $Y = ( (X_1 + X_2) ~+~
(X_3 + X_4) ) ~+~ + ( ( \cdots ) )$. Importantly, this keeps the
support of combined random variables from growing too disparate as the
algorithm progresses (as happens in the $O(n^2 k^2)$ dynamic
programming method). The backward pass is slightly more complicated,
and relies on the fact that subtraction can be performed via addition
and negation, and that negation of a random variable can be performed
by reversing the discrete distribution and shifting the support.

Thus given prior distributions $\pmf_{X_1}, \pmf_{X_2}, \ldots$ and a
likelihood distribution $\pmf_{D|Y}$, all priors, likelihoods (and
therefore, posteriors, which are the products of priors and
likelihoods) can be computed: The prior distribution on $Y$ is
computed via a forward pass (aggregating pairwise sums of
distributions, performed using convolution). In the process of
computing $\pmf_Y$, the forward pass also computes all pairwise sums as
variables are successively merged; these pairwise sums will be used
again in the backward pass. The likelihoods $\pmf_{D | X_1}, \pmf_{D |
  X_2}, \ldots$ are computed via a backward pass. \emph{E.g.},
subtracting the prior $\pmf_{X_{n/2+1}+X_{n/2+2}+\ldots X_n}$ from the
likelihood $\pmf_{D|Y}$ yields the likelihood $\pmf_{D | X_1+X_2+\ldots
  X_{n/2}}$. Note that this is distinct from deconvolution. Consider a
case with $n=2$: In deconvolution, we are given priors on $Y$ and
$X_1$ and we seek a prior on $X_2$ consistent with $Y=X_1+X_2$. Here,
we are given a likelihood on $D|Y$ and priors on $X_1$ and $X_2$, and
we seek all priors and likelihoods (which would be computed in a naive
case by marginalizing subject to the constraint that $Y=X_1+X_2$).

Given $n$ prior distributions each with $k$ unqiue support values, and
1 likelihood distribution of arbitrary support, the probabilistic
convolution trees can solve all posteriors simultaneously $O(n k
\log(n k) \log(n))$. Note that this is $\subset O( {(n
  k)}^{1+\epsilon} )$ for any $\epsilon >0$; in comparison,
constructing the $n$ prior PMFs would take runtime $O(n k)$, meaning
the convolution tree algorithm is not very more difficult than loading
the data.

\subsection*{Sum-product inference, the suppression of higher moments, and the cumulative aggregation of noise}

A key problem with additive models when $n \gg 1$ is that when adding
several random variables, the central limit theorem results in smooth,
Gaussian-like distributions. This means that asymptotically as $n$
becomes large (the target use-case), only the means and variances of
each of $X_1, X_2, \ldots, X_n$ will influence the resulting prior
distribution on $Y$ (the reverse pass is likewise influenceed by the
central limit theorem, but in a more subtle manner, because internal
nodes of the tree are only expected to have Gaussian-like priors after
a minimum number of mergers have been performed). Although this could
be used to speed up the algorithm (seeing as the family of Gaussians
are closed under convolution and multiplication, and thus once the
distributions become approximately Gaussian, convolution can be
performed in $O(1)$ by simply adding the means and variances), the
greater concern is that when $n$ is large, marginal distributions are
simply not very informative. Significantly, these marginals and
posteriors may be uninformative even when only a narrow solution space
is possible (such as when only a single joint event $X_1=x_1, X_2=x_2,
\ldots X_n=x_n$ would be consistent with the likelihood $D|Y$). The
information in the higher moments (skew, kurtosis, \emph{etc.}) is
suppressed as $n$ becomes large.

This is because the dynamic programming algorithms above (including
the convolution tree) rely on sum-product inference: that is, at every
step, they aggregate all possible paths that pass through a node. This
style of aggregation may simultaneously entertain events that are
mutually exclusive. For example, in hidden Markov models (HMMs) with
latent variables $S_1, S_2, \ldots, S_n$, the forward-backward
algorithm, the latent variable $S_1$ contributes all possible ways to
transition to the next latent variable $S_2$; however, when $S_2$
contributes all possible ways to transition to $S_3$, it transmits
some mutually exclusive paths: $S_1=0, S_2=0$ and $S_1=0, S_2=1$ will
both contribute to $S_3=1$. In large problems, this cumulative
aggregation of information can obscure the signal (\emph{i.e.}, the
true values of the latent variables), particularly when the input
distributions are more noisy or uncertain.

This is one reason that HMMs are typically analyzed using the Viterbi
path rather than the forward-backward algorithm. Where the
forward-backward algorithm aggregates the all paths (in sum-product
space), the Viterbi path computes the single \emph{maximum a
  posteriori} (MAP) path (in max-product space); rather than aggregate
multiple paths, only the best path reaching a node is considered from
then on, and this eliminates the aggregation of mutually exclusive
signals.

\begin{figure}
\centering
  \begin{tabular}{cc}
    \includegraphics[width=2.2in]{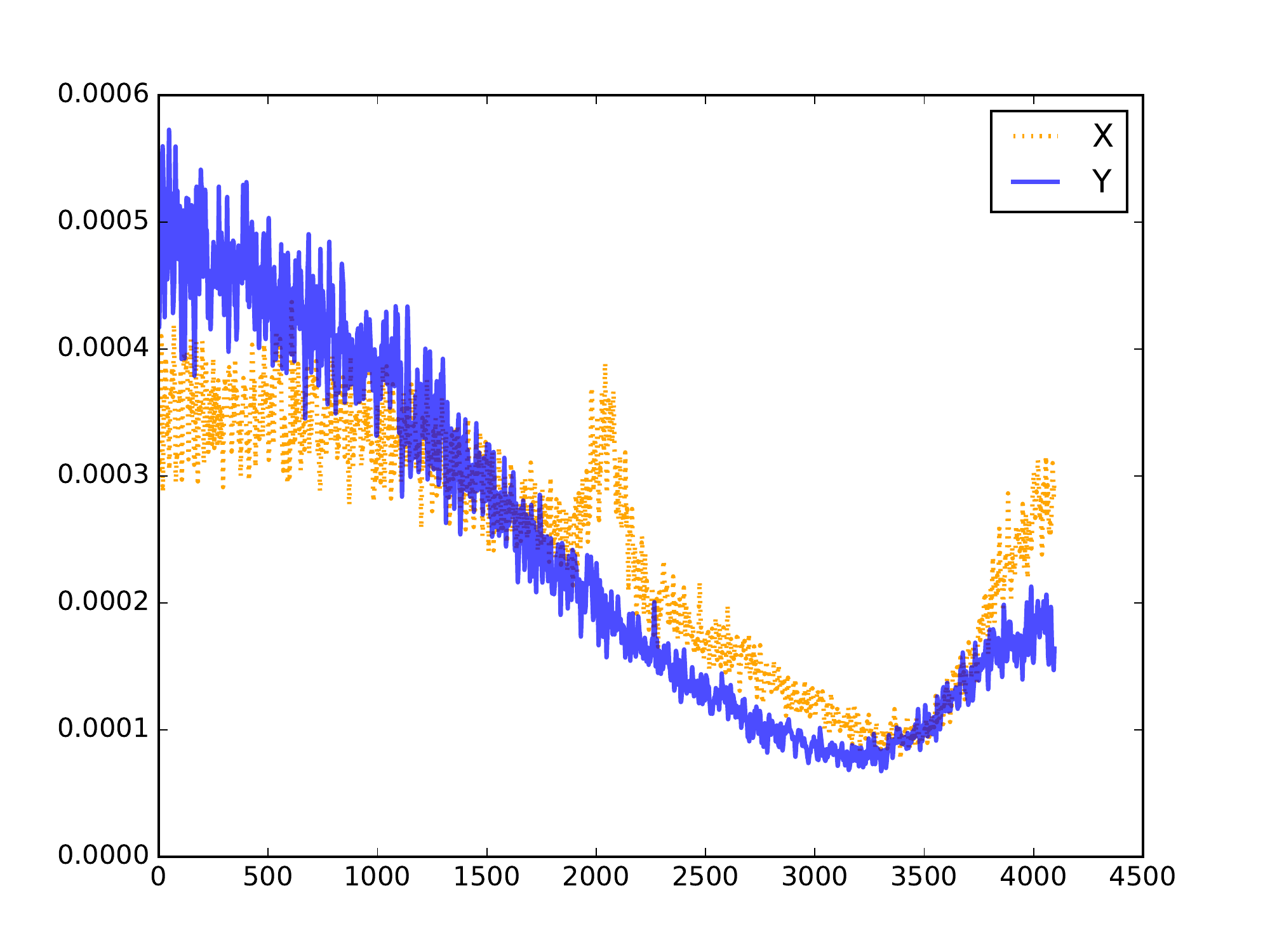} & \includegraphics[width=2.2in]{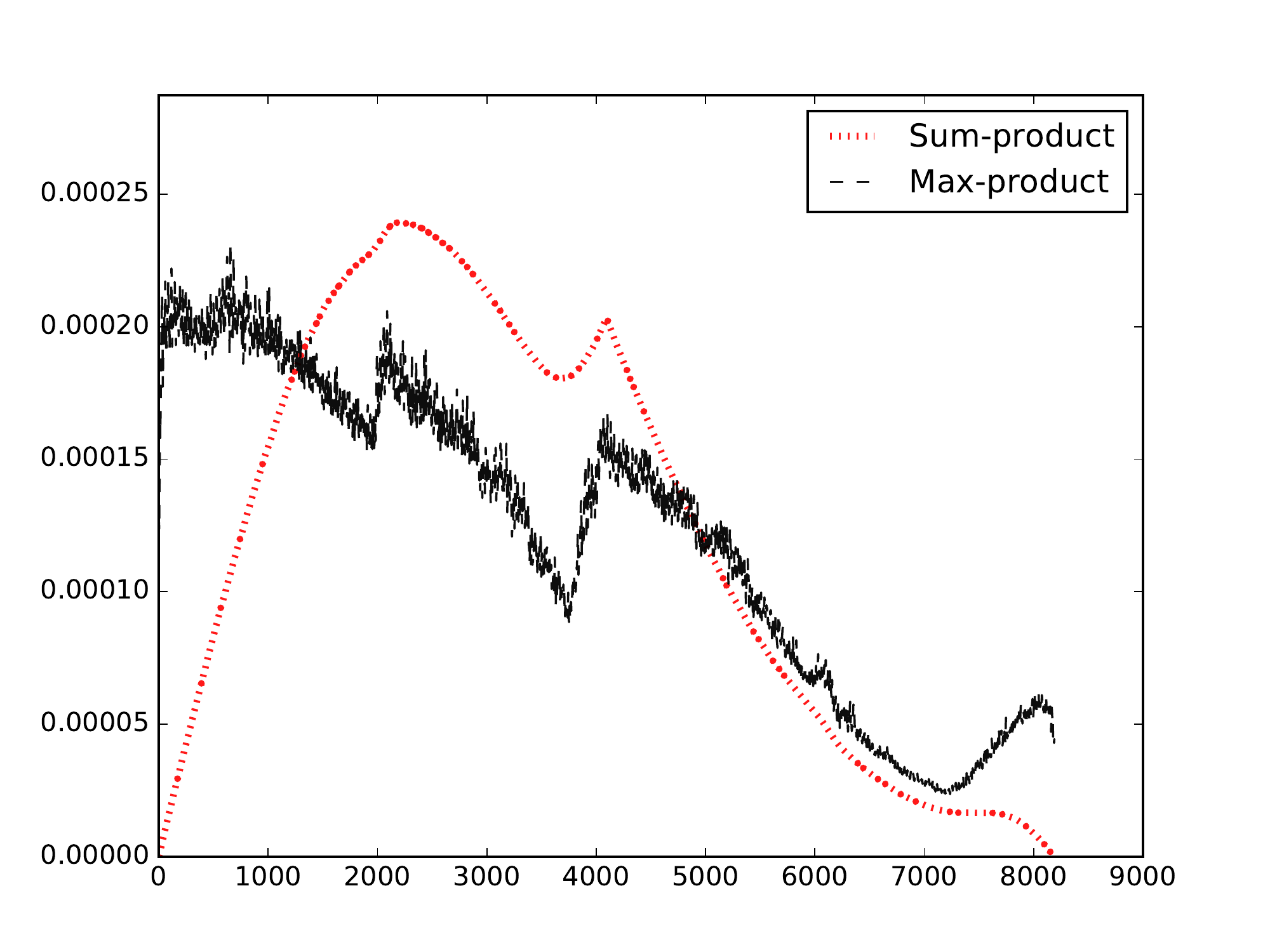} \\
  \end{tabular}
\caption{{\bf Adding X+Y with sum-product and max-product inference.}
  {\bf Left panel:} The PMFs of two discrete random variables $X$ and
  $Y$ are plotted. {\bf Right panel:} The PMF of $X+Y$ is plotted
  using sum-product inference (computed in $O(n \log(n))$ using
  standard FFT convolution) is quite different from the result using
  max-product inference (computed in $O(n^2)$ using naive
  max-convolution).
  \label{fig:sum-vs-max}}
\end{figure}

The difference between standard convolution and max-convolution can be
quite significant. For example, Figure~\ref{fig:sum-vs-max} shows the prior
PMFs for two discrete random variables $X$ and $Y$, and the resulting
prior for $Z=X+Y$ as computed in a sum-product space and as computed
in a max-product space. The sum-product and max-product results for
$Z$ are strikingly different.

\subsection*{Fast numeric max-convolution}

PMFs can easily be adapted to work in max-product inference (by simply
replacing $+$ operations with $\max$ operations when computing
marginals). These max-marginals bear a mechanistic similarity to the
\emph{maximum a posteriori} (MAP) estimate, but where the MAP estimate
yields only a single point estimate, max-marginals produce full
marginal distributions, but where the most probable configuration is
taken over all other variables of interest (rather than the sum over
all other variables of interest, as performed by sum-product
inference). The max-marginal can be used to produce an MAP estimate,
but the converse is not true.

However, adapting convolution trees to work in max-product space is
not trivial: Subquadratic FFT convolution works by moving to a
transformed space (the frequency domain), multiplying in the
transformed space (which corresponds to convolution in the time
domain), and then moving back from the transformed space (via the
inverse FFT). When replacing $+$ operations with $\max$ operations,
inverse operations are no longer defined; therefore, there is no way
to return from the max equivalent of the frequency domain. This
trouble boils down to the fact that fast convolution on the ring
$(\times, +)$ has (at the current time), faster algorithms than fast
convolution on the semiring $(\times, \max)$. The first subquadratic
max-convolution algorithm was published by Bremner \emph{et al.} in
2006, and despite its profound theoretical significance, it only
achieved a slightly subquadratic runtime \cite{bremner:necklaces} when
compared to the $O(n \log(n))$ runtime of FFT
convolution. Furthermore, where FFT can be performed in place, the
method from Bremner \emph{et al.} relies on a reduction to max matrix
multiplication, which is further reduced to an all-pairs shortest
paths (APSP) problem. Performing both of these reductions in a
cache-optimized in-place manner would likely be difficult.

Likewise, another method for computing max-convolution of two vectors
based on sorting the vector arguments and visiting them in descending
order \cite{bussieck:fast} has a runtime that depends somewhat
cryptically on the input data, and thus is also not reliably $\in
o(n^2)$; however, that sorting-based approach has been used quite
successfully in practice for work calculating the most intense isotope
peaks in mass spectrometry \cite{lacki:isospec}, which is quite
interesting given the additive nature of isotope problems (which will
be exploited here with max-convolution rather than sorting) and the
fact that L\k{a}cki \emph{et al.} are not explicitly using
max-convolution. This suggests the possibility of more unified
approaches to additive problems, which would connect max-convolution
on one hand and a priority queue of the top values in the cartesian
product.

The lack of inverse operations in max-product space can approached by
using rings that behave similar to semirings. Specifically, the $L_p$
ring space defines $x \oplus y = {(x^p + y^p)}^{1/p}$, and when $p \gg
1$, $z = x \oplus y \approx \max(x, y)$, but with the option of an
inverse operation: given $x$ and $z$, it is possible to solve for $y$
(this would not be possible in a genuine semiring). By using $L_p$
ring spaces, it is possible to numerically approximate
max-convolution. Moreover, it is possible to either directly compute
or approximate (depending on the $p$ desired) a continuum between
sum-product inference (equivalent to $p=1$) and max-product inference
(equivalent to $p=\infty$) \cite{serang:fast}, which we denote here as
numeric $p$-convolution. This continuum is useful in its own right,
and $p$ can be thought of as a hyperparameter. $p=1$ is democratic and
places a high value on popularity, $p=\infty$ is more like a
dictatorship where only the strongest solution is weighed, and finite
$p>1$ resembles a republic, where the results reflect a compromise
between popularity and quality of the solutions. This numeric
$p$-convolution approach generalizes to convolution on tensors,
whereas the approach in Bremner \emph{et al.} is as of now only
applicable to 1D vectors.

Underflow concerns sometimes limit the choice of $p$ for which
$p$-convolution can be stably computed, particularly when many values
in the input arrays are close to zero; therefore, a collection of a
small or constant number of $L_p$ ring spaces can be used. Rather than
using a single $L_p$ space (\emph{e.g.}, the one corresponding to the
largest $p$ that is numerically stable for a result of interest), the
shape of the collection of $L_p$ spaces can be used to more accurately
approximate the result \cite{pfeuffer:bounded}. Fast numeric
$p$-convolution has a runtime that is, in practice, roughly
$<10\times$ that of a fast implementation of FFT convolution; while
not quite as fast or exact as FFT convolution (which is itself a
numeric method), numeric $p$-convolution is fast enough to make large
problems trivial (while in contrast the naive approach would simply be
too slow). This approximation strategy generalizes to all problems on
semirings isomorphic to $(\times, \max)$, such as
APSP \cite{serang:fast2}.

\subsection*{$p$-convolution trees}

The forward pass of a standard convolution tree can be used to solve
the subset-sum problem on the integers. This is performed by
converting the set values to prior probabilities on $X_1, X_2, \ldots
X_n$, performing the forward pass to compute the prior on $Y = X_1 +
X_2 + \cdots + X_n$, and then locating the support where $Y$ has
nonzero probability. This has been rediscovered independently by
Koiliaris and Xu \cite{koiliaris:faster} and by
Bringman \cite{bringmann:near}. Likewise, the forward pass of a
max-convolution tree can be used to solve the knapsack problem on the
integers by preserving not only which sums have nonzero probability,
but also what is the highest probability attainable for each of those
nonzero probability supports. This knapsack variant from the forward
pass has also been rediscovered independently by Cygan \emph{et
  al.} \cite{cygan:problems}, by K{\"u}nnemann \emph{et
  al.} \cite{kunnemann:fine}, assuming the availability of a fast
max-convolution algorithm. Similarly, Backurs \emph{et al.} have used
a tree of convolutions on the semiring $(+,\min)$ (which is isomorphic
to $(\times, \max)$) to solve the tree sparsity
problem \cite{backurs:better}.

By combining convolution trees with fast numeric $p$-convolution
(rather than standard convolution or max-convolution), it is possible
to unify and generalize the special-case dynamic programming
algorithms for solving subset-sum and knapsack. This generalized
approach is denoted here as the ``$p$-convolution tree''. With $p=1$,
the prior on $Y$ can be used to solve subset sum, whereas $p=\infty$
allows the prior on $Y$ to solve knapsack. However, more significant
is the backward pass, which as before makes it possible to compute all
priors and likelihoods (and thus all posteriors) simultaneously in
$O(n k \log(n k) \log(n))$, regardless of whether sum-product space,
max-product space, or the continuum between them is sought. Where the
forward pass can be employed to ask whether a particular restaurant
menu can be used to build an order costing exactly $1073.25$ (or in
the $p=\infty$ knapsack case, to find the most satisfying order
costing exactly $1073.25$, where preferences of each person ordering
are included in the priors), the backward pass efficiently finds the
precise orders (or distributions on those orders, if multiple
solutions exist) that produce the total bill of $1073.25$. With $p=1$,
the backward pass is equivalent to aggregating all possible order
configurations that would reach total $1073.25$ and then marginalizes
by summing out all customers but the one of interest to find their
likelihood or posterior distribution. With $p=\infty$, marginalization
maxes out all customers but the one of interest.

\subsection*{Into the convolution forest}

This manuscript introduces the ``convolution forest'', a method that
combines loopy belief propagation \cite{weiss:correctness} with large
numbers of $p$-convolution trees. Each convolution tree is queried
iteratively using a variety of possible message passing schemes. Each
convolution tree computes messages out, which can be fed into other
convolution trees, and so on until convergence is reached.

New methods are introduced that improve the performance of inference:
Trimmed $p$-convolution trees are able to automatically detect narrow
solution spaces of particular data before computing convolutions and
thus on some data decreases the runtime below the currently known
limit of $O(n k \log(n k) \log(n))$. These trimmed convolution trees
rely on a lazy caching strategy for propagating through the tree.

These methods are implemented in the {\tt C++11} graphical models
library, ``EvergreenForest'', which is specifically tailored for
solving and prototyping additive probabilistic models. The library
includes modular, from-scratch implementations of several tools used
in inference: these include real and complex FFT (using a
template-recursive approach), p-convolution (using a lazy approach
that may terminate early without computing the full family of
convolutions in $L_p$ spaces), PMFs, and message passing methods for
graphical models. The template-recursive TRIOT tensor library is used
for manipulating distributions of arbitrary dimension (and dimension
unknown at compile time) \cite{heyl:triot}.

This implementation of the convolution forest method is demonstrated
on a few important applied problems, detection of GC-rich nucleotide
regions, molecular decomposition from approximate mass and
hydrophobicity, and estimation of elemental abundances in the presence
of overalapping isotope peaks.

\section*{Methods}
The code in the {\tt EvergreenForest} repository is split into
modules: Tensor (the TRIOT library), BitReversedShuffle (for
performing bit-reversed permutations in FFT), FFT, Convolution, PMF,
Engine (containing the core components for message passing in graphs),
and Evergreen (containing the header for user interface with the
engine and all other components). Noteworthy features modules are
described below.

\subsection*{Bit-reversed permutation}

With the use of TRIOT for manipulating tensor data and with a fairly
optimized FFT implementation, a significant percentage of the FFT
runtime is performed in the bit reversed shuffle. A novel,
template-recursive cache-oblivious method was used. This method is
described in greater detail in Knauth \emph{et
  al.} \cite{knauth:practically}.

\subsection*{Template-recursive FFT}

The FFT module includes implementations of both decimation in time
(DIT) and decimation in frequency (DIF) FFTs, implemented using a
template-recursive version of the Cooley-Tukey method. These are
implemented in a manner reminiscent of GFFT \cite{myrnyy:simple}, but
written from scratch in an object oriented manner (which, thanks to
improvements in {\tt C++11} such as {\tt constexpr}, no longer
sacrifices efficiency for readability as it would when GFFT was first
published). Template recursion is used to essentially generate all
recursive FFT calls at compile time; these nested recursive calls will
have {\tt constexpr} length, which enables perfect loop unrolling and
enables trigonometric constants to be generated at compile
time. Furthermore, the compiler may detect similarities between the
recursive calls (each length $n$ FFT reduces to two FFTs of length
$\frac{n}{2}$), including the complex twiddle values used.

The fixed-length 1D template-recursive FFTs are generated up to a
fixed maximum size. The maximum log-length can be set using the
constant {\tt const unsigned char FFT1D\_MAX\_LOG\_N}, which has
default value 32. TRIOT, the bit-reversed permutation tools, and the
FFT implementations are built using {\tt unsigned long} indices, so
that FFTs of length $>2^{32}$ can be used if enough RAM is available
to store the data. A double precision complex-valued array of length
$2^{32}$ requires 64GB; however, seeing as even large FFTs can be
performed efficiently (and the FFT library supports in-place FFTs), it
may be beneficial to have the option to perform longer FFTs.

The FFT module implements an multidimensional FFT via the row-column
algorithm, which can be called directly on {\tt Tensor<cpx>} types,
meaning the interface is quite simple. This can be performed in place
or out of place. Axes are transposed using the optimal cache-oblivious
strategy from Prokop \cite{prokop:cache} to perform row-order FFTs for
greater cache performance (it should be noted that when the number of
dimensions is larger than one, a buffer may be used even for in-place
FFTs in order to help perform these transpositions). The trigonometry
to compute the complex number corresponding to the twiddle factor for
an FFT of a given size is evaluated at runtime by employing the {\tt
  constexpr} trigonometric functions, and an in-house complex class
{\tt cpx}, with the template parameter specifying the size of the 1D
FFT (these template parameters will be known at compile time and
therefore, can be evaluated to {\tt static const cpx} values at
compile time).

Unlike GFFT's template-recursive Cooley-Tukey implementation, FFTs of
unknown length at runtime are no longer invoked via a table of base
class pointers (GFFT calls a virtual function on the object in index
$i$ to invoke an FFT of length $2^i$); instead, greater performance
was achieved by simply performing template recursion to produce an
{\tt if-else} ladder to map the runtime log-length to the appropriate
template parameter. This strategy effectively checkes larger and
larger lengths until one matches or until {\tt FFT1D\_MAX\_LOG\_N} is
surpassed (producing an assertion error). Although this results in an
additional $\log(n)$ steps when computing an FFT of length $n$, this
additional cost is amortized out by the $O(n \log(n))$ steps required
by FFT. In practice, this actually achieves superior performance to
the table used by GFFT, because the compiler better optimizes these
simple, non-{\tt virtual} functions.

A numerically stable recurrence relation is used to compute all
necessary complex values from the twiddle factor. A simple recurrence
uses the fact that the sequence of complex polars $e^{-j \theta},
e^{-j 2 \theta}, e^{-j 3 \theta}, \ldots$ can be found by starting
with the twiddle factor $e^{-j \theta}$ and iteratively performing
{\tt *=} by the twiddle factor, using the property that $e^{-j (a+1)
  \theta} = e^{-j a \theta} \cdot e^{-j \theta}$. Even though the
complex values are stored in Cartesian form (rather than polar), the
property is nonetheless valid, and so all necessary trigonometric
values can be computed in terms of the {\tt static const cpx} values
known at compile time. The in-house {\tt cpx} class also has forced
inlining of the {\tt *=} operator via {\tt
  \_\_attribute\_\_((always\_inline))} (supported by both {\tt g++}
and {\tt clang++}), making it possible for a clever compiler to fully
unroll the loops for smaller FFTs at compile time.

For large FFTs, the above recurrence relation begins to lose
precision. The $\theta$ value of the twiddle factor will be close to
zero (being that there are $n$ evenly spaced values around the unit
circle, where $\theta$ is of the form $\frac{2 \pi k}{n}$), and
therefore $e^{-j \theta} = \cos(\theta) ~-~j \sin(\theta)$ will have a
real component $\cos(\theta) \approx 1$. Floating point values are
very effective at distinguishing zero from quantities close to zero,
but are not effective at distinguishing one from quantities close to
one. For this reason, the above recurrence can be reconfigured in
terms of $\cos(\theta)-1 ~-~j \sin(\theta)$: Rather than compute
$e^{-j (a+1) \theta} = e^{-j a \theta} \cdot e^{-j \theta}$, it is
instead possible to compute $e^{-j (a+1) \theta} = e^{-j a \theta} +
e^{-j a \theta} \cdot (e^{-j \theta} - 1)$. Thus at the cost of an
extra complex addition (which is quite small), the recurrence can be
described in terms of the value $e^{-j \theta} - 1$, which has both
real and imaginary components close to zero when $n \gg 1$. This is
implemented in a simple object oriented manner in the {\tt Twiddles}
class via the {\tt static} function {\tt void Twiddles<N>::advance(cpx
  \& current)} (where $N$ is the length of the FFT being performed).

Both complex and real FFT are implemented. Real FFTs achieve greater
performance, reducing to an FFT of half the size by packing real
values such as {\tt [1,2,3,4, \ldots]} into complex values {\tt [1+2j,
    3+4j, \ldots]} and undoing the final butterflying
step \cite{proakis:introduction}.

The FFT can be invoked with options to ignore shuffling, to ignore
undoing the transpositions (for multidimensional FFT), and to exploit
a freshly zero-padded tensor (for convolution) for greater
performance. These options have practical implications that enable
faster convolution (described below).

Although the Cooley-Tukey approach is not quite as good for large
numbers of dimensions (because each dimension must be zero padded, and
reaching the next power of two in each dimension may result in a
$\approx 2^d$ slowdown where $d$ is the number of
dimensions \cite{serang:exact}), this implementation is lightweight,
produced completely in house, and is fast in practice for small
numbers of dimensions.

\subsection*{Standard convolution (for complex and real tensors)}

The Convolution module implements naive (exact) convolution and
numeric $p$-convolution. Even though the FFT is quite efficient, naive
convolution can be substantially faster on small tensors, particularly
because the naive convolution is implemented as a TRIOT expression;
therefore, the numeric $p$-convolution algorithm automatically defers
to the naive version on small problems.

Convolution can be performed on tensors of type {\tt Tensor<cpx>} and
{\tt Tensor<double>}, with the latter being more efficient, as it
employs the real FFT (which in turn calls a complex FFT of half the
length). By combining the DIT and DIF FFTs, some shuffling can be
avoided. For instance, when convolving two complex tensors, the
arguments will be zero padded, FFTed, multiplied element-wise, and
then inverse FFTed. DIT FFTs perform the bit-reversed shuffle before
butterflying and DIF FFTs apply the shuffle after butterflying. Thus,
if the FFTs of the zero-padded arguments are performed via the DIF
FFT, the element-wise multiplication will not be affected if shuffling
is ignored (because both FFT results will be identically permuted, and
so the correct elements will still be multiplied with one
another). When computing the inverse FFT, the element-wise multiplied
result will still be shuffled; however, if the inverse FFT is
performed via the DIT FFT, then it can simply ignore the shuffling
(seeing as it would shuffle first, and the data are already
bit-reversed shuffled). Currently, some shuffling is still necessary
for the real FFTs, but this speedup is nonetheless significant.

Likewise, when convolving multidimensional tensors, some
transpositions can be ignored. Let the axes of the tensor be denoted
$(x,y,z)$, where lower-case letters are used when the axis has not yet
been FFTed and where upper-case letters mean the axis has been
FFTed. Performing row FFTs will result in $(x,y,Z)$. By treating $x$
and $y$ as a single flat index (whose length is the product of the
lengths of the axes for $x$ and $y$) rather than two separate indices,
a single cache-oblivious matrix transposition will reorder the axes to
$(Z, x, y)$. Applying row FFTs will result in $(Z, x, Y)$. At this
point, transposing back will produce $(x, Y, Z)$, which can in turn be
transposed to yield $(Y,Z,x)$ and then $(Y,Z,X)$ and then transposed
back to the finished $(X,Y,Z)$. If the FFT is performed for the
purposes of convolution, undoing the transposition is unnecessary for
similar reasons to shuffling: axes will be reversed during the forward
FFT process, and then reversed again during the inverse FFT, thereby
yielding the correct result with half the transpositions. The forward
FFT process while ignoring undoing the transpositions will be as
follows: $(x,y,z)$ to $(x,y,Z)$ to $(Z,x,y)$ to $(Z,x,Y)$ to $(Z,Y,x)$
to $(Z,Y,X)$.

\subsection*{Lazy numeric $p$-convolution}

Approximate $p$-convolution is implemented as described in Pfeuffer \&
Serang, using a two-term projection to a multiset followed by affine
correction for postprocessing \cite{pfeuffer:bounded}. This method is
motivated by a projection to convolutional problems in a
lower-dimensional space in a method distinct from but qualitatively
similar to sparse FFT  \cite{hassanieh:simple,
  hassanieh:nearly}. However, unlike the previous {\tt python}, the
{\tt C++11} introduced here begins with the largest $p$ of interest
(if $p$ is finite, otherwise, the value of $p$ beyond which there are
diminishing returns on the accuracy of the approximation) and then
decreases downward. This enables processing to terminate prematurely
if computation in only a subset of the considered $L_p$ spaces is
necessary for the approximation. For example, when performing
$p$-convolution with $p=16$, if directly computing $x^{16} \circledast
y^{16}$ (where $\circledast$ represents the convolution operator and
$x^{16}$ indicates taking to the power 16 element-wise) is numerically
stable, then $(x^{16} \circledast y^{16})^{1/16}$ achieves the desired
result directly without using multiple $L_p$ spaces. Likewise, if
$p=16384$ is desired, and all result indices are stable with $p=512$,
$p=384$, $p=256$, and $p=128$, then the two-term projection can be run
without bothering to compute the convolution at $p=64$, $p=32$,
etc. This can result in a significant time savings in practice, not
only because it decreases the number of convolutions performed, but
also because it decreases the number of tensors allocated, which
prevents data in the cache from being contaminated by temporary
results.

\subsection*{Trimmed $p$-convolution trees}

Here the method of ``trimmed convolution trees'' is
presented. Consider $Y=X_1+X_2+X_3+X_4$, where the priors on the $X_i$
variables have support $X_1 \in \{0,1,2\}$, $X_2 \in \{0,1\}$, $X_3
\in \{1,2\}$, and $X_4 \in \{1,2,3\}$, and where the likelihood on $Y$
has support $Y \in \{1,2,3\}$. The forward pass of the convolution
tree algorithm will first compute priors on $X_1+X_2$ and $X_3+X_4$,
then compute the prior on $Y=X_1+X_2+X_3+X_4$. Then the backward pass
will compute the likelihoods on $X_1+X_2$ and $X_3+X_4$, and finally
the likelihoods on $X_1$, $X_2$, $X_3$, and $X_4$. After both passes
have been performed, all priors and likelihoods will be available,
meaning that all posteriors can be computed.

As the forward pass progresses, the support of the distributions
grows, leading to the prior on $Y$ with support $Y \in \{ 2, 3,
\ldots, 8 \}$. In a large tree, the cost of this growing support is
non-trivial because the cost of FFT convolution is
super-linear. Moreover, in practice, the cache effects of storing
several large distributions (rather than several distributions with
trivial support such as $\{0,1\}$) can be quite pronounced.

However, the likelihood on $Y$ has support $Y \in \{1,2,3\}$;
therefore, given the observed data, the event $Y=8$, which is
entertained by the prior on $Y$ computed in the forward pass, is
impossible. We seek to ``trim'' the distributions during processing to
narrow their support so that only events in the intersection of the
prior support and likelihood support are considered. Unfortunately,
the prior support on $Y$ will only be known once the forward pass is
completed, and thus it cannot be used to avoid the unnecessarily large
convolutions (which are caused by distributions that could be trimmed,
but for which information on the intersecting support is not yet
known).

An alternative approach would be to simultaneously trim all
distributions in a layer of the convolution tree by considering the
bounding box containing their minimum and maximum supports. For
example, consider the first layer (which contains the priors on the
$X_i$). $X_4 \in \{1,2,3\}$, but for $X_4 = 3$ to be possible given
the data (and thus the likelihood on $Y$), then $X_1+X_2+X_3+X_4 \leq
3$ (using the maximum support from the likelihood on $Y$). By using
the minimum possible values of $X_1$, $X_2$, and $X_3$, this requires
$0+0+1+3 \leq 3$, which is a contradiction, meaning that $X_4=3$ is
impossible when taking into context the priors on $X_1$, $X_2$, and
$X_3$ and taking into account the likelihood on $Y$.

Performing this variable-by-variable will cost $O(n)$ per variable,
resulting in an $O(n^2)$ runtime. Alternatively, the sum of minimum
(or, w.l.o.g., maximum) supports excluding a given variable can be
computed by caching the sum of the minimum (or maximum) supports of
all variables, and then subtracting out the minimum (or maximum) of
the variable excluded. This will permit trimming the distributions in
$O(n)$ before the forward pass is even run.

Unfortunately, that strategy for trimming is not very easy to adapt to
online processing (where priors and posteriors for individual
variables are updated iteratively). This online use-case is the main
use-case of the convolution forest: the driving notion behind the
convolution forest is that iteratively computing and passing marginal
distributions will result in sparsity in the solution space, and will
yield high-quality results without resorting to the full joint
distribution. The reason it's challenging is that changing the support
of one variable (\emph{e.g.}, $X_1$) will necessarily propagate
through all internal nodes, which will have a cost $\in \Theta(n)$. In
an online setting, successively receiving messages from all $n+1$
inputs will result in a runtime that is $\in \Omega(n^2)$, and thus
loses a great deal of performance compared to the non-online setting.

\begin{figure}
\centering
\includegraphics[width=4.7in]{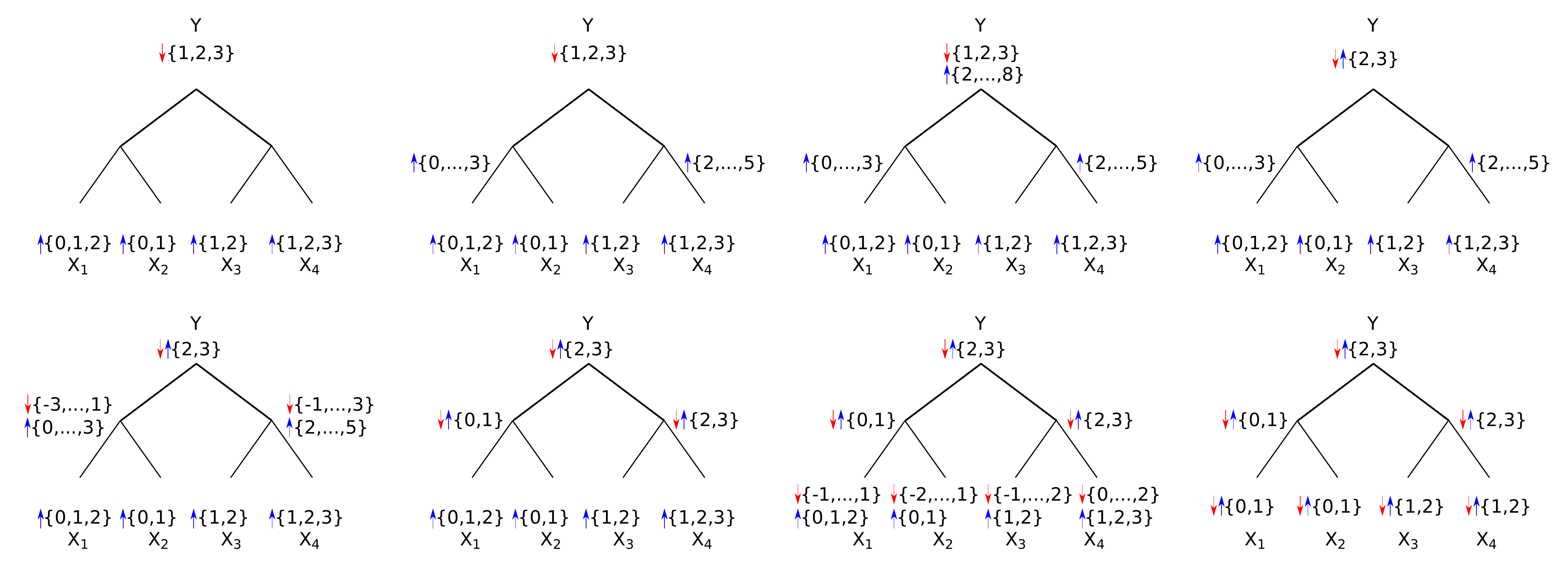}
\caption{{\bf Trimmed convolution tree.} The first forward and
  backward pass are illustrated, wherein the possible supports at each
  node are computed. Possible prior supports are labeled using blue up
  arrows and possible likelihood supports are labeled using red down
  arrows. When both supports are available, the intersection is
  labeled with both arrow types. Progressing left to right and then
  top down: {\bf 1:} a convolution tree immediately after construction
  with supports of leaves and root known, but no convolutions
  propagated, nor any supports at trimming or supports at internal
  nodes computed. {\bf 2:} The forward pass begins, computing the
  possible prior support of the second layer of the tree. {\bf 3:} The
  forward pass reaches the root node. {\bf 4:} The root node has both
  the possible prior support and the possible likelihood support
  available; the intersection is stored. {\bf 5:} The backward pass
  begins. {\bf 6:} As the backward pass progresses, internal nodes
  have both prior support and likelihood support known; the
  intersection is computed before propagating further. {\bf 7:} The
  possible likelihood supports of the inputs are now known. {\bf 8:} A
  bounding box of possible supports for each node in the tree is now
  known. At this point, convolutions would be propagated (in a forward
  pass and then a reverse pass), and each node would narrow any
  message passed through it to the intersection of the PMF at that
  node and the possible support at the node (the PMFs and the supports
  both change one another to use the narrowest possible intersecting
  support). This intersection is applied to the prior PMFs and to the
  likelihood PMFs reaching the node.
  \label{fig:trimmed-convolution-tree}}
\end{figure}

An alternative approach to narrowing the support, but which can be
easily mated with online receipt of messages, is to perform four
passes through the convolution tree: The first two passes are forward
and backward passes that compute only the support of the prior at the
given node and the support of the likelihood at the given node (and
intersect these whenever either changes). The second two passes are
forward and backward passes that compute the convolution results and
then narrow the distributions by intersecting with the supports
computed in the first two passes
(Figure~\ref{fig:trimmed-convolution-tree}). Like the other approach
(wherein the sum of all minimum supports and the sum of all maximum
supports are stored), this update strategy costs $O(n)$ when updating
all supports in a non-lazy manner; however, now that the
responsibility of keeping track of the intersecting supports is moved
to each internal node (rather than centrally keeping track of the sum
of all minimum and maximum supports), it is now relatively simple to
propagate changes in selectively, updating only when a message out is
requested.

Solving a trimmed convolution tree where each input to the sum $X_i
\in \{0, 1, \ldots k-1\}$ and the result of the sum $Y \in \{0, 1,
\ldots, k-1\}$ will cost $O(n k \log(k))$, which is substantially
faster than the $O(n k \log(n k) \log(n))$ required by the untrimmed
convolution tree. Furthermore, even when the inputs and output do not
have the same support nor the same support size, trimming can be quite
beneficial to performance. This is demonstrated in the Results
section.

By trimming the PMFs passed through the tree, the sizes of the PMFs
may be kept much smaller, permitting faster convolution. In
Figure~\ref{fig:trimmed-convolution-tree}, the forward pass of a
non-trimmed convolution tree would result in a distribution with seven
distinct support values, whereas trimming can decrease this to only
two support values that would be consistent with all prior supports
and the likelihood support. In trees where all $X_i$ and the sum $Y$
have binary support $\{0,1\}$ (\emph{i.e.}, $k=2$), the forward pass
of a non-trimmed convolution tree would cost $O(n k \log(n k) \log(n))
= O(n \log(n) \log(n))$; however, trimming prevents the state space at
internal nodes from growing, making the cost of solving all posteriors
$O(n)$ (because the state space of each node will be trimmed to
$\{0,1\}$ and therefore there will be $O(n)$ convolutions, each
costing a constant number of steps).

\subsection*{Lazy, trimmed $p$-convolution trees for online processing}

\begin{figure}
\centering
\includegraphics[width=3.5in]{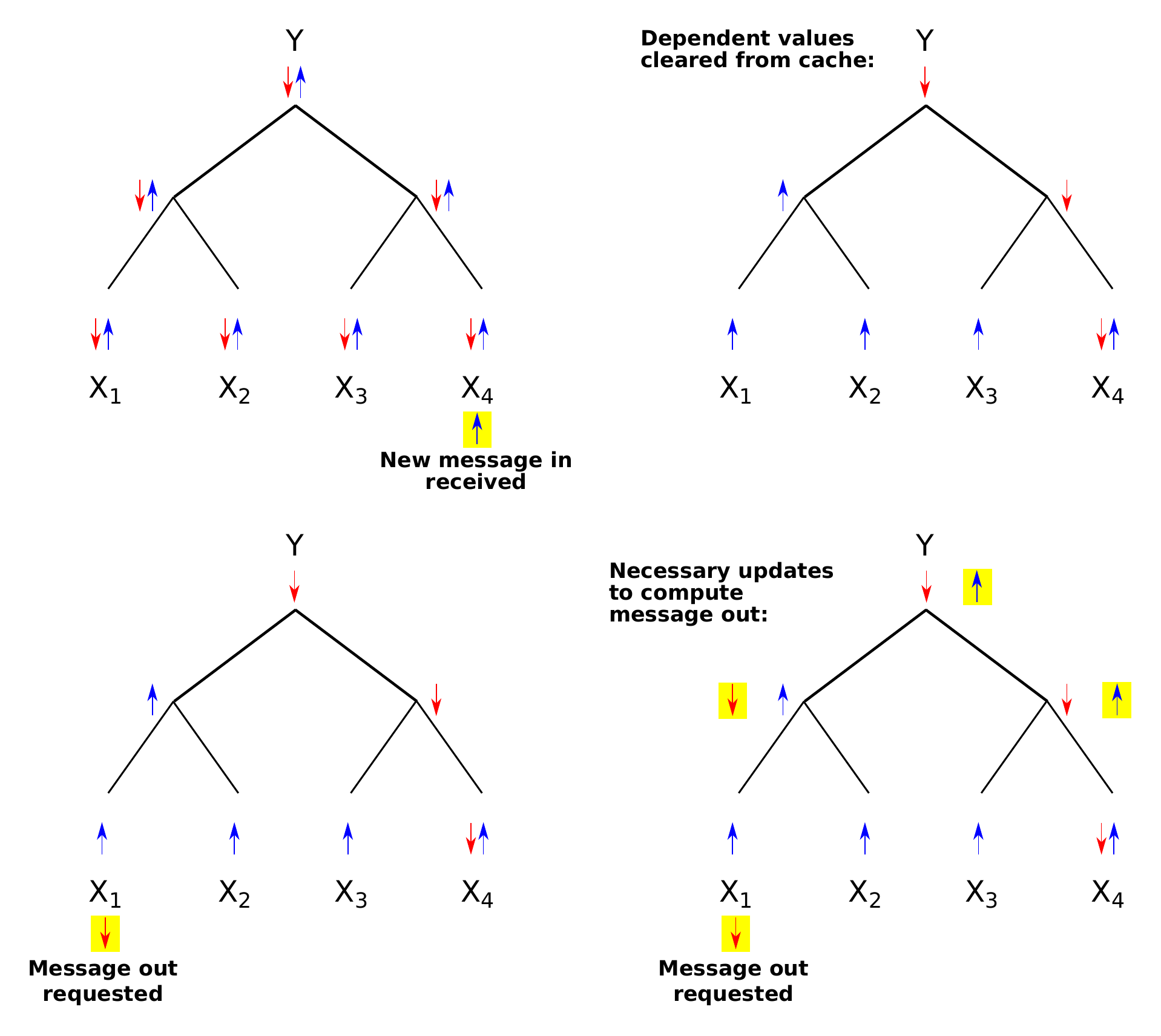}
\caption{{\bf Lazy, trimmed convolution tree.} Progressing left to
  right and top down: {\bf 1:} A convolution tree in which all
  internal nodes have computed their possible prior and likelihood
  supports as well as their prior and likelihood PMFs receives a new
  message in (an updated prior for $X_4$). {\bf 2:} Values depending
  on the prior of $X_4$ are dirtied in the cache to indicate that they
  are not current. This costs $O(n)$. But receiving a new prior on
  $X_3$ will now take only $O(1)$ steps, because the process of
  dirtying the cache can be terminated once another node with a dirty
  prior is reached. {\bf 3:} A message out (the likelihood of $X_1$)
  is requested. {\bf 4:} The nodes where either a prior or likelihood
  is requested are marked. These requests form a path for repairing
  the cache. This process does not need to visit every node; instead,
  in this case, it need only visit $O(\log(n))$ nodes.
  \label{fig:lazy-trimmed-convolution-tree}}
\end{figure}

To best enable a trimmed convolution tree to receive all relevant
support information, it is best to not compute any convolutions until
necessary (in case further information is received that will narrow
the support). For this reason, cached supports and PMFs throughout the
tree are recomputed only when a message out is reqeusted
(Figure~\ref{fig:lazy-trimmed-convolution-tree}).

The first message out will cost $\Omega(n)$ (because it must at least
touch each node in the tree). In terms of convolutions, it will
require a full forward pass and a partial backward pass along the path
from the root to the node of interest (or if prior of the root is
requested, then no backward pass is necessary, because that prior will
be computed by the foward pass). After the first message out,
subsequent messages out will be significantly faster, having many
nodes in the tree with up-to-date support and PMF
information. Likewise, after all nodes in the tree are cached (in both
directions), the first message sent into the tree will cost
$\Theta(n)$ (because it must mark one direction on all but one node as
not cached). This can be done in $\Theta(n)$ rather than $\Omega(n)$
because no convolutions will be performed (because the tree is lazy
and only performs convolutions when a message out is requested). But
after the first message received, subsequent messages received will
cost $O(\log(n))$, because they are guaranteed to reach the root in
$O(\log(n)$ steps and then reverse direction, and there is at most one
path down from the root that has not yet been marked (the path exactly
opposite the path used to dirty the first message in). 

Updating the cache as $t$ successive messages in are received will
have amortized cost $\in \widetilde{O}(1)$. Let $\phi$ be a potential
function (using ``potential'' in the context of amortized analysis,
not the context of graphical models) that counts the number of cached
supports, including both the booleans for whether a prior is cached
from below and whether a likelihood is cached from above. The runtime
required by successive messages in received at any iteration $i$ will
be a constant plus the number of caches dirtied by the message
received, or formally $r_i = O(1) + \phi_{i-1} - \phi_i$. The sum of
costs of $t$ successive messages receieved will be $\sum_{i=1}^t r_i$,
which will $= t \cdot O(1) + \phi_0 - \phi_t$ because of the
telescoping sum. $\phi_0 - \phi_t \in O(n)$ because $\phi \in O(n)$;
therefore, the cost of these $t$ operations will be $O(t) +
O(n)$. Furthermore, the $O(n)$ cost can be amortized out by including
it in the cost of constructing the tree (which costs
$\theta(n)$). Thus the cost per message received will be amortized to
$\widetilde{O}(1)$.

Identical reasoning (but where $\phi$ represents the count of nodes
that are not cached rather than the number cached) can be used to show
that the cost of updating the cache when $t$ successive messages out
are requested will likewise be $\in \widetilde{O}(1)$. Alternately
sending and receiving messages out is more complicated and would merit
further investigation on its own. The balanced construction of
probabilistic convolution trees \cite{serang:probabilistic} means that
the longest non-cyclic path between any nodes in the tree will be $\in
O(\log(n))$, which would likely benefit the worst-case amortized or
average analysis. In practical application, the method of caching the
trimmed support sizes is demonstrated to perform quite well in the
Results section.

In addition to the faster runtimes, trimmed convolution trees have the
added benefit of greater accuracy. One reason for this is because
shorter FFT convolutions (which are used when $p=1$ and which are used
as part of numeric $p$-convolution when $p>1$) grow slightly less
accurate as the size of the tensors grows \cite{pfeuffer:bounded}. But
also, this is because the implementation of $p$-convolution relies on
the numeric approach (via FFTs) on long tensors, but the naive
approach (especailly when implemented in TRIOT \cite{heyl:triot}) is
faster on small problems and also achieves the exact result (rather
than a numeric approximation).

\subsection*{Message passing dynamics and hyperedge connectivity}

\begin{figure}
\centering
\includegraphics[width=3.5in]{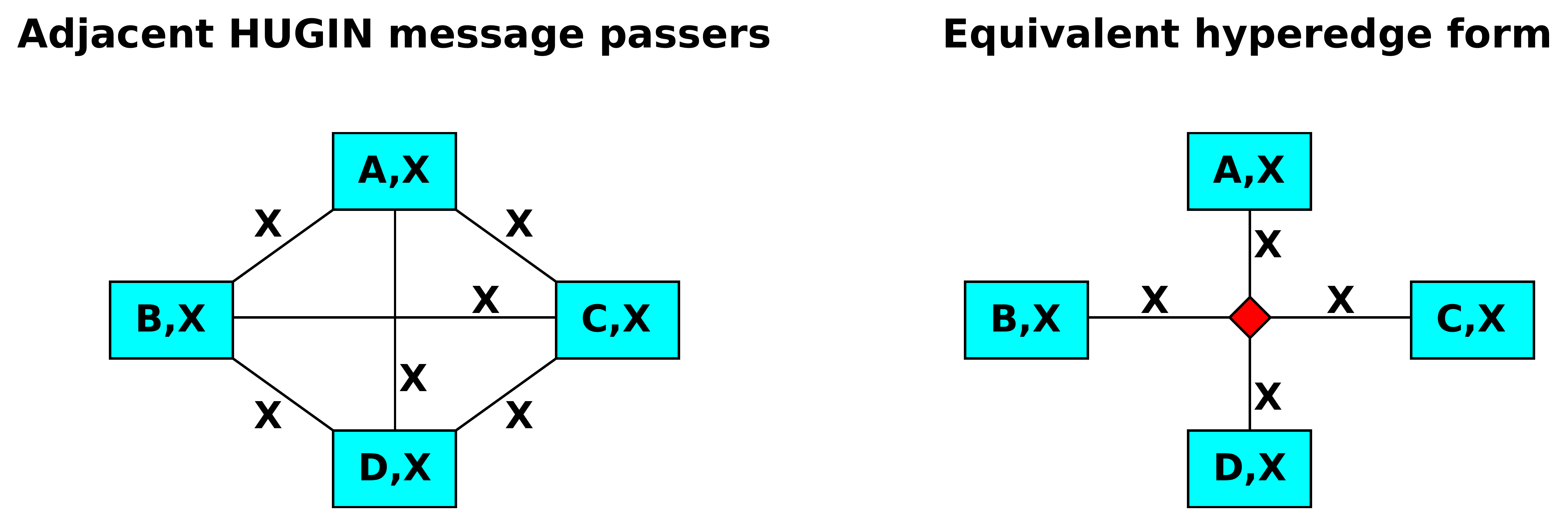}
\caption{{\bf Adjacent HUGIN nodes and their equivalent form via a
    hyperedge.} Four HUGIN message passers, all which share the
  variable $X$, are connected directly with edges containing the
  variable $X$. The hyperedge form is equivalent but with far fewer
  edges.
  \label{fig:hyperedge}}
\end{figure}

The core {\tt MessagePasser} types in the {\tt EvergreenForest}
implementation include HUGIN nodes \cite{andersen:hugin} (which may be
constructed with prior joint distributions of arbitrary dimension and
cache products of distributions on messages in in order to prevent
recomputation), $p$-convolution tree nodes (which are trimmed and
cached for online processing), and hyperedge nodes. Although message
passers may be connected directly or via HUGIN nodes, hyperedges
provide a means by which cliques can be represented in $o(n^2)$
edges. This is key to achieving a subquadratic runtime in cases where
several message passers are connected via the same variables (which
will happen in Bethe graphs, for example). Figure~\ref{fig:hyperedge} depicts
HUGIN nodes with prior distributions sharing a common variable and one
means of connecting them as opposed to the simpler and more efficient
hyperedge form.

In all other message passers (\emph{e.g.}, HUGIN and convolution tree
message passers), a directed edge $e$ is elligible to send a message
out when the message passer has received messages in along all other
directed edges in. That is, receiving a message along the edge
$\mbox{rev}(e)$ is unnecessary to send a message out along $e$. There
are two exceptions to this: The first exception occurs when message
passers are elligible to pass along an edge \emph{ab initio} (indeed,
some message passer will need to pass first, and none may be able if
they are all awaiting messages to be received). An example of this is
the HUGIN message passer, which is able to pass along an edge $e$
\emph{ab initio} if the HUGIN node has a prior distribution that is a
superset of all variables along $e$. If a user is writing custom
message passers, the ability to modify these default behaviors are
provided in the functions {\tt virtual bool
  MessagePasser::ready\_to\_send\_message} and {\tt virtual bool
  MessagePasser::ready\_to\_send\_message\_ab\_initio}, which can be
overridden in derived classes that declare new message passer types.

The second exception to the message passing dynamics is for
hyperedges. Hyperedges are essentially a shorthand for several direct
edges. For this reason, a hyperedge does not need to receive messages
on all incoming directed edges except $\mbox{ref}(e)$; instead,
hyperedges should be elligible to pass messages as soon as they have
received any messages in. In {\tt EvergreenForest}, this is
implemented in a more general manner, which allows construction of
hyperdge types even in the case where all edges do not carry the same
variable sets. The {\tt Hyperedge} class is a descendent of the {\tt
  MessagePasser} class and by overriding the {\tt
  ready\_to\_send\_message} function, {\tt Hyperedge} is elligible to
pass message out along edge $e$ when the messages it has received are
a superset of the variables along edge $e$.

Hyperedges produce greater efficiency for two reasons: The first of
these reasons (which is mentioned above) is the ability to represent
an $n$-clique in $O(k)$ edges (rather than the $O(k^2)$ needed by
direct connections). The second reason that hyperedges produce greater
efficiency is because of the ability to cache products in a manner
reminiscent of HUGIN message passers: In the left panel of
Figure~\ref{fig:hyperedge}, the HUGIN node with variables $(A,X)$ will
receive messages from the $(B,X)$, $(C,X)$, and $(D,X)$ nodes, and
will need to mulitply the product of those messages into
itself. Likewise, the node $B,X$ will receive a product of messages
over the $(A,X)$, $(C,X)$, and $(D,X)$ nodes. Rather than compute $n$
individual products (each over $n-1$ messages) in a time of
$\Omega(n^2)$, the hyperedge caches the full product of all messages
and then divides out the message that should be left out to send out
along a particular edge. This permits the runtime to be subquadratic
in the clique size $n$. Because of the similarity to HUGIN message
passers in the way that the product of these messages are cached, the
{\tt Hyperedge} class inherits from {\tt HUGINMessagePasser}, which in
turn inherits from {\tt MessagePasser}.

Rather than determine whether a message passer is elligibile to pass
by checking whether each of the $n-1$ relavent edges in have received
messages, elligibility to pass is performed by first checking the
count of unique edges in that have received messages (which will be in
$\{0, 1, \ldots, n\}$). If the count is $n$, then all messages in have
been received and so any edge out is elligible to pass. If the count
is $<n-1$, then no edges out are elligible to pass. If the count is
$n-1$, then the edge $e$ is elligible to pass if and only if the edge
in $\mbox{rev}(e)$ has not been received (implying that the other
$n-1$ edges are responsible for the count of $n-1$, and so all other
edges in have received messages). This is crucial to unlocking the
full subquadratic capability of {\tt ConvolutionTreeMessagePasser}
types, because otherwise simply querying which edges are elligible to
pass messages would cost $\Omega(n^2)$.

A similar caching strategy is employed when determining whether {\tt
  Hyperedge} types are elligible to pass a message out along a
particular edge. There, the superset computation is cached as a
boolean (so that it is not computed multiple times) and the {\tt
  Hyperedge} instance is marked when all edges out have been marked as
ready to pass. In the case where all edges incident to a hyperedge
contain the same variables (such as in Bethe graphs, where the edges
incident to a hyperedge contain only one variable, which is identical
between all of those incident edges), this guarantees amortized
$\widetilde{O}(1)$ number of subset queries.

Although all additive dependencies could be encoded as compositions of
three variables of the form $Y=X_1+X_2$ (so $Z=X_1+X_2+X_3$ would be
encoded as $Y=X_1+X_2$ and $Z=Y_1+X_3$ by introducing a dummy variable
$Y$), the {\tt ConvolutionTreeMessagePasser} type are used to directly
encode additive dependencies with an arbitrary number of random
variables $Z=X_1+X_2+\cdots+X_n$. This is not only more appealing from
the perspective of software engineering and usability (\emph{i.e.},
not needing to declare so many dummy variables), it also has important
effects on performance. This modular solution is important for
``trimming'' the convolution trees (described below). It also eases
the burden on the scheduler: An implementation using binary additions
for all additive dependencies would have far more message passers, and
therefore, there would be more candidate edges that could pass
messages during inference. These edges will be multiplexed by the
scheduler to determine the next message to be passed.

\subsection*{Object oriented scheduling}

Messages between {\tt MessagePasser} types may be sent and received
manually, but they may also be automatically handled via {\tt
  Scheduler} types. {\tt EvergreenForest} implements a small number of
schedulers, and like the {\tt MessagePasser} types, users can create
their own custom schedulers by inheriting and overriding the necessary
{\tt virtual} functions.

The included schedulers have complementary strengths and weaknesses
and are suited to different applications. The {\tt FIFOScheduler}
stores edges elligible to pass in a FIFO queue. In each iteration, the
front elligible edge is dequeued, the message out along that edge is
computed (by requesting it from source message passer), and the
message is received by the destination message passer for that
edge. Then, any edges coming out from the destination message passer
that are now elligible to pass and are also not in the queue are
passed. For greater performance, edges's membership in the queue is
implemented via a {\tt bool} belonging to the {\tt Edge} type (via
inheritance from the {\tt Queueable} mixin), rather than by using {\tt
  std::set<Edge*>}; this enables graphs that have many edges but which
should be solveable in linear time (such as HMMs) to not have runtime
$\in \Omega(n \log(n))$. Also important is the lazy message
computation of this scheduler: edges are enqueued into the scheduler
before their messages have been computed. This enables a message along
edge $e$ to be computed at the last possible moment, which can permit
greater trimming by {\tt ConvolutionTreeMessagePasser} types (which
may have receied new messages in the time since $e$ was first
enqueued). This scheduler is simple and lightweight, and is well
suited to small to moderatly sized loopy graphs with loops.

The {\tt PriorityScheduler} type keeps track of the the deviation (via
mean squared error) of the last message passed along the edge and the
current message passed along the edge. Edges are visited in the order
of most changed edges first, which prevents cycling. This can benefit
performance in the case where after some iterations of message
passing, part of the graph has converged, while another part of the
graph has not yet converged; the {\tt PriorityScheduler} type will
spend greater time on the not yet converged parts of the graph,
largely ignoring the regions of the graph that have come close to
convergence \cite{koller:probabilistic}. Although the priority
scheduler benefits from focusing on the regions of the graph that are
least converged, it does so at a cost: First, a heap is needed to
store the edges, which introduces a logarithmic cost to graphs that
are tree-like (and which could otherwise be solved in $O(n)$;
\emph{e.g.}, HMMs). Second, computing the priority of an edge requires
computing the new message (to compare with the old message along the
edge), which means that trimming in {\tt ConvolutionTreeMessagePasser}
types may be less effective. For example, when used with the {\tt
  PriorityScheduler} a {\tt ConvolutionTreeMessagePasser} type will
compute a message out as soon as $n-1$ messages have been received (in
order to enqueue the now elligible edge out into the {\tt
  PriorityScheduler}); this means that trimming cannot be performed
for this first message out, because not all messages will have been
received.

Lastly, the {\tt RandomSubtreeScheduler} is well suited to tree-like
graphs, and is another scheduling heuristic mentioned by Koller \&
Friedman \cite{koller:probabilistic}. At construction, it computes two
subtrees of the graph (via a random depth-first search), and then
iteratively passes messages along one full tree, and then along the
other full tree. When visited during the tree traversal, each {\tt
  MessagePaser*} type (which constitute the nodes of the tree) will
pass messages along every elligible edge out. By using multiple random
subtrees, even if the graph does not resemble a tree, information may
be passed efficiently. For example, in an Ising grid, a popular
scheduling heuristic is to pass among all rows and then pass among all
columns \cite{tarlow2012fast}. This subtree approach uses a
qualitatively similar approach, but generalized for arbitrary
graphs. A superior approach may build several such random trees at
construction (thereby ensuring a greater chance that highly different,
complementary subtrees are found).

Message passing in {\tt EvergreenForest} supports dampening, where the
older message along an edge is mixed with the new message to produce a
message that is passed \cite{koller:probabilistic}. This can be used to
improve convergence in graphs with many loops (it is qualitatively
reminiscent of a momentum term in neural network
backpropagation). Dampening can be performed manually (if passing
messages manually), and is alternatively built into the schedulers, so
that all messages passed will be dampened.

\section*{Results}
All benchmark results are compiled with {\tt g++} version 6.3.1 and
using the compiler options {\tt -std=c++11 -O3 -march=native
  -mtune=native}. All benchmarks were run on an Intel {\tt i7} running
Fedora and with 8GB of RAM.

\subsection*{Complex FFT benchmarks}

\begin{figure}
\centering
\includegraphics[width=3.5in]{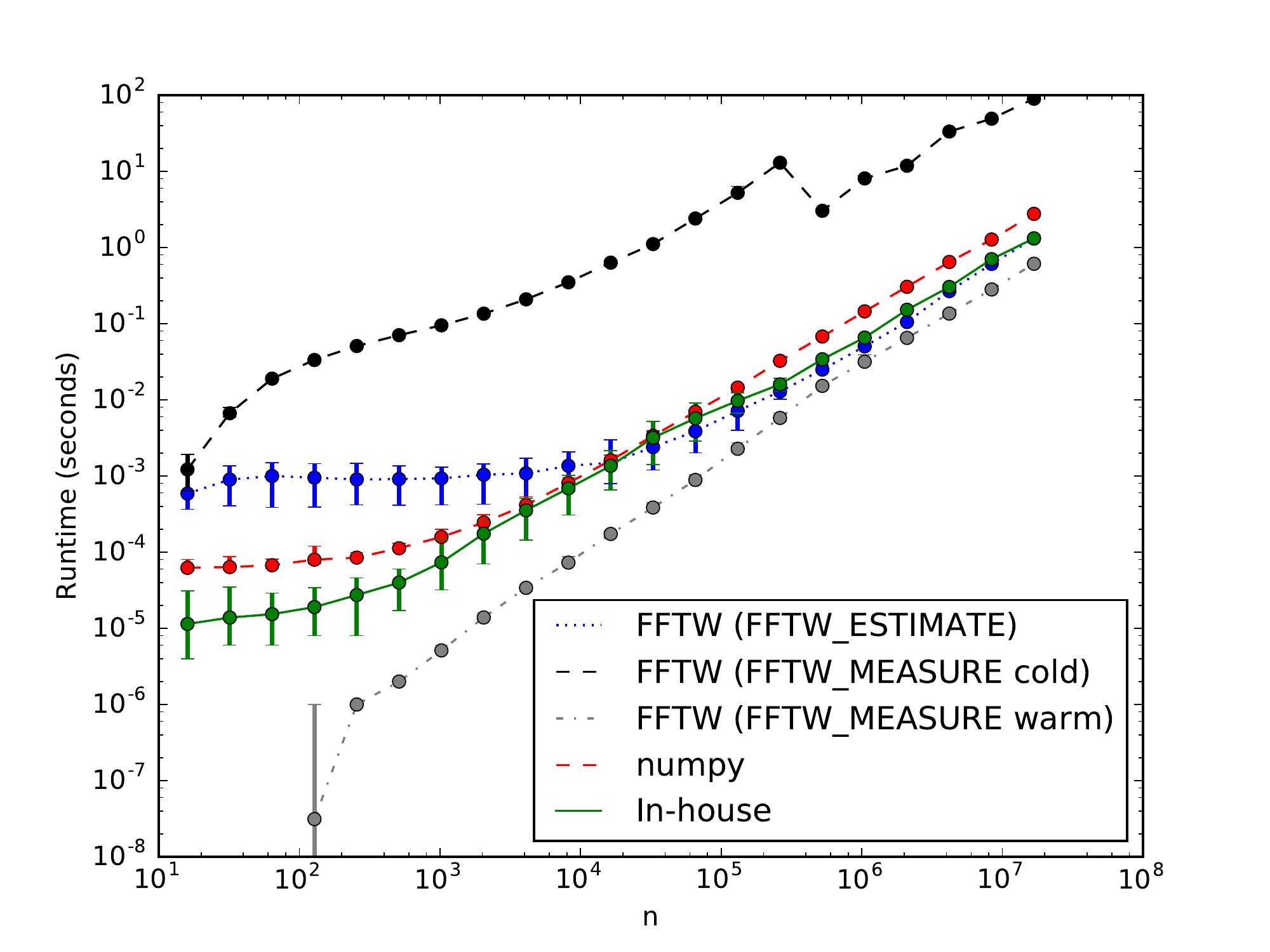}
\caption{{\bf FFT benchmarks.} Warm-start {\tt FFTW\_PLAN} is the most
  efficient, but cold-start {\tt FFTW\_PLAN} is much less
  efficient. {\tt numpy.fft}, {\tt FFTW\_ESTIMATE}, and the in-house
  implementation all perform similarly to one another. Error bars show
  the minimum and maximum runtimes over 32 replicate trials.
  \label{fig:fft-benchmarks}}
\end{figure}

The complex 1D FFT was benchmarked using vectors of different sizes
(Figure~\ref{fig:fft-benchmarks}). The in-house FFT library used by
{\tt EvergreenForest} was compared to {\tt numpy.fft} in {\tt python}
and {\tt FFTW} \cite{frigo:fast} version 3 ({\tt FFTW} programs were
compiled in {\tt C++ using {\tt g++} the same compiler options}.

FFTW includes multiple modes: {\tt FFTW\_ESTIMATE} is the most
lightweight, and has very low overhead for just-in-time JIT
compilation. {\tt FFTW\_PLAN}, on the other hand, trades more time
spent on optimization of JIT code and consequently less time spent on
FFT computation. For this reason, {\tt FFTW} was benchmarked using
both options, and the {\tt FFTW\_PLAN} option was benchmarked from a
cold start (including the time for JIT compilation to produce the FFT
``plan'') and warm start (not including the time for JIT
compilation). Broadly speaking, {\tt FFTW\_ESTIMATE} is the best fit
for the use-case of convolution of large vectors of arbitrary length
on the fly, because the runtime to produce the higher-quality FFT plan
is substantial and will be wasted unless that plan can be cached and
reused several times. Storing plans or ``wisdom'' for future use may
require a substantial amount of storage when the length of the FFTs
approaches the total amount of RAM available, and therefore may be
necessary to save on disk. Furthermore, {\tt FFTW} plans are
associated with a particular block of memory, and so the space for
these buffers is essentially married to the plans. For these reason,
{\tt FFTW\_PLAN} is by far a better fit for the use-case where many
FFTs of the same length are used in succession.

While the sophisticate JIT compiler powering {\tt FFTW} is complex and
thus has a fairly complex interface, the in-house complex FFT module
can be applied in-place by simply running {\tt apply\_fft<DIF, true,
  true>(x)}, where {\tt x} is of type {\tt Tensor<cpx>}. The two
boolean template arguments in this example specify that shuffling must
be performed and transpositions must be undone.

\subsection*{1D complex convolution benchmarks}

\begin{figure}
\centering
\includegraphics[width=3.5in]{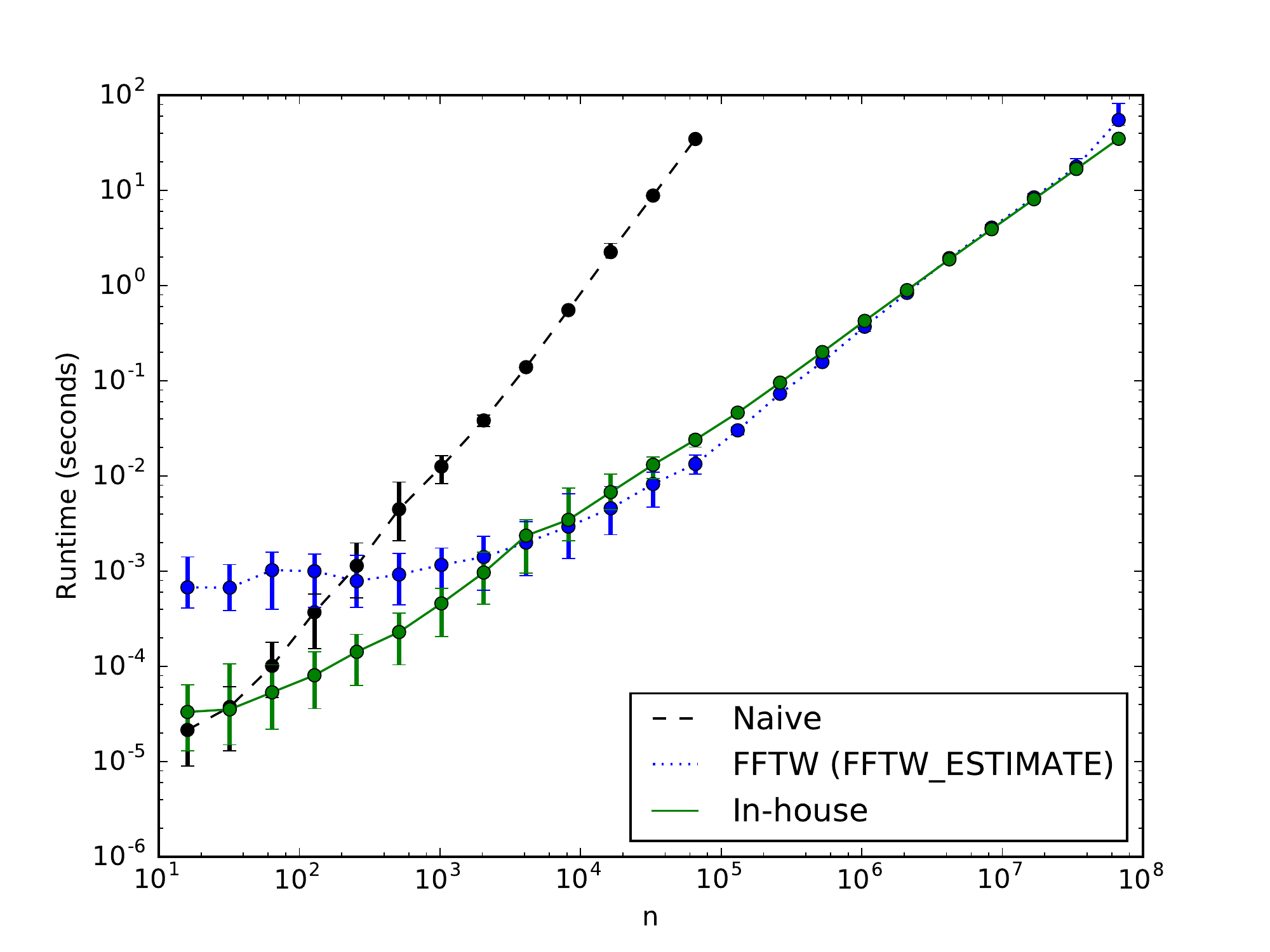}
\caption{{\bf Complex convolution benchmarks.} Naive complex
  convolution was compared to {\tt FFTW} (with {\tt FFTW\_ESTIMATE})
  and to the in-house convolution module. The in-house complex
  convolution performs better for small problems, slightly worse for
  moderately sized problems, and then begins performing slightly
  better for very large problems. Error bars show the minimum and
  maximum runtimes over 32 replicate trials.
  \label{fig:complex-convolution-benchmarks}}
\end{figure}

1D complex convolution runtimes were compared using a naive tensor
convolution implementation (via {\tt TRIOT}), an {\tt FFTW}
implementation (with the most relevant {\tt FFTW\_ESTIMATE} option),
and complex convolution with the in-house FFT package
(Figure~\ref{fig:complex-convolution-benchmarks}).

The {\tt FFTW} code was optimized by reusing the {\tt FFTW\_ESTIMATE}
plan for both forward FFTs and the inverse FFTs by using the property
that the $FFT^{-1}(x) = \mbox{conj}(FFT(\mbox{conj}(x))) / n$ and by
inlining these convolutions into subsequent loops where possible.

\subsection*{Numeric $p$-convolution benchmarks}

\begin{figure}
\centering
\includegraphics[width=3.5in]{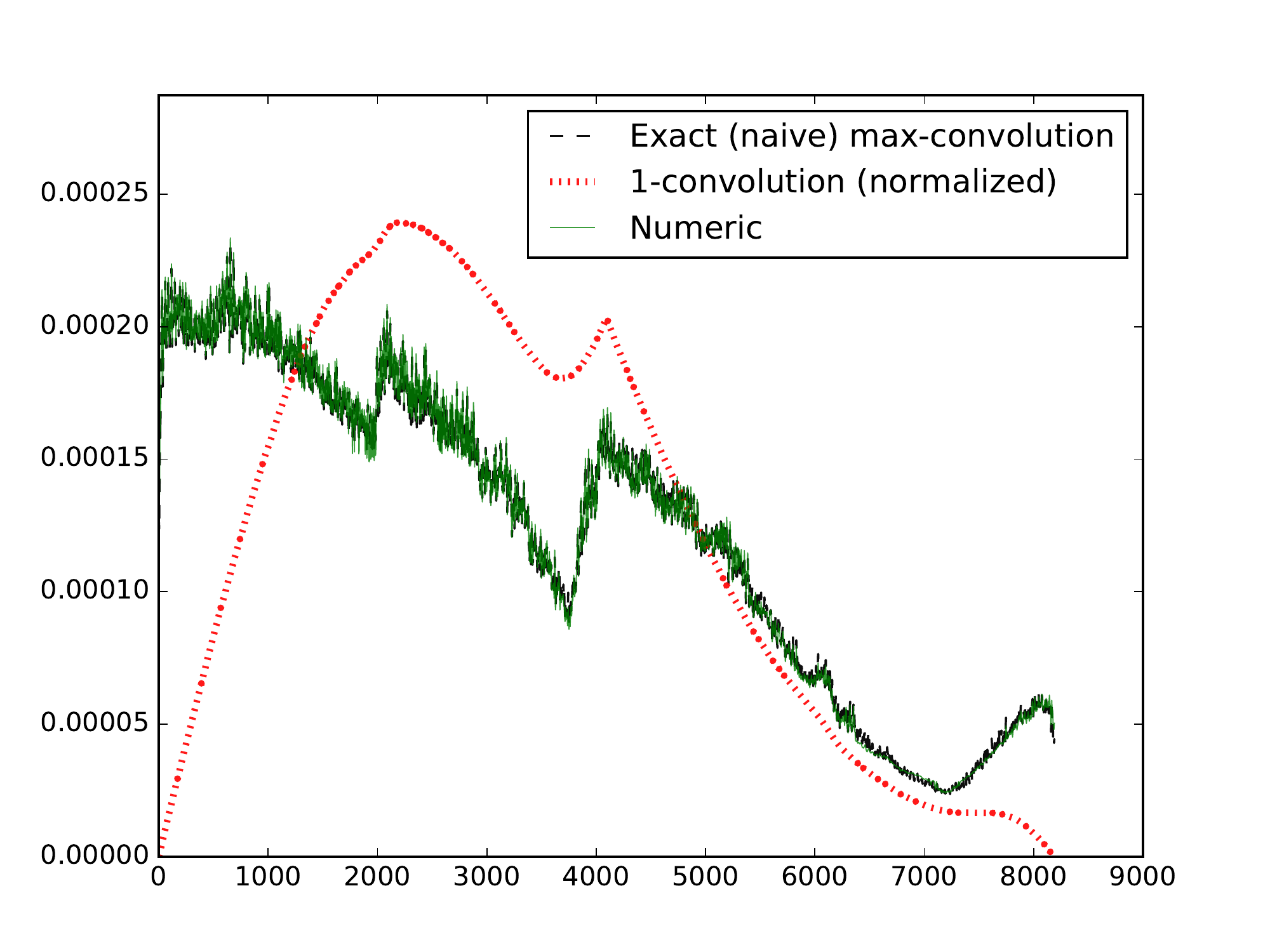}
\caption{{\bf Exact vs. numeric max-convolution.} Exact
  max-convolution is compared with numeric max-convolution on a
  problem with inputs of size 4096. The numeric produces a highly
  similar result to the exact, naive method, but in a subquadratic
  runtime. Both of these methods yield starkly different results
  compared to using standard FFT convolution (\emph{i.e.}, relaxing to
  sum-product inference, or $p=1$), which is labeled as
  ``1-convolution (normalized)''. The un-normalized 1-convolution
  result would have y-values much larger than max-convolution (because
  standard convolution sums all contributions to each index while
  max-convolution counts only the maximum).
  \label{fig:p-convolution}}
\end{figure}

\begin{figure}
\centering
\includegraphics[width=3.5in]{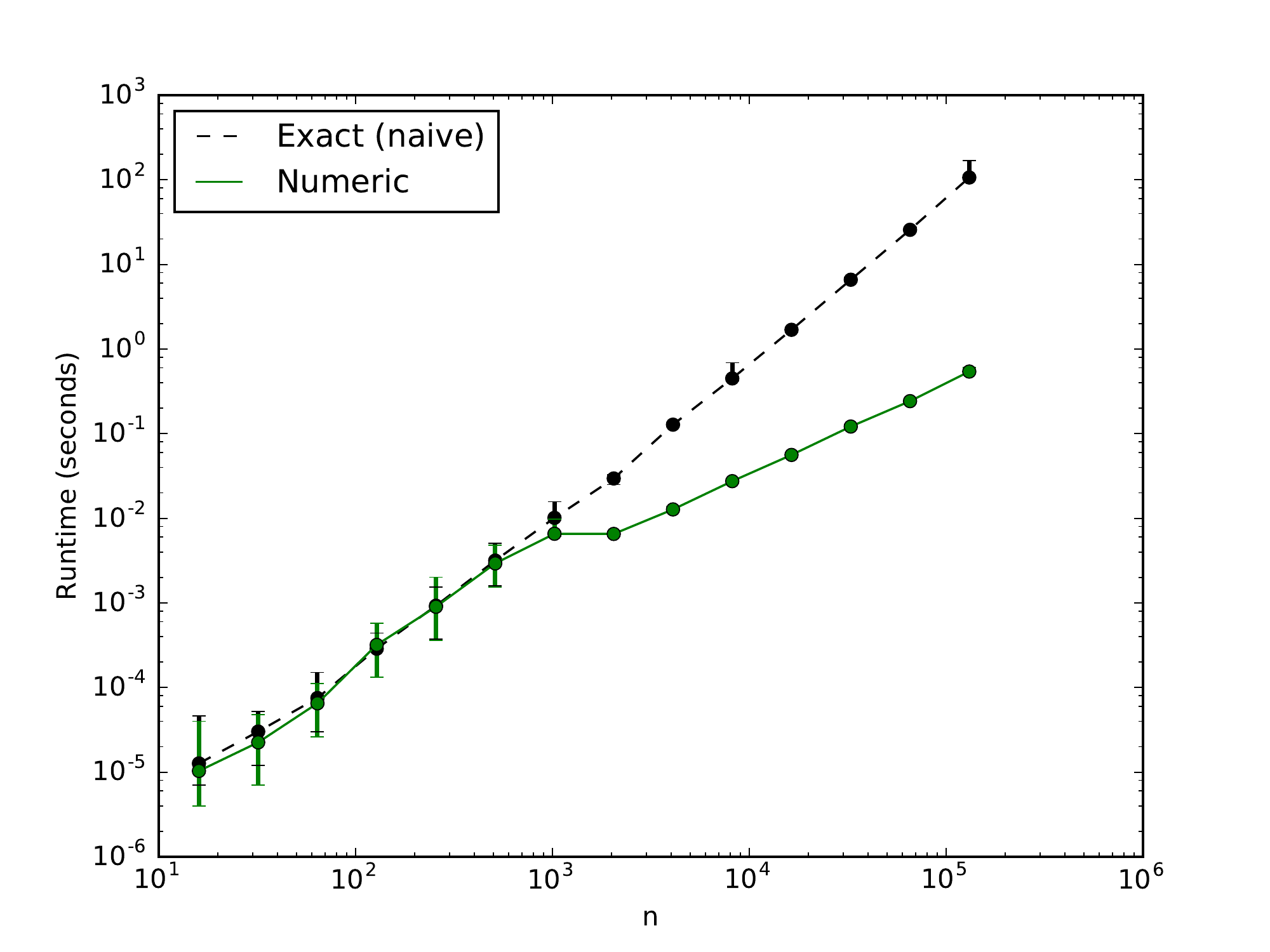}
\caption{{\bf Exact vs. numeric max-convolution benchmarks.} The
  runtime of exact, naive max-convolution is compared to the runtime
  of numeric max-convolution via the $p$-convolution with
  $p=\infty$. The numeric approach becomes substantially faster as
  problem sizes grow moderate to large. Error bars show the minimum
  and maximum runtimes over 32 replicate trials.
  \label{fig:p-convolution-benchmarks}}
\end{figure}

Numeric $p$-convolution was compared with $p=\infty$ (\emph{i.g.},
max-convolution) against exact, naive convolution using the data from
Figure~\ref{fig:sum-vs-max}. The numeric $p$-convolution result is
highly similar to the exact, naive result
(Figure~\ref{fig:p-convolution}). Although the numeric $p$-convolution
produces high-quality results, it does so with a substantially faster
runtime for large problems (Figure~\ref{fig:p-convolution-benchmarks}). 

\subsection*{Computing posteriors when $Y = X_1 + X_2 + \cdots + X_n$, with $Y \in \{0,1\}$ and $X_i \in \{0,1\}$}

\begin{figure}
\centering
\includegraphics[width=3.5in]{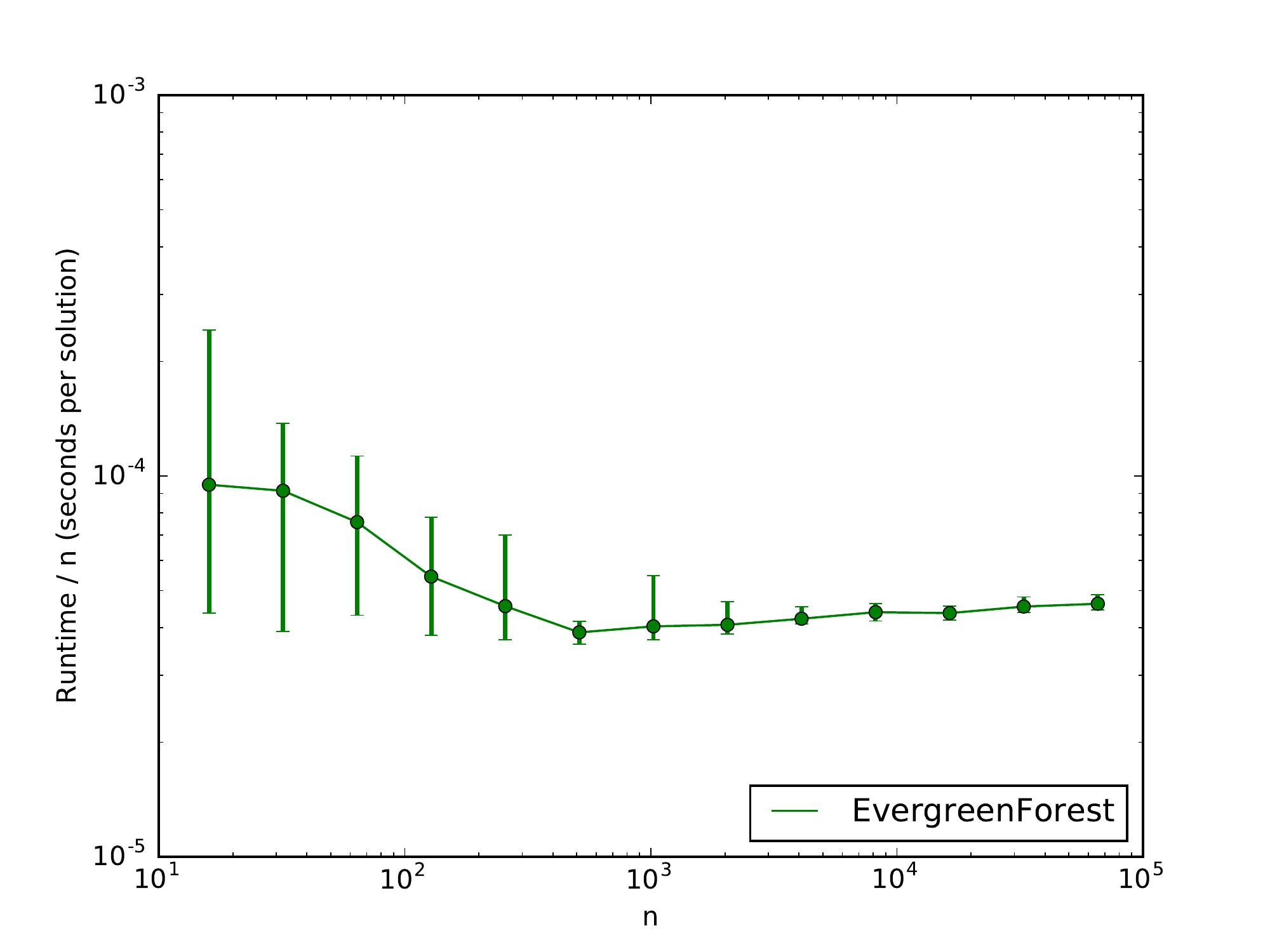}
\caption{{\bf Runtime per posterior computed.} A model with all $X_i
  \in \{0,1\}$ and $Y \in \{0,1\}$ was solved using {\tt
    EvergreenForest}, and runtime per posterior computed is plotted as
  a function of $n$, the number of $X_i$ in the model. After $n \geq
  256$ or so, the runtime is no longer dominated by overhead
  (\emph{e.g.}, problem construction, \emph{etc.}), and the runtime
  per posterior distribution computed is roughly flat. A very slight
  upward slope is expected due to caching effects, which produce a
  nonlinear slowdown as more memory is used by a program.
  \label{fig:binary-convolution-tree-runtimes}}
\end{figure}

A model with a single additive dependency was created and solved using
{\tt EvergreenForest}. The model was solved using {\tt
  FIFOScheduler}. The model had random priors in $\{0,1\}$ for each
$X_i$ and a random likelihood in $\{0,1\}$ for $Y$. All posteriors
were computed (for all $X_i$ and for $Y$). The lazy, trimmed
$p$-convolution trees achieve nearly $\widetilde{O}(1)$ amortized
performance per posterior retrieved, as indicated by the nearly
constant runtime of the slope in
Figure~\ref{fig:binary-convolution-tree-runtimes}.

\subsection*{Restaurant bill illustration}

The example using an additive model to describe the total restaurant
bill is illustrated. Ice cream prices from Big Dipper were retrieved
from \url{http://bigdippericecream.com} on May 28, 2017. Prices are
all divisible by $\$0.25$, and so the menu was discretized into
$\$0.25$ increments. A collection of $n$ individual possible order
preferences (as random distributions of preferences among possible
items selected at random from the menu), and each is used as a prior
$X_i$ for one of $n$ people ordering. A random order is generated by
sampling independently from each distribution $\pmf_{X_i}$, and a
likelihood on $Y=X_1 + X_2 + \cdots + X_n$ is chosen as a Kronecker
delta with 100\% of its mass at the total value of the order. Using
the total order, posteriors are computed among each of the
customers. That is, the posterior distribution on each person's order
is found conditional on the prior ordering preference from them, the
prior ordering preference from the other $n-1$ individuals, and the
total restaurant bill. The model was solved using {\tt FIFOScheduler}.

\begin{figure}
\centering
\includegraphics[width=3.5in]{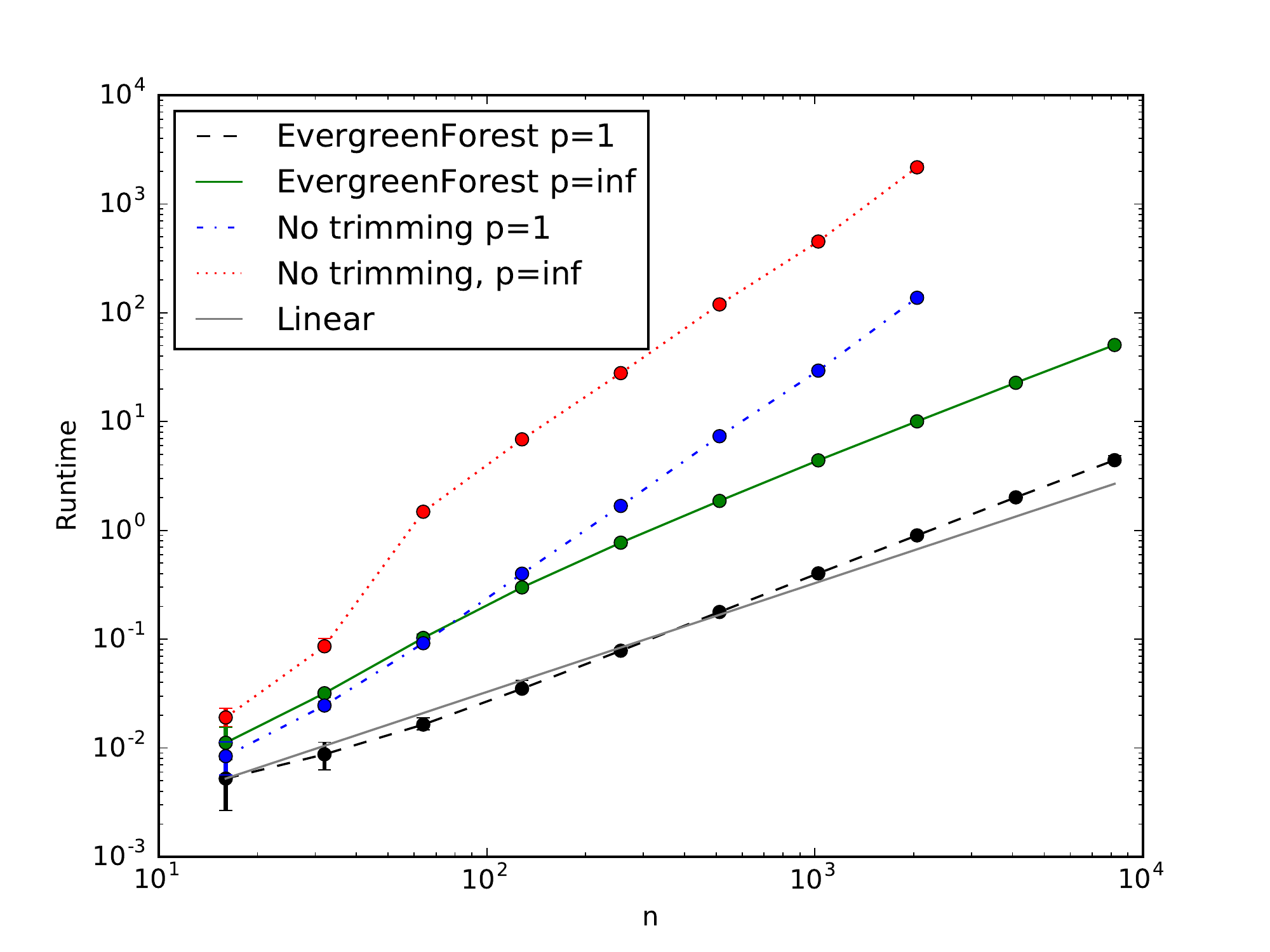}
\caption{{\bf Total runtime to solve all posteriors on restaurant
    bill.} The runtimes on several problems of different size $n$ are
  compared, where $n$ people order from a restaurant menu according to
  their individual preferences. The slope of each series on this
  log-log plot determines its asymptotic runtime, and a linear runtime
  slope is plotted for reference. Trimming (which is enabled by
  default and is only disabled for illustration here where
  specifically stated) contributes a significant speedup. Furthermore,
  the trimmed solutions with $p=1$ and $p=\infty$ have just runtimes
  that are asymptotically only slightly superlinear in $n$. Large
  problems can be solved and the choice of max-product inference only
  produces a constant slowdown compared to sum-product inference (as
  shown by the asymptotically non-widening gap between the $p=1$ and
  $p=\infty$ runtime series); naive max-product inference would be an
  order of magnitude slower than this number max-product
  implementation, as the naive method uses a quadratic algorithm.
  \label{fig:restaurant-runtimes}}
\end{figure}

This problem is solved multiple times with and without trimming
enabled in the convolution trees and with $p=1$ (sum-product
inference) and $p=\infty$ (max-product inference). Runtimes for
problems of different sizes $n$ are plotted in
Figure~\ref{fig:restaurant-runtimes}. Even though the problem is not
trivial to trim (because the sub-total bills must first pass the known
total restaurant bill before trimming can be guaranteed), with
trimming it is solved in only slightly superlinear time. This is true
regardless of which $p$ is used ($p=1$ is only a constant time faster
than $p=\infty$).

\subsection*{Applied results}

\subsubsection*{GC-rich HMM}

\begin{figure}
\centering
\includegraphics[width=3.5in]{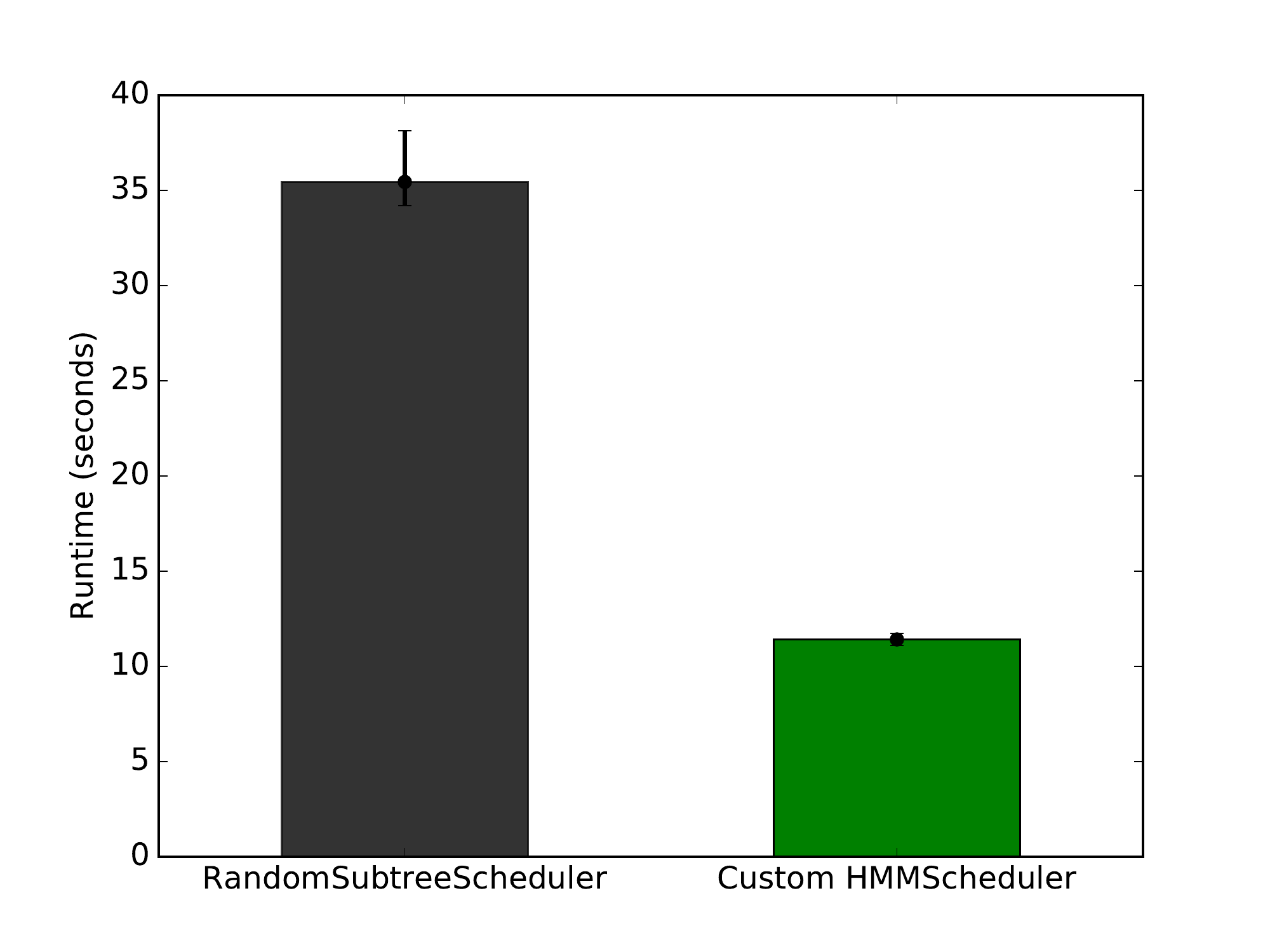}
\caption{{\bf Runtimes for solving a two-state HMM with length
    126697.} The runtimes to compute max-marginal posterior
  distributions on an HMM for GC-enrichment are shown using {\tt
    RandomSubtreeScheduler} and a small, custom scheduler for HMM
  graphs.
  \label{fig:gc-hmm-runtimes}}
\end{figure}

A two-state HMM for classifying DNA bases into GC-rich and non-GC-rich
states (which has been demonstrated to find noncoding RNA genes in
hyperthermophiles  \cite{klein:noncoding}) is written in {\tt
  EvergreenForest} as a manually constructed graph. All posteriors on
the Shigella boydii genome (126697 base pairs) are computed by solving
with $p=\infty$. The {\tt RandomSubtreeScheduler} computes all
posteriors in roughly 35 seconds, while a custom made HMM scheduler
performs the same task in just over 11 seconds
(Figure~\ref{fig:gc-hmm-runtimes}).

No additive dependencies are used (the principal focus of convolution
forests), but this demonstrates that the core algorithms in {\tt
  EvergreenForest} are robust and scale well even when users are
interfacing through PMFs indexed by {\tt std::string} (\emph{i.e.},
the {\tt LabeledPMF<std::string>} type). In contrast, a hand-made
Viterbi path (imlemented in {\tt C++}) computed on this problem runs
in roughly one second; however, this from-scratch Viterbi path (not
using {\tt EvergreenForest} gives only the point estimate of the MAP
path, while {\tt EvergreenForest} computes posterior distributions for
each variable according to the max-product marginals. As such, a
from-scratch implementation allocates and writes to significantly less
memory. As stated above, max-product marginal distributions can be
used to compute the MAP, but the converse is not necessarily true.

The forward-backward posteriors could be computed by simply switching
to $p=1$ throughout the entire model. Achieving a high-level interface
without straying far from the performance of hard-coded
implementations on problems like this (which are dominated by their
runtime constants) makes a strong case for the implementation as an
engine for prototyping models. Furthermore, the result distributions
are indexed by variable names, which can make tasks like implementing
expectation maximization training substantially easier for
non-experts.

The custom HMM scheduler is faster because the {\tt
  RandomSubtreeScheduler} type visits every outgoing edge on each node
visited, whereas a custom HMM scheduler need only visit each directed
edge once (the degree of latent variable nodes in this HMM will be
three-- an edge to the previous base pair, an edge to the next latet
base pair, and an edge to the observed DNA base pair from the genom--
and hence, a $\approx 3\times$ speedup is achieved by a custom
scheduler).

\subsubsection*{Loopy graphs of convolution trees and 2D convolution trees}

The implementation presented here solves multidimensional convolution
trees as well. Where 1D convolution trees accept priors on $B$, $C$,
$D \ldots$ and the likelihood on $A=B+C+D+\cdots$, multidimensional
convolution trees take joint priors on $(B,W)$, $(C,X)$, $(D,Y),
\ldots$ and the likelihood on $(A,V) = (B,W) + (C,X) +(D,Y) +
\cdots$. This means that $A=B+C+D+\cdots$ and $V=W+X+Y+\cdots$, but
with the covariance of the joint distributions properly respected to
achieve an exact result (via multidimensional $p$-convolution). The
principle extends to convolution trees of arbitrary dimension (even
when the dimension is not known at compile time).

\begin{figure}
\centering
\begin{tabular}{m{0.1in}m{2.2in}}
  \begin{sideways}Bethe 1D (loopy)\end{sideways} & \includegraphics[width=2.2in]{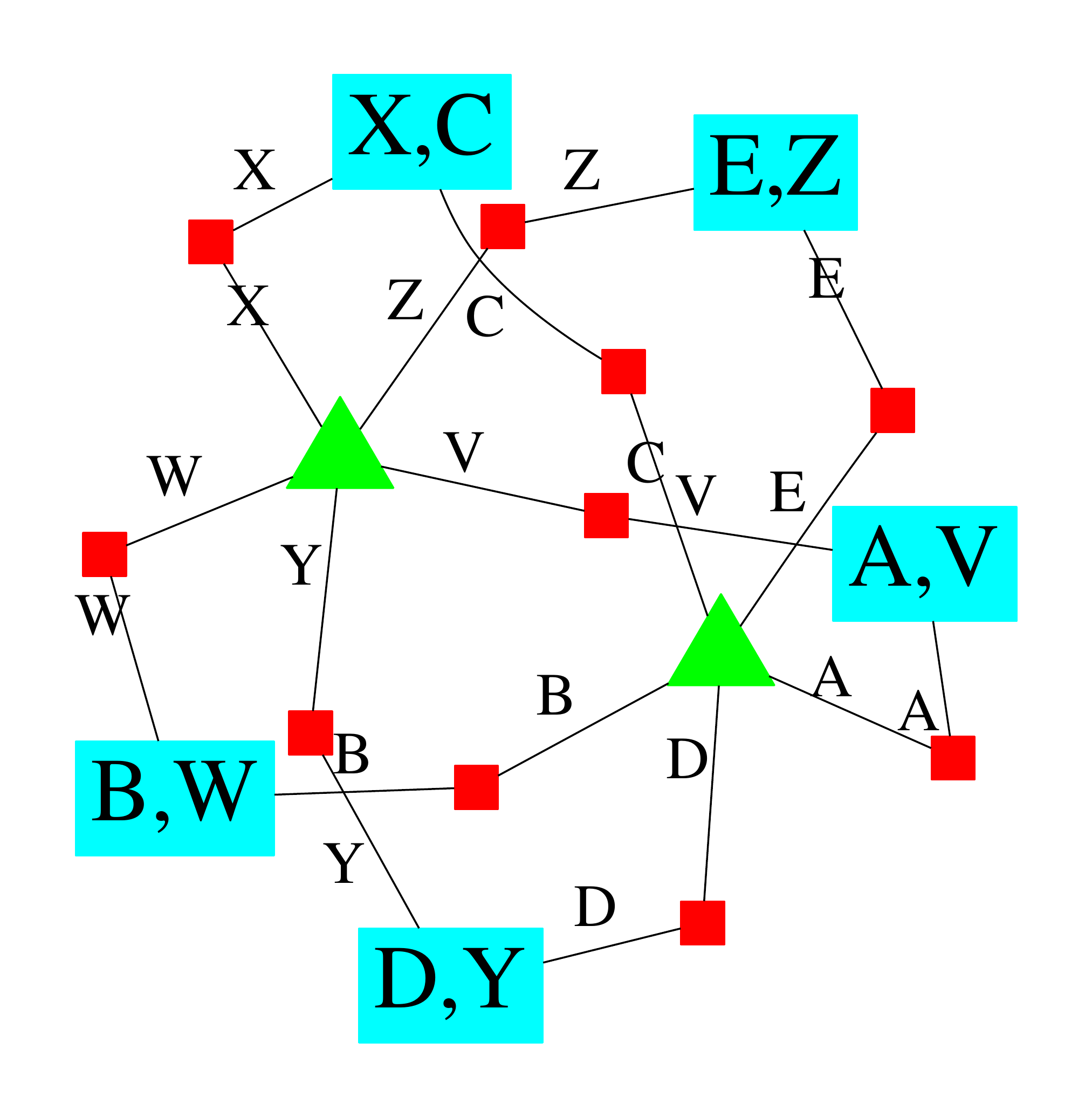}\\
  \begin{sideways}Bethe 2D\end{sideways} & \includegraphics[width=2.2in]{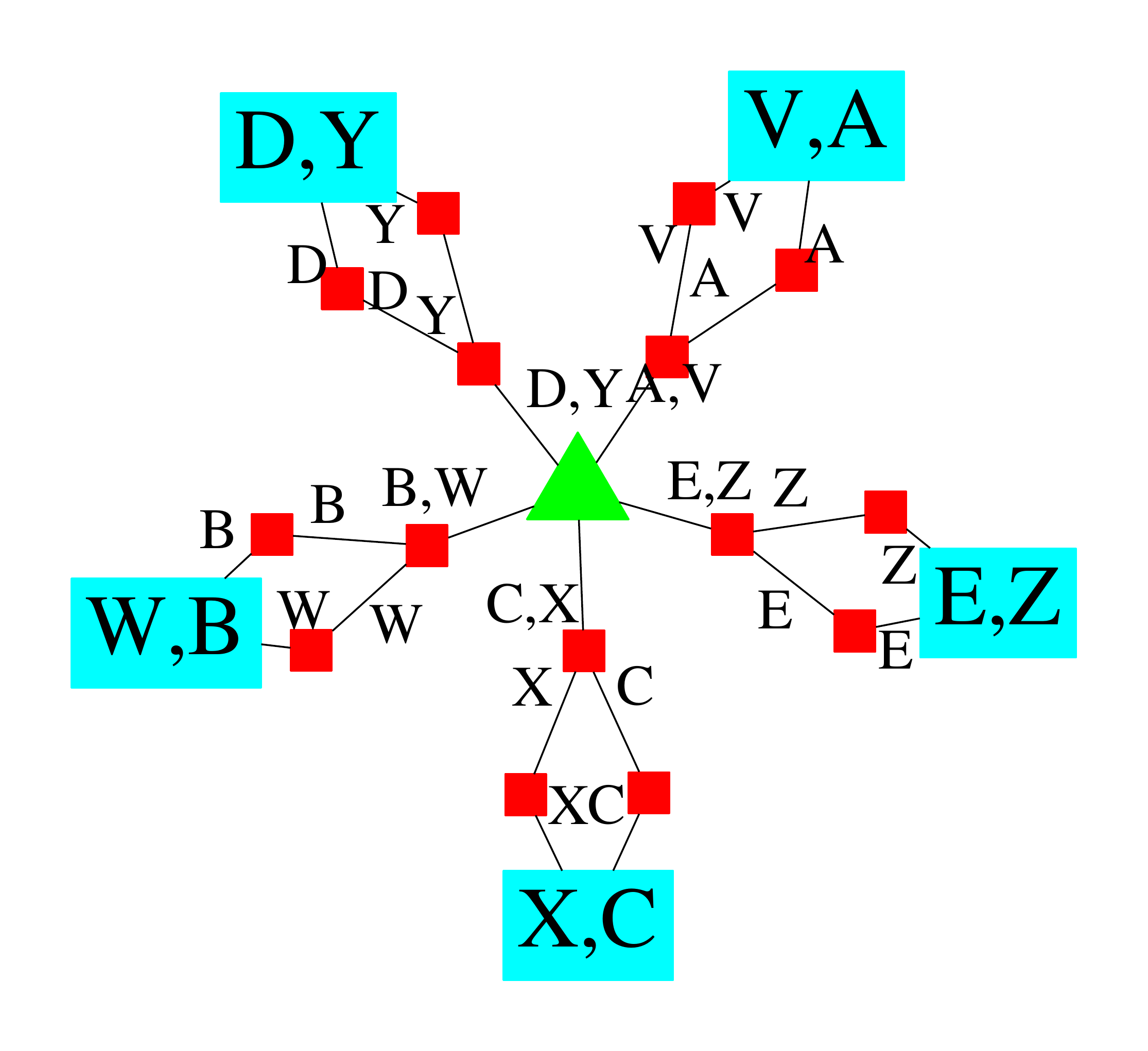}\\
  \begin{sideways}Exact 2D\end{sideways} & \begin{minipage}{2.2in}\centering \includegraphics[width=1.4in]{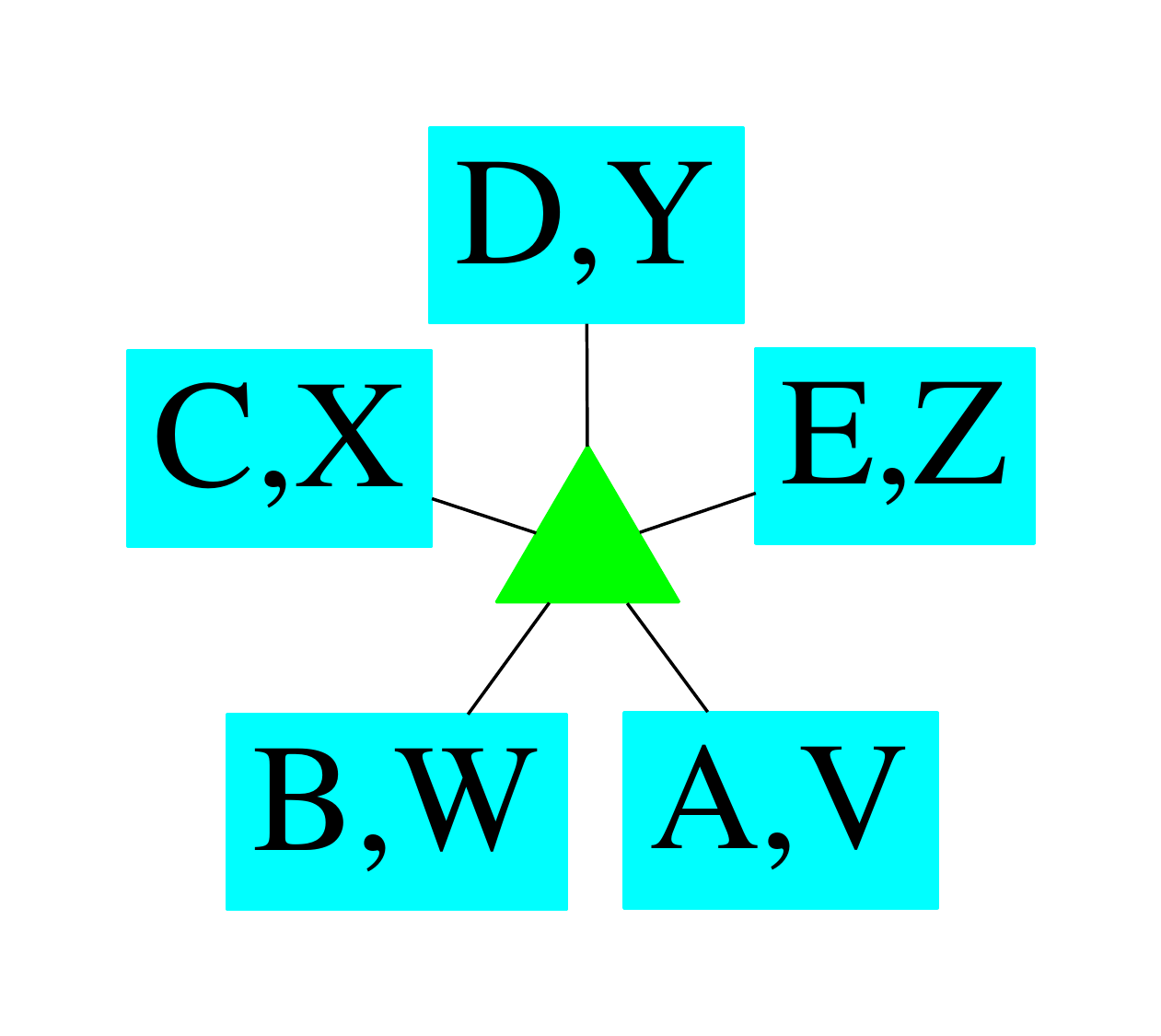} \end{minipage}
\end{tabular}
\caption{{\bf Three representations of two-dimensional additive
    dependencies.} The relationship $(A,V)=(B,W)+(C,X)+(D,Y)+(E,Z)$ is
  encoded using 1D convolution trees (which can be solved
  approximately with loopy inference), as a Bethe graph with a 2D
  convolution tree (the Bethe graph introduces 1D bottlenecks), and as
  a true 2D convolution tree. HUGIN nodes are drawn as cyan
  rectangles, hyperedges are drawn as red squares, and convolution
  trees are drawn as green triangles.
  \label{fig:simple2d-graphs}}
\end{figure}

The results on a two-dimensional additive problem are shown using
three approaches: The first approach constructs two 1D convolution
trees (in a loopy manner). The second approach constructs a graph with
a 2D convolution tree automatically (using the {\tt
  BetheInferenceGraphBuilder}). The third approach manually constructs
a 2D convolution tree graph without the 1D bottlenecks introduced by
Bethe construction. These graphs are shown in Figure~\ref{fig:simple2d-graphs}

On an simple sample problem, the three approaches compute similar
results. The posterior for $(A,V)$ with $2\times$ 1D convolution trees
is
\[
(A,V) = \left[
  \begin{array}{ccc}
    0.0103893 &  0.309359 &  0.318217\\
    0.0234906 &  0.16321 &  0.0532962\\
    0.00445803 &  0.026549 &  0.0910306
  \end{array}
  \right],
\]
while the Bethe graph with a 2D convolution tree yields
\[
(A,V) = \left[
  \begin{array}{ccc}
    0.0104458 & 0.309293 & 0.31815\\
    0.0236186 & 0.163177 & 0.0532856\\
    0.004482 & 0.0265418 & 0.0910059
  \end{array}
  \right],
\]
and the exact result via a 2D convolution tree is
\[
(A,V) = \left[
  \begin{array}{ccc}
    0.00869492 &  0.302207 &  0.310852\\
    0.045331 &  0.151367 &  0.0530654\\
    0.0107954 &  0.0266539 &  0.0910336
  \end{array}
  \right].
\]

Understandably, the Bethe construction suppresses the benefits of the
2D convolution tree (because it forces the 2D distributions through 1D
bottlenecks in a manner reminiscent of using $2\times$ 1D convolution
trees in a loopy manner). The loopy method based on 1D convolution
trees converges after passing 138 messages, while the exact 2D method
converges after passing 10 messages, but where those 10 messages are
more expensive to compute. All runtimes were similar for this demo,
being that the overhead of such a small inference task is nearly as
high as the negligible cost of inference itself.

\subsubsection*{Molecular decomposition from approximate mass and hydrophobicity}

In this demo, the amino acid composition is inferred by knowing only
the approximate total mass and hydrophobicity of a peptide. The mass
of the intact molecule is sum of the masses of the amino acid residue
in the molecule (thus neglecting relativistic effects). Likewise, the
assumption is made that the observed hydrophobicity of the intact
molecule is the sum of the hydrophobicities of the contained amino
acids. This second assumption is less realistic, but is nonetheless
reasonable if the discretization of hydrophobicities is coarse enough
(or, alternatively, if the uncertainty in the likelihood distribution
on the total hydrophobicity has great enough uncertainty).

A new message passer type, the {\tt ConstantMultiplierMessagePasser},
is introduced. This message passer has one input and one output, and
simply scales the distributions passed through it by stretching the
axes of the messages. {\tt ConstantMultiplierMessagePasser} types
scale by some constant {\tt Vector<double>} when passing messages
forwards and by one over that constant vector when passing messages
backwards). {\tt ConstantMultiplierMessagePasser} scale by floating
point values, and so they produce distributions on floating point
values; however, at present {\tt EvergreenForest} only natively
includes distributions with integral support, and so these floating
point outcomes are dithered into their neighboring integer bins. When
scaling by values $>1$ along some axis, multiple interpretations can
be made: In the first interpretation, the input distribution is truly
on the integers, and so a distribution with support $\{0, 1, 2,
\ldots, k\}$ scaled by $\times 7$ will produce a distribution with
support $\{0, 7, 14, \ldots, 7k\}$. The alternate interpretation is
when a discrete distribution is really used as a proxy for a
continuous distribution (\emph{i.e.}, essentially a primitive
quadrature). In that case, the input distribution on $\{0, 1, 2,
\ldots, k\}$ is really a proxy for the distribution $[0, k]$, and so
scaling by $7$ should produce a distribution with support $[0, 7k]$,
which would be sampled over the integers as $\{0, 1, 2, 3, \ldots
7k\}$. For this reason, when creating a {\tt
  ConstantMultiplierMessagePasser} (or its corresponding {\tt
  ConstantMultiplierDependency}, if the message passers are to be
constructed automatically), it is necessary to specify in each
direction whether or not scaling should interpolate. Essentially, {\tt
  ConstantMultiplierMessagePasser} types extend convolution forests to
not only solving additive models, but also to solving all possible
solutions of discretized linear diophantine
equations \cite{hilbert:mathematical} for discrete random variables of
bounded support. 

\begin{figure}
\centering
\begin{tabular}{m{0.1in}m{2.2in}}
  \begin{sideways}Mass\end{sideways} & \includegraphics[width=2.2in]{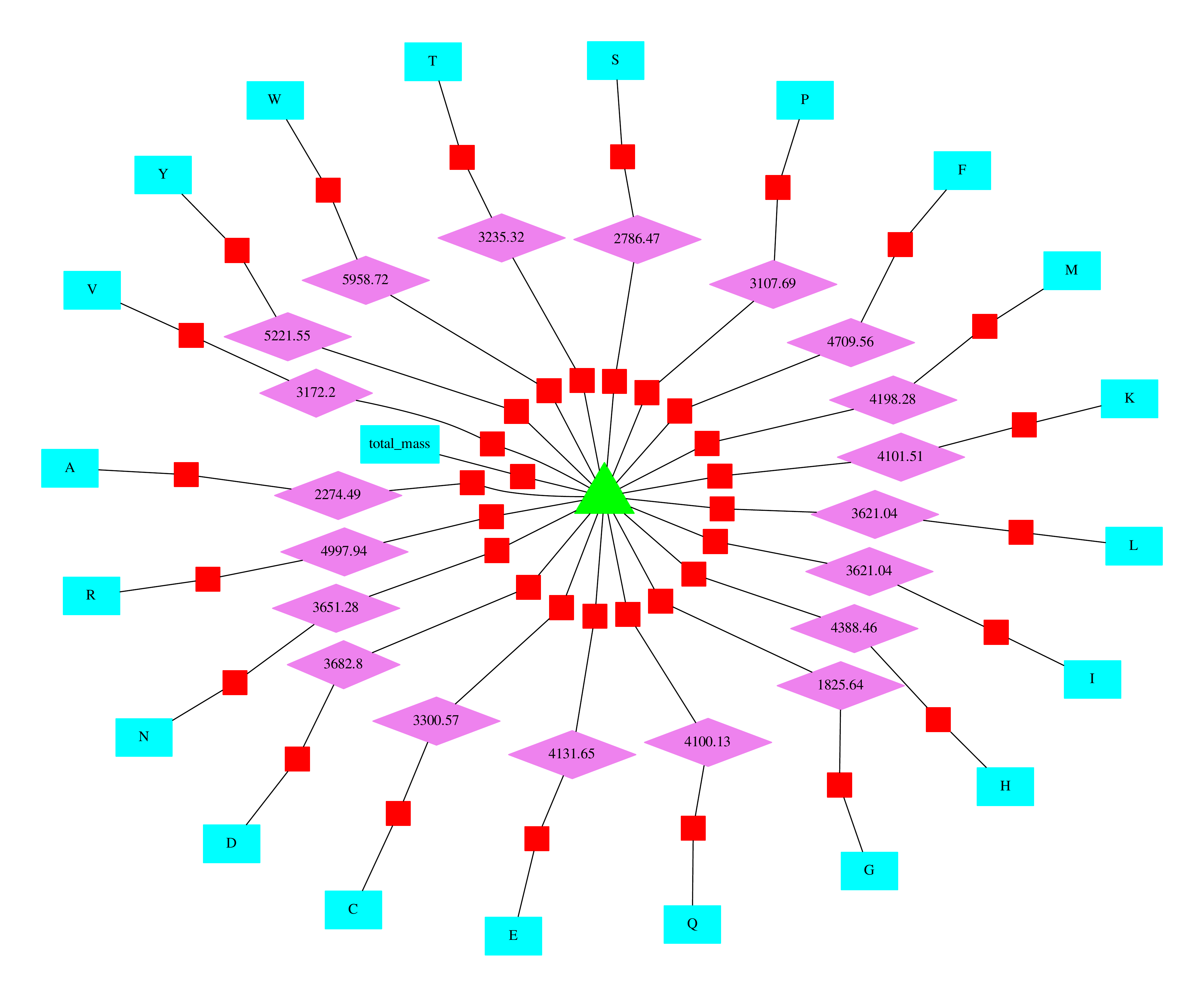}\\
  \begin{sideways}Hydrophobicity\end{sideways} & \includegraphics[width=2.2in]{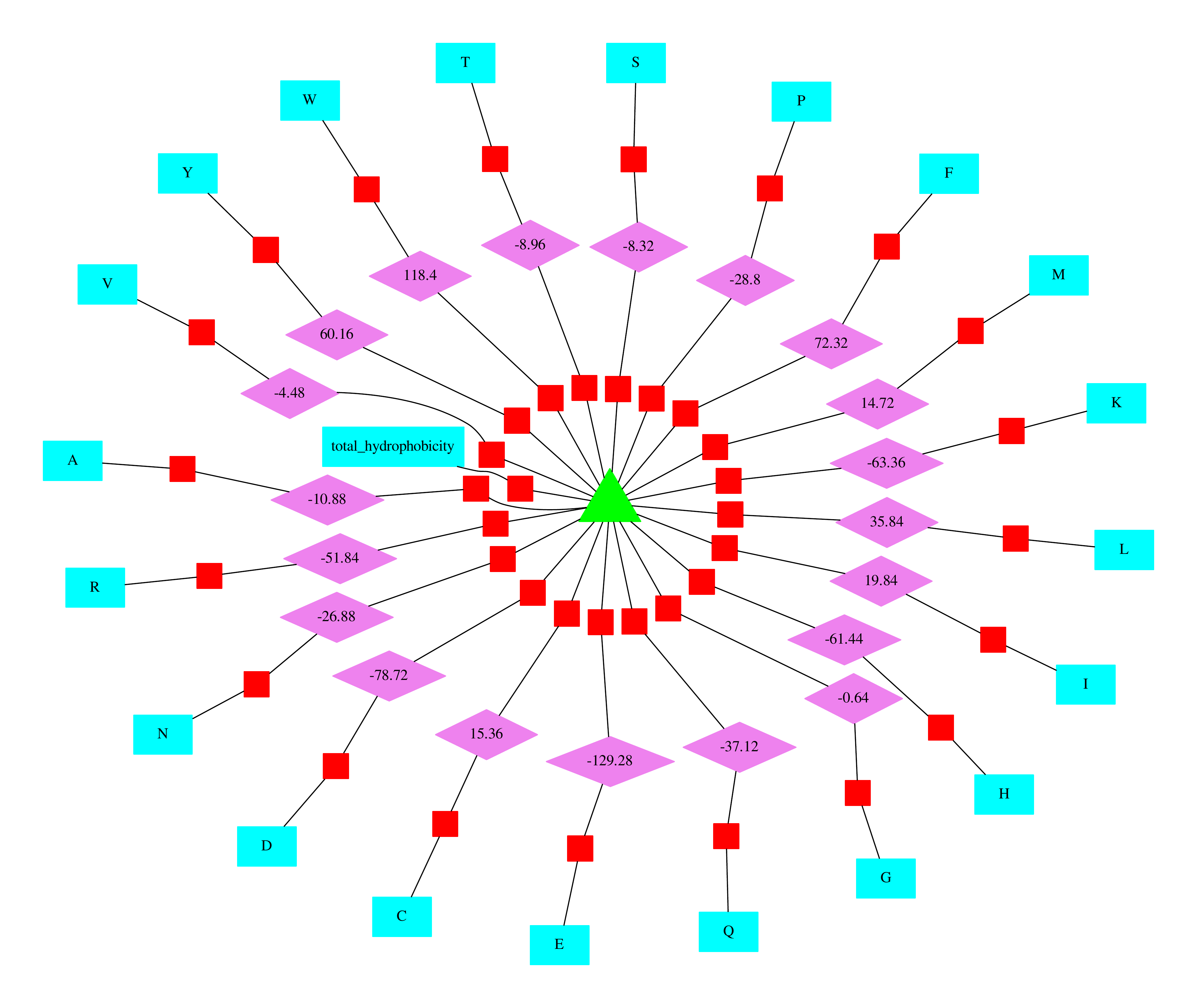}\\
  \begin{sideways}Both (loopy)\end{sideways} & \includegraphics[width=2.2in]{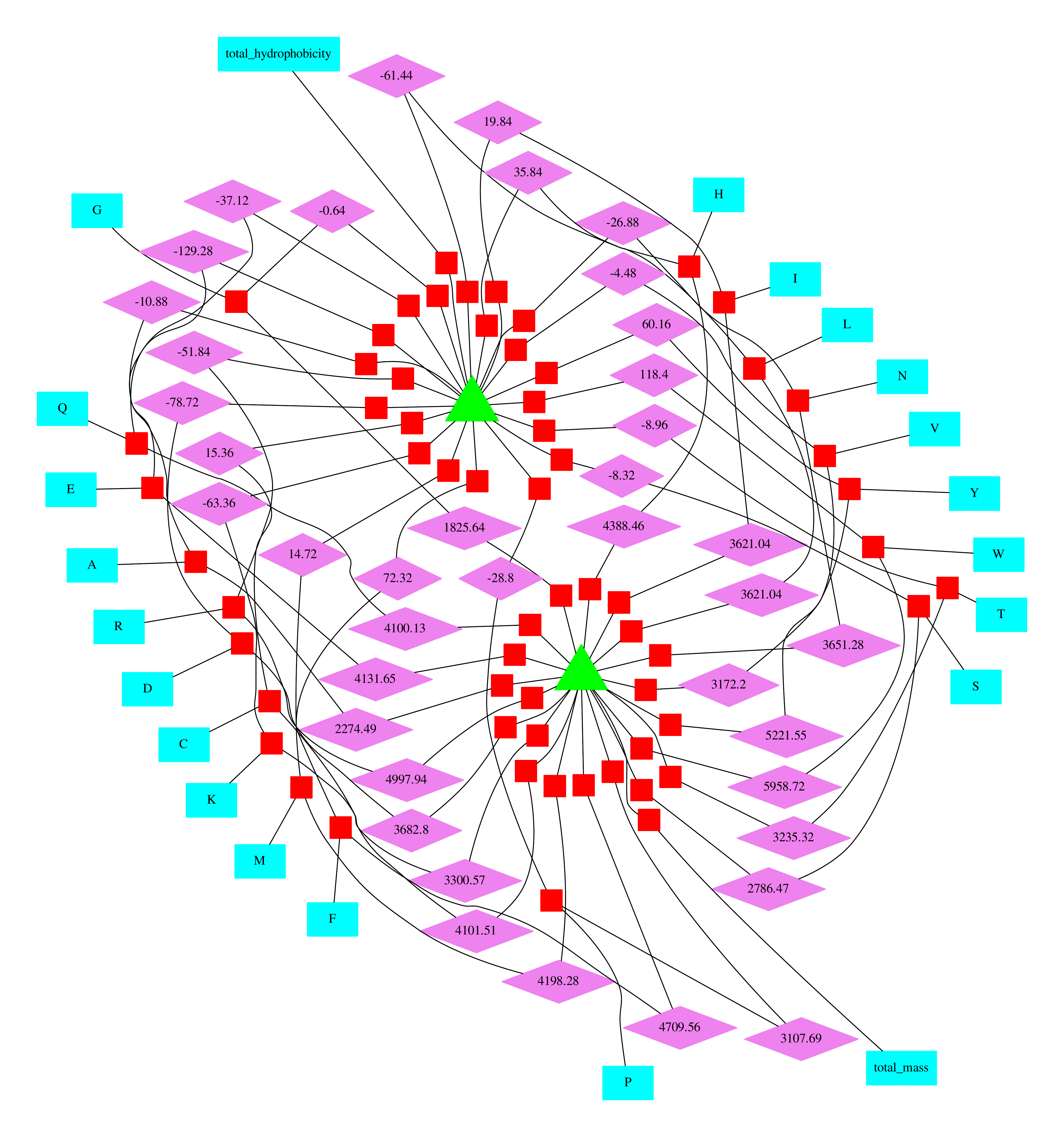}
\end{tabular}
\caption{{\bf Mass, hydrophobicity, and joint peptide graphs.} These
  graphs were built for mass-only inference, hydrophobicity-only
  inference, and joint inference with a loopy graph. HUGIN nodes are
  drawn as cyan rectangles, constant multipliers are drawn as violet
  diamonds, hyperedges are drawn as red squares, and convolution trees
  are drawn as green triangles.
  \label{fig:peptide-graph}}
\end{figure}

With these {\tt ConstantMultiplierMessagePasser} types, it is now
possible to translate the number of instances of amino acid lysine
($K$) to the mass contribution from lysine $K_{\mbox{mass}} = 128.1723
\times K$ daltons. It is likewise possible to write the hydrophobicity
contribution from lysine in terms of a constant multiplication
$K_{\mbox{hydrophobicity}} = -0.99 \times K$. Two convolution trees
(whose message passers are constructed automatically from {\tt
  AdditiveDependency} types) are constructed to sum the overall mass
contributions from all amino acids and to sum the overall
hydrophobicity contributions from all amino acids. The resulting loopy
graph is shown in Figure~\ref{fig:peptide-graph}. Graphs are plotted
with a {\tt python} script included in {\tt
  EvergreenForest/src/Utilities}, which uses the {\tt pygraphviz}
package.

In order to improve the granularity of the discretization, masses are
discretized not in 1 dalton bins, but instead by multiplying by a
small constant (in this case 32, so that each bin spans $\frac{1}{32}$
dalton). Similarly, hydrophobicities were discretized using bins of
size $\frac{1}{64}$. This is simply achieved by multiplying all amino
acid masses (which only appear in the {\tt
  ConstantMultiplierDependency} types) by 32 and multiplying all amino
acid hydrophobicities (which also only appear in the {\tt
  ConstantMultiplierDependency} types) by 64. Masses for each amino
acid were taken from
\url{http://www.matrixscience.com/help/aa_help.html} and
hydrophobicities were taken using the ``wwHydrophobicity'' measure
from
\url{https://www.cgl.ucsf.edu/chimera/docs/UsersGuide/midas/hydrophob.html}
(originally experimentally estimated by Wimley \&
White \cite{wimley:experimentally}). The posterior distributions for
such discretized solutions of linear diophantine probabilistic
equations can be quite sensitive to the scaling constants used for
binning (\emph{e.g.}, 32 is used here to scale the mass axis and 64 is
used to scale the hydrophobicity axis). Essentially, this is because
different constants will result in different collision-like behavior
as floating point values are mapped to integer bins. This phenomenon
can make the use of {\tt ConstantMultiplierDependency} types more
challenging in practice. Regardless, even when the posterior
distributions do not reflect certainty about the molecular
composition, the imperfect information may still be used in a ``big
data'' context and used to further narrow the solution space from
still more evidence (\emph{i.e.}, neither mass nor hydrophobicity)
about the molecule.

When the goal peptide mass or goal peptide hydrophobicity are floating
point values, they are divided uniformly between the adjacent bins
(\emph{i.e.}, the ceiling and the floor).

The peptide EEAMPK (with total residue mass 685.79 and total
hydrophobicity -5.42) is run with only the mass, only the
hydrophobicity, and both (Figure~\label{fig:peptide-results}). The
graphical model was constructed automatically from {\tt Dependency}
types using {\tt BetheInferenceGraphBuilder} and inference was
performed using {\tt FIFOScheduler}. Interestingly, the 1D marginal
convolution trees correctly infer the result using loopy belief
propagation. This line of thinking could be used with not two but
several features, and while each of those may only yield approximate
information, the joint solution space could be quite sparse.

This is only presented as an illustration; the motivating notion
behind this approach is not so well matched to molecules like peptides
(whose linear structure affords straightforward sequencing with mass
spectrometry using either databases \cite{eng:approach} or \emph{de
  novo} approaches \cite{kim:spectral}). What is more interesting is
the ability to ravel large amounts of weak information on molecules
with complex, nonlinear structures (such as those formed by sugars or
small molecules investigated in drug discovery). These nonlinear
structures are far more difficult to solve  \cite{bhatia:constrained,
  serang:sweetseqer}, and the aggregation of weak information
(\emph{e.g.}, from several different separation techniques) could
prove valuable.

\begin{figure}
\centering
\begin{tabular}{m{0.1in}m{2in}m{2in}}
  & \begin{minipage}{2in}\centering $p=1$\end{minipage} &  \begin{minipage}{2in}\centering $p=\infty$\end{minipage}\\
  \hline\\
  \begin{sideways}Mass\end{sideways} & \includegraphics[width=2.2in]{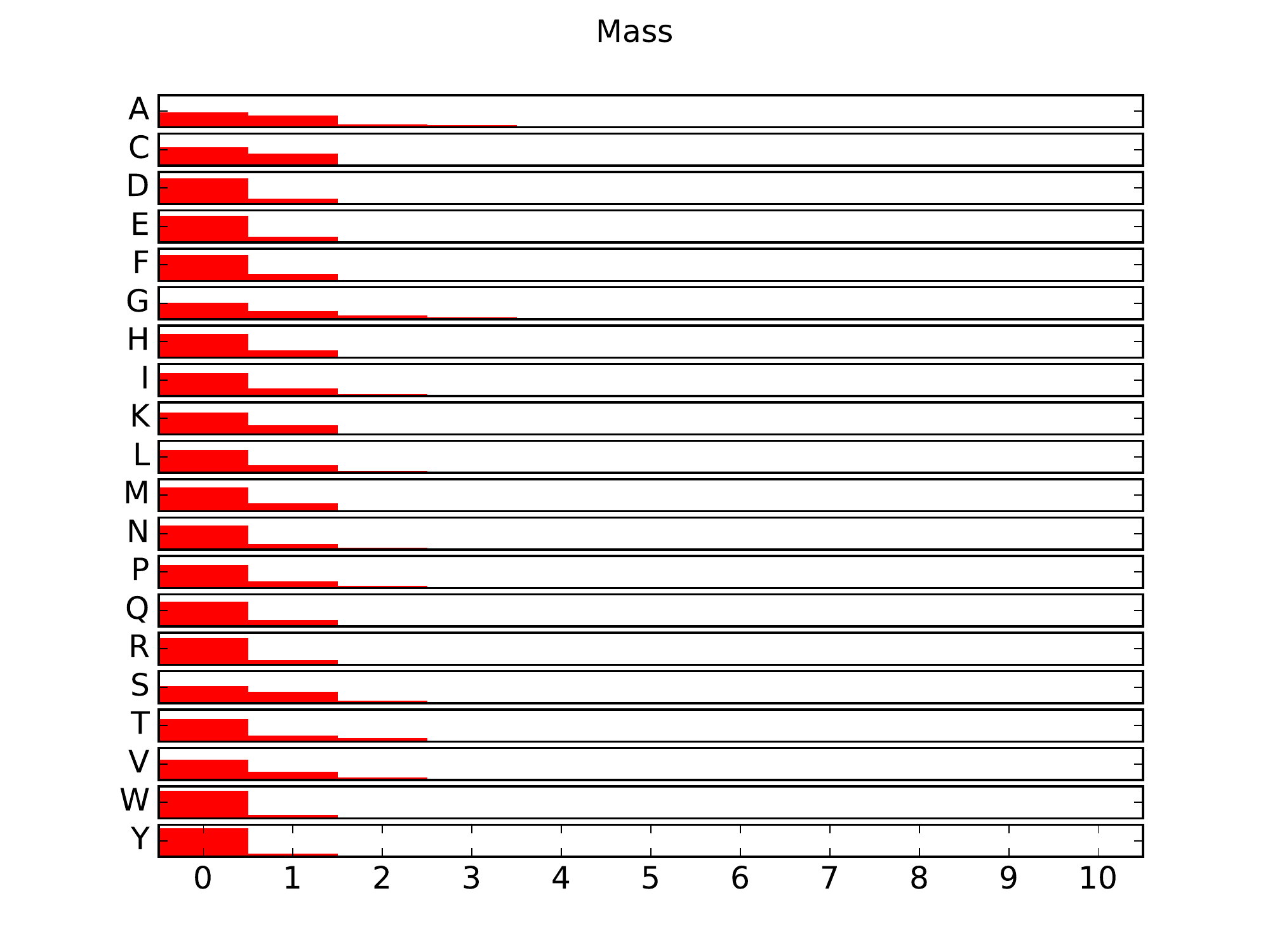} & \includegraphics[width=2.2in]{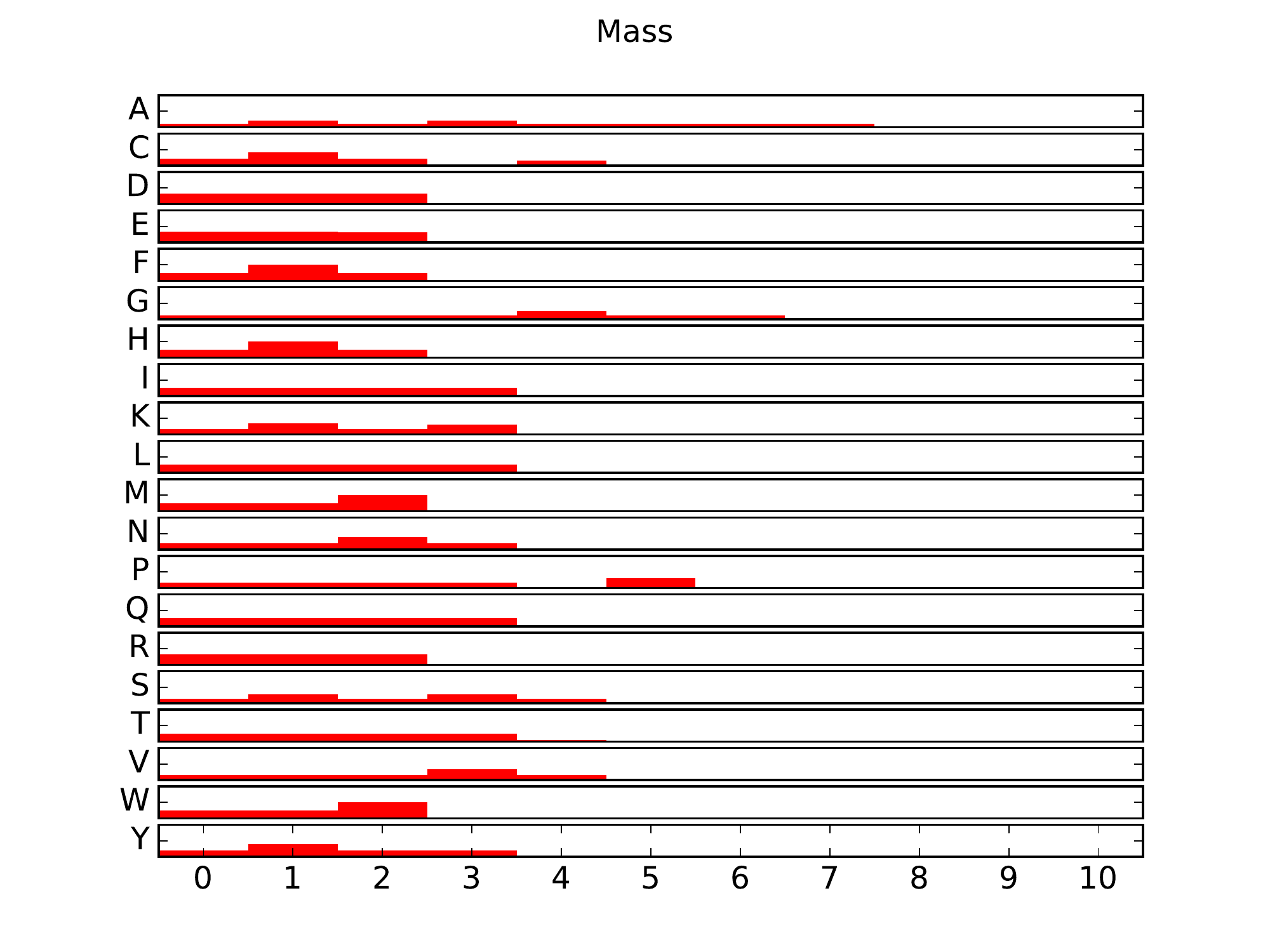}\\
  \begin{sideways}Hydrophobicity\end{sideways} & \includegraphics[width=2.2in]{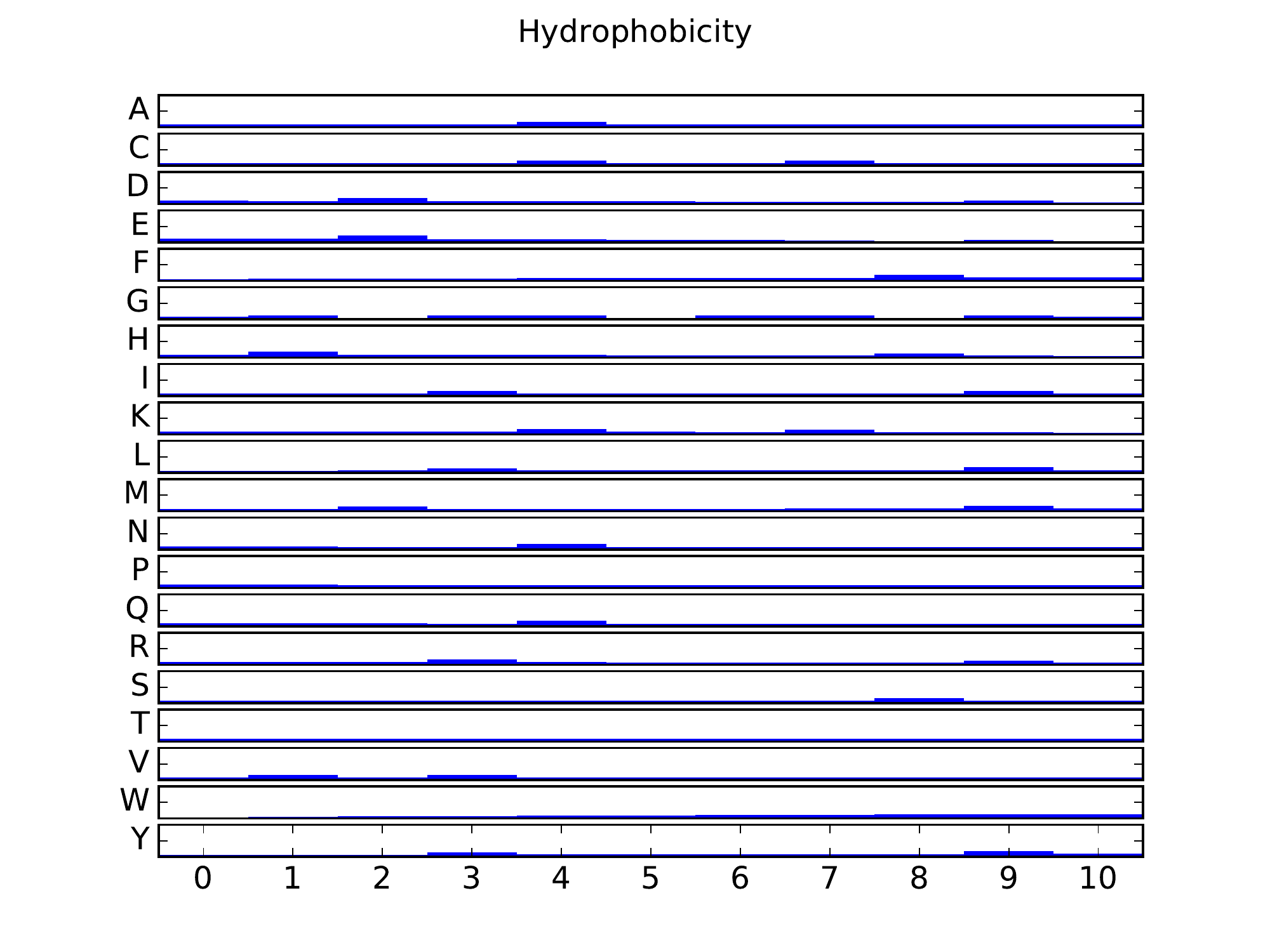} & \includegraphics[width=2.2in]{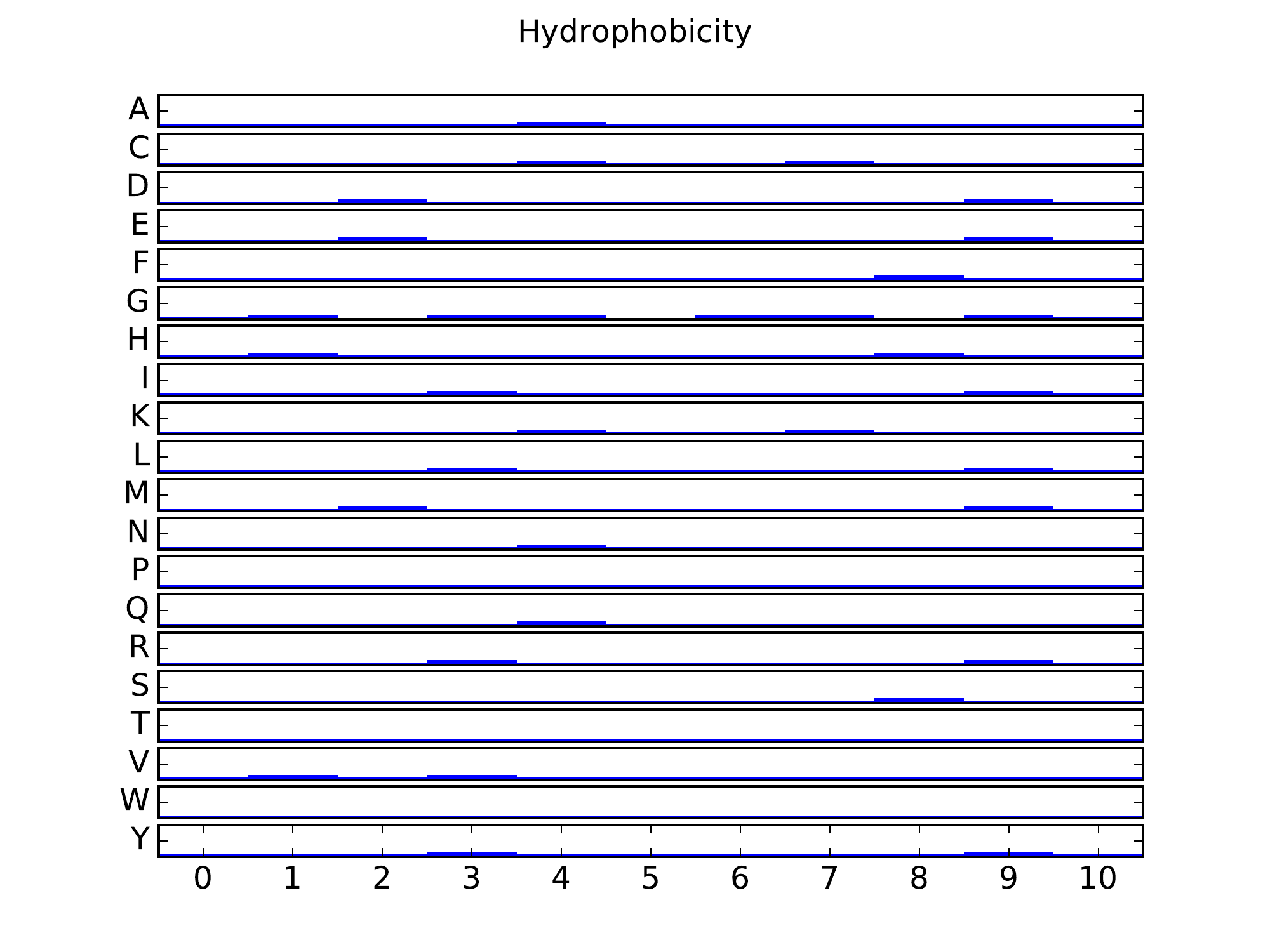}\\
  \begin{sideways}Both (loopy)\end{sideways} & \includegraphics[width=2.2in]{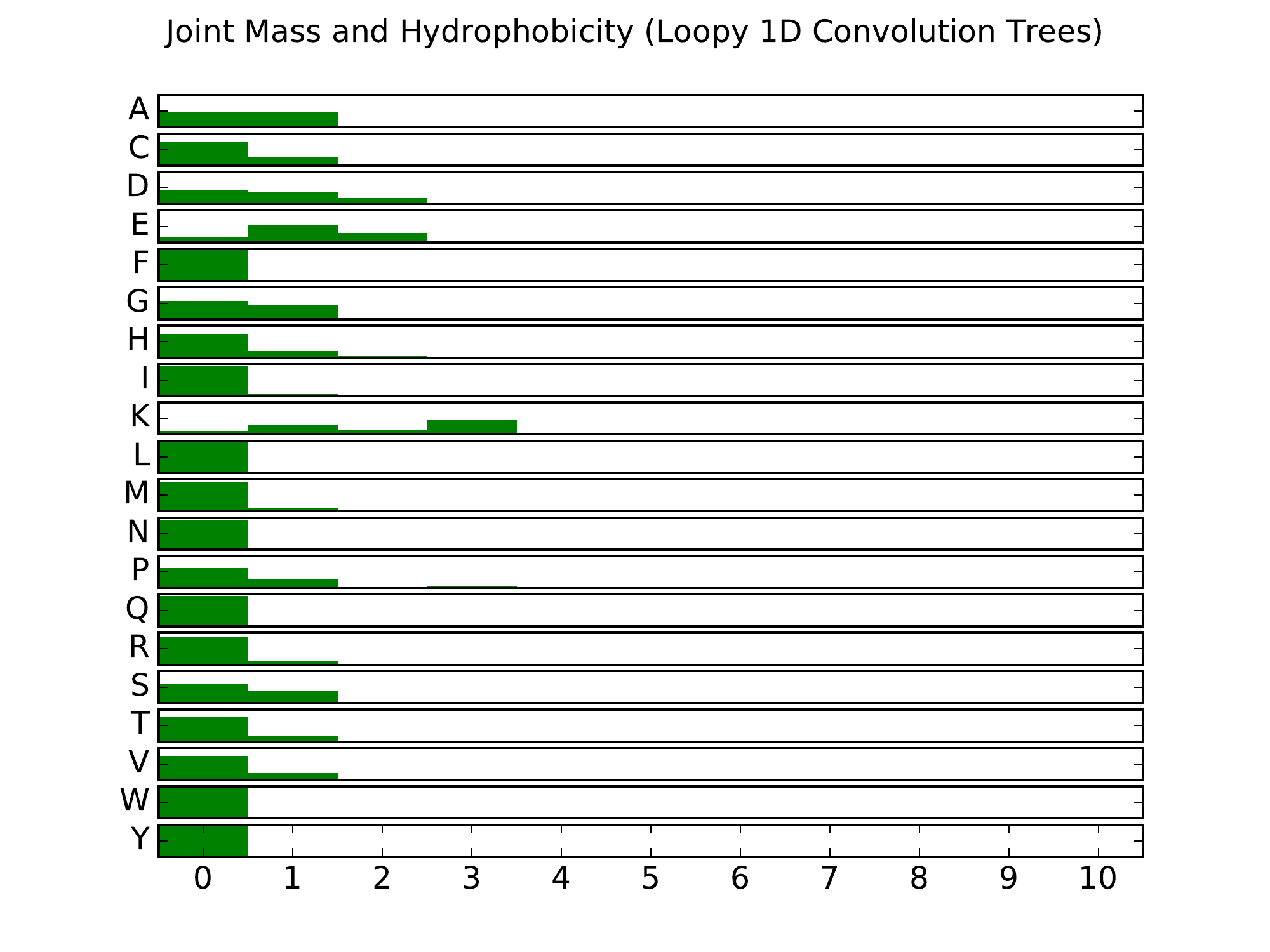} & \includegraphics[width=2.2in]{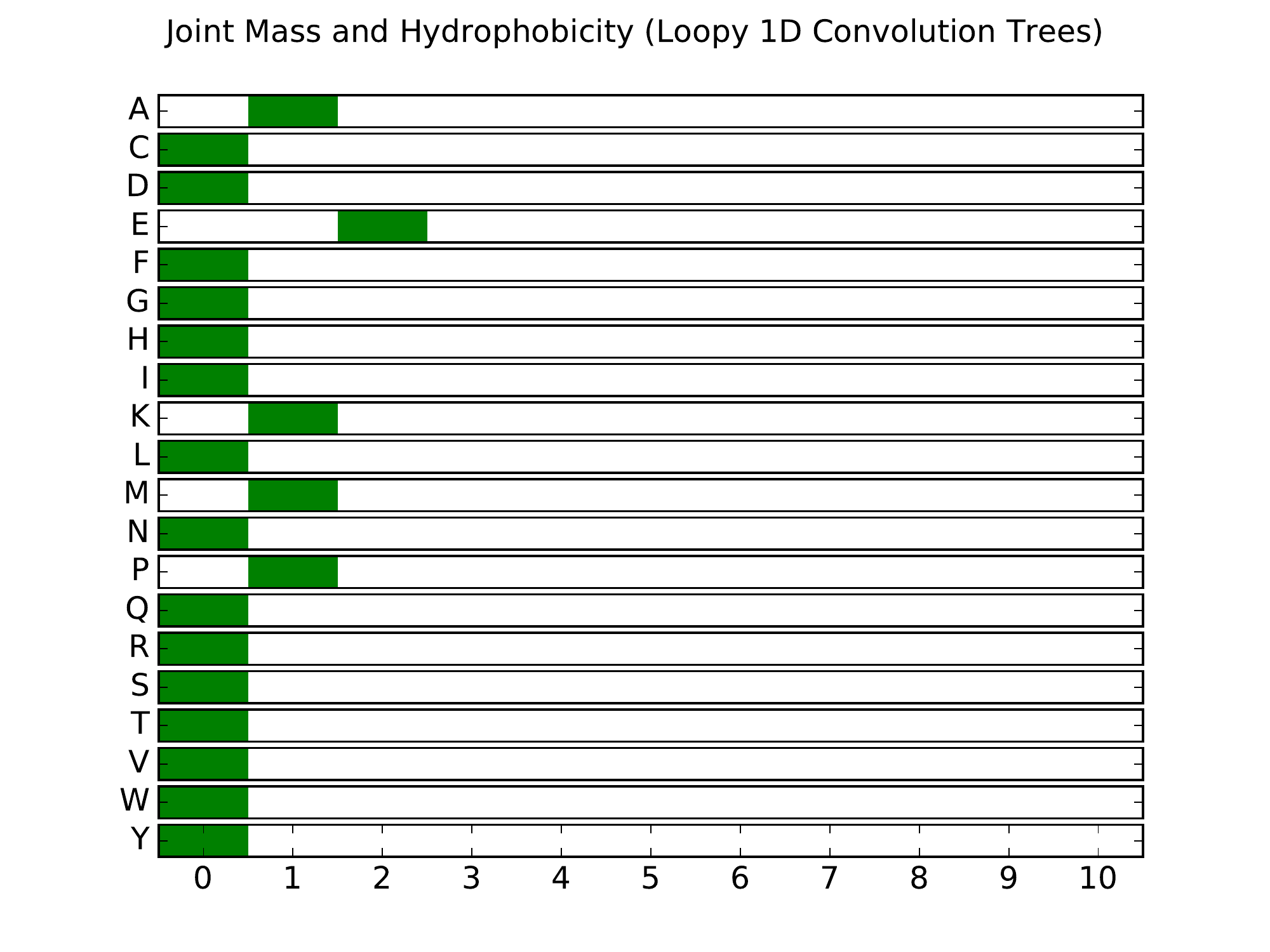}\\
\end{tabular}
\caption{{\bf Mass, hydrophobicity, and joint amino acid posteriors.}
  Posterior distributions are each drawn with $y$-axis in $[0,1]$.
  Independently, mass and hydrophobicity narrow down the solution
  space of possible counts for each amino acid; however, the product
  of those posterior distributions would still not produce a unique or
  even sparse result. On the other hans, the approximate joint
  posteriors (as estimated using a loopy graph) correctly estimate
  100\% probability of $E=2$, $M=1$, $P=1$, $K=1$ using only the mass
  and hydrophobicity of peptide EEAMPK when using $p=\infty$. The
  benefit of having efficient inference on additive dependencies when
  $p=\infty$ is demonstrated.
  \label{fig:peptide-results}}
\end{figure}

\subsubsection*{Elemental quantification with shared isotope peaks}

A final demo of the engine is presented by performing elemental
quantification in the presence of overlapping isotope peaks.  Isotopic
masses and abundances were taken from
\url{http://www.chem.ualberta.ca/\%7Emassspec/atomic_mass_abund.pdf}
(which in turn uses data from Rosman \& Taylor  \cite{rosman:table},
from Audi \& Wapstra  \cite{audi:1993}, and from Audi \emph{et
  al.} \cite{audi:ame2003}).

Masses were discretized into an array, roundig masses into bins of
size 0.1 daltons, forcing some elements to some observed peaks to map
ambiguously to isotopes from multiple elements. Observed peaks were
treated as measuring sums of abundances of all isotopes matching that
0.1 dalton mass window. Isotope abundances were modeled using {\tt
  ConstantMultiplierDependency} types (\emph{e.g.}, the abundance of
$^{36}Ar$ is $0.3365\%$ of the abundance of argon in the
sample. Observed abundances were modeled as having (discretized)
Gaussian distributions centered around the true abundance. Note that
for convenience in this simple illustration, abundances and
intensities are conflated, essentially assuming that the mass
spectrometer will measure all peaks equally well. This approach was
first outlined in 2014 \cite{serang:probabilistic}. 

\begin{figure}
\centering
\includegraphics[width=4in]{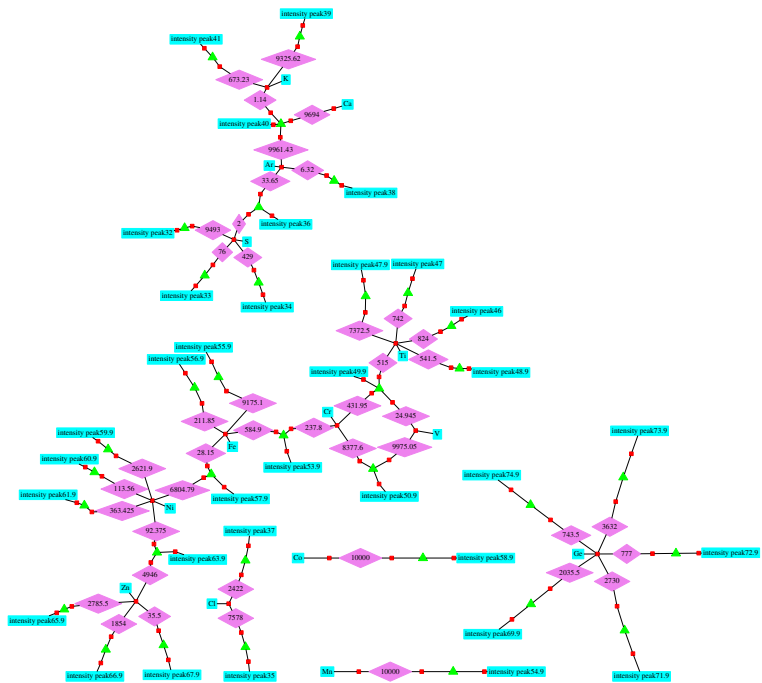}
\caption{{\bf Graphical model for estimating elemental abundance from
    overlapping isotope peaks.} A spectrum is generated from
  composition $Ni_3 V_2 Co_3 Cl_4 Zn_6 Ca_{10} Mn_5 S_{10} Ge_7 Ar_{5}
  Fe_5 Ti_9 K_8$, and then that spectrum is used to construct the
  graph. HUGIN nodes are drawn as cyan rectangles, constant
  multipliers are drawn as violet diamonds, hyperedges are drawn as
  red squares, and convolution trees are drawn as green triangles.
  \label{fig:isotope-graph}}
\end{figure}

\begin{figure}
\centering
\begin{tabular}{cc}
  \includegraphics[width=2.3in]{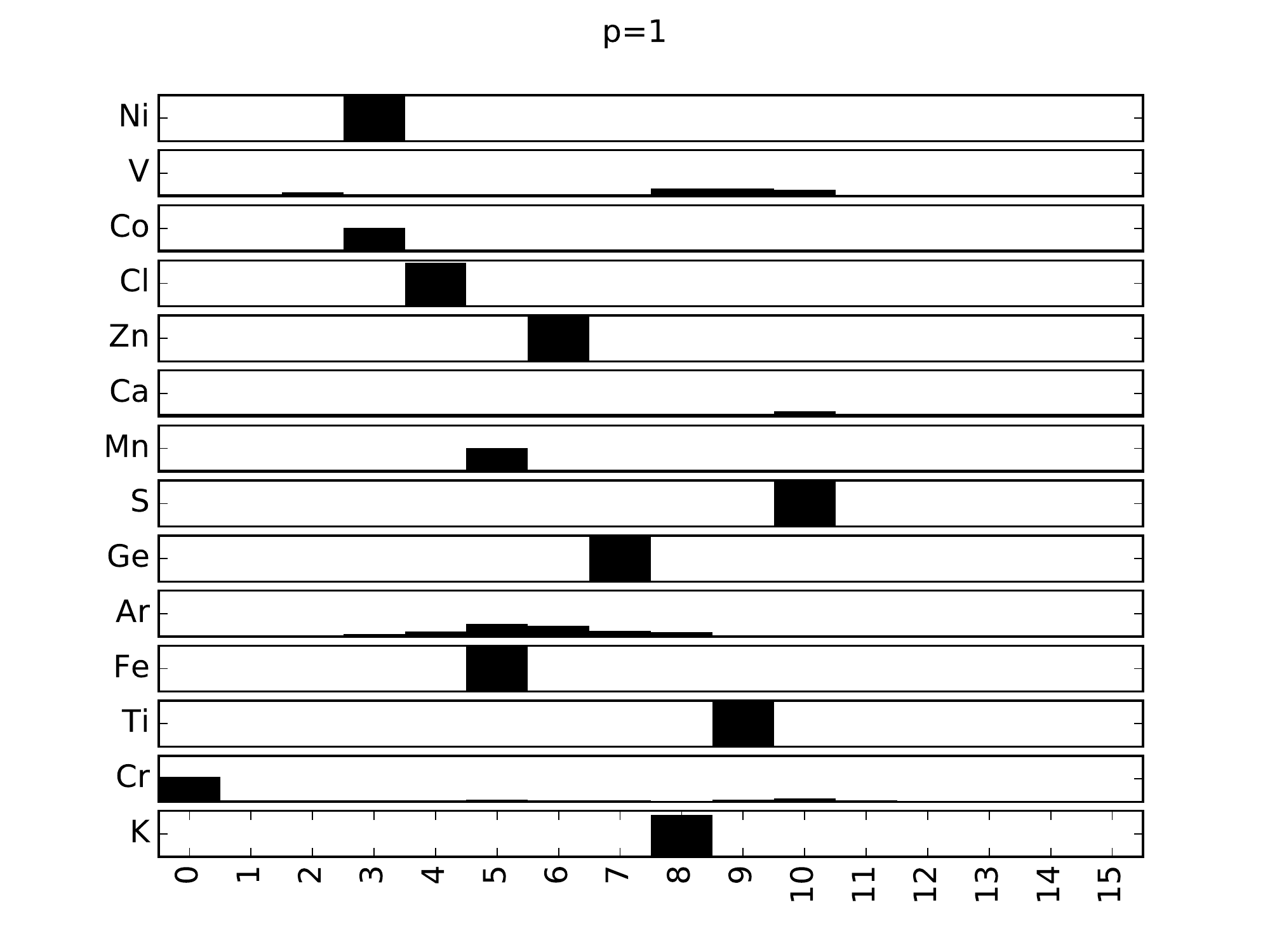} & \includegraphics[width=2.3in]{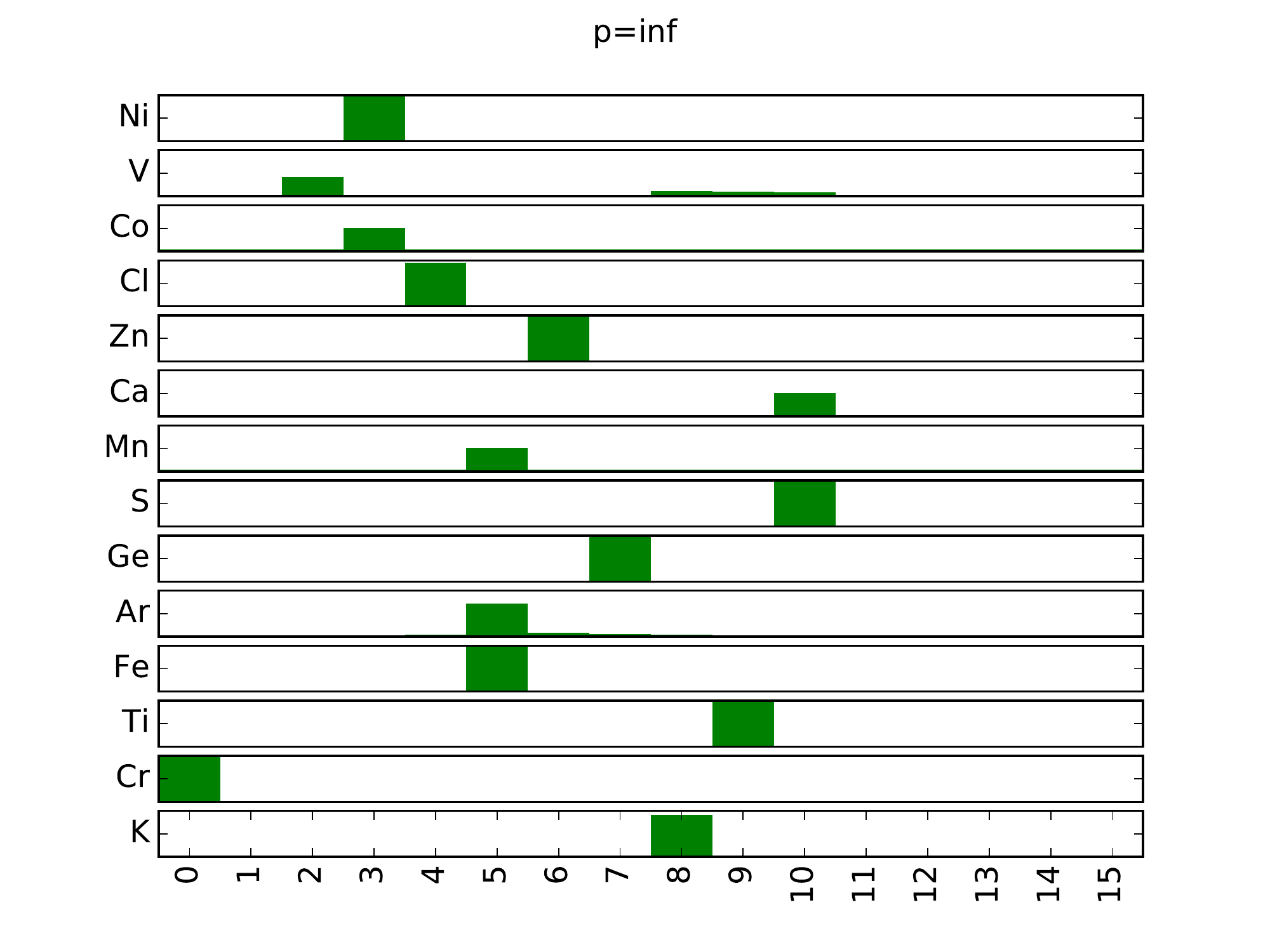}\\
\end{tabular}
\caption{{\bf Estimating elemental abundance from overlapping isotope
    peaks.} The correct answers (matching $Ni_3 V_2 Co_3 Cl_4 Zn_6 Ca_{10} Mn_5 S_{10} Ge_7 Ar_{5} Fe_5 Ti_9 K_8$, the chemical composition
  used to generate the spectrum) are always shown with nonzero
  probability; however, with $p=\infty$, the correct abundances of
  vanadium, calcium, argon, and chromium are all much higher than with
  $p=1$.
  \label{fig:isotope-posteriors}}
\end{figure}

A test problem generated a spectrum from a sample with composition
$Ni_3 V_2 Co_3 Cl_4 Zn_6 Ca_{10} Mn_5 S_{10} Ge_7 Ar_{5} Fe_5 Ti_9
K_8$. The graph produced from this problem is shown in
Figure~\ref{fig:isotope-graph}. The graphical model was constructed
automatically from {\tt Dependency} types using {\tt
  BetheInferenceGraphBuilder} and inference was performed using {\tt
  FIFOScheduler}. Posterior distributions from that sample problem are
given in Figure~\ref{fig:isotope-posteriors}.

The model itself is quite simple and meant only for
illustration. Better models would weigh more intense peaks as more
reliable. For example including a small amount of addtive noise in the
spectrum would easily suggest a $1.2$-fold change in a low-intensity
peak, whereas only a larger amount of additive noise could make a
high-intensity peak $1.2 \times$ its expected value. But as a
mechanism for prototyping such models, convolution forests are quite
useful, because discretizations of any prior or likelihood
distribution families can be easily made.

The much more interesting use-case would be for processing the 2D
heatmaps showing intensity as a function of precursor mass and
retention time. Convolution forests could offer a unified
probabilistic approach for demixing overlapping peaks. Graphical
information could likewise tie the underlying analytes to include
biologically driven covariation information (\emph{e.g.}, when protein
A is abundant, protein B should also be).

\subsubsection*{Elemental quantification with regularization}

\begin{figure}
\centering
\includegraphics[width=3in]{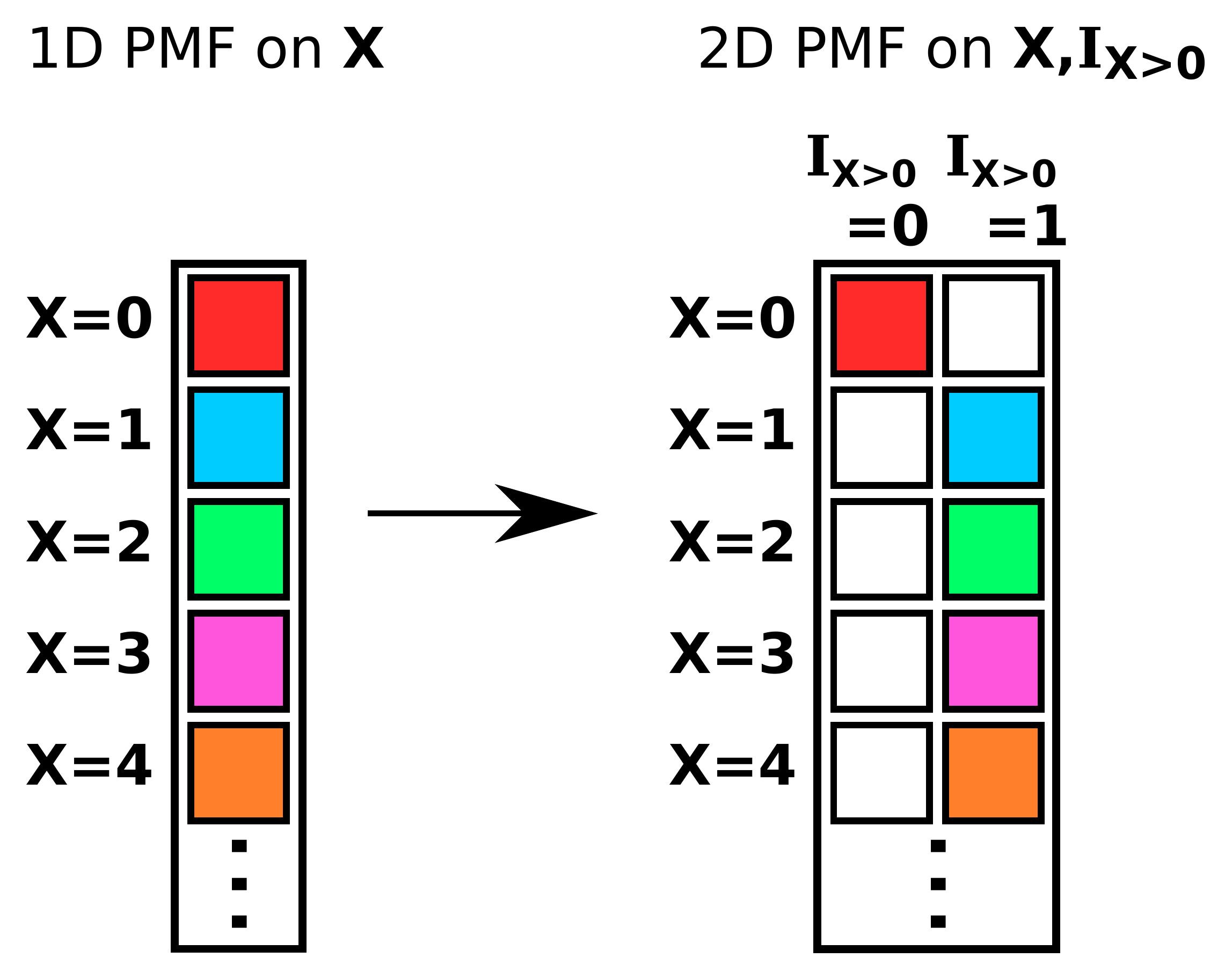}
\caption{{\bf Creating an indicator variable for $I_{X>0}$.} A joint
  distribution between the original variable of interest, $X$, and the
  indicator variable $I_{X>0}$ is created by placing the probabilities
  from $\pmf_X$ in the correct row and in the corresponding column for
  whether $X>0$ or not.
  \label{fig:regularization}}
\end{figure}

A second example of elemental quantification is presented to show how
additive models can be used to enforce regularization. Here
regularization can be performed by transforming a 1D PMF on random
variable $X$ into a 2D PMF with an indicator variable $I_{X>0}$, which
measures whether $X>0$ (Figure~\ref{fig:regularization}). An additive
depedency can be used to restrict the sum of these indicator variables
to be uniform in $\{0, 1, \ldots k\}$, which will enforce that $\leq
k$ elements are permitted to have nonzero abundance.

\begin{figure}
\centering
\includegraphics[width=4in]{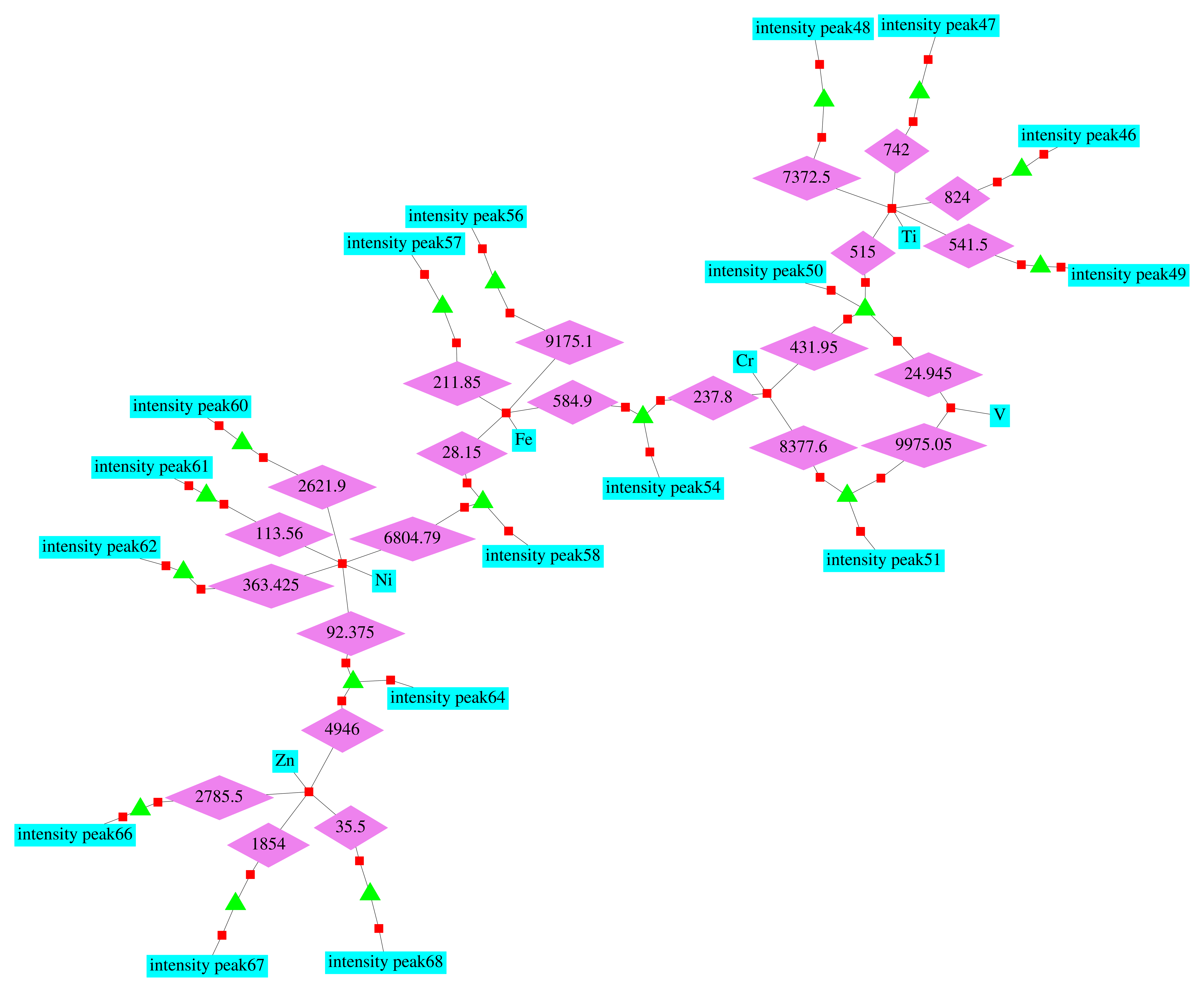}
\caption{{\bf Graphical model for elemental abundance.} A spectrum is generated from
  composition $Ni_2 V_7 Zn_2 Fe_4 Ti_3$, and then that spectrum is
  used to construct the graph. HUGIN nodes are drawn as cyan
  rectangles, constant multipliers are drawn as violet diamonds,
  hyperedges are drawn as red squares, and convolution trees are drawn
  as green triangles.
  \label{fig:isotope-graph-2}}
\end{figure}

\begin{figure}
\centering
\includegraphics[width=4in]{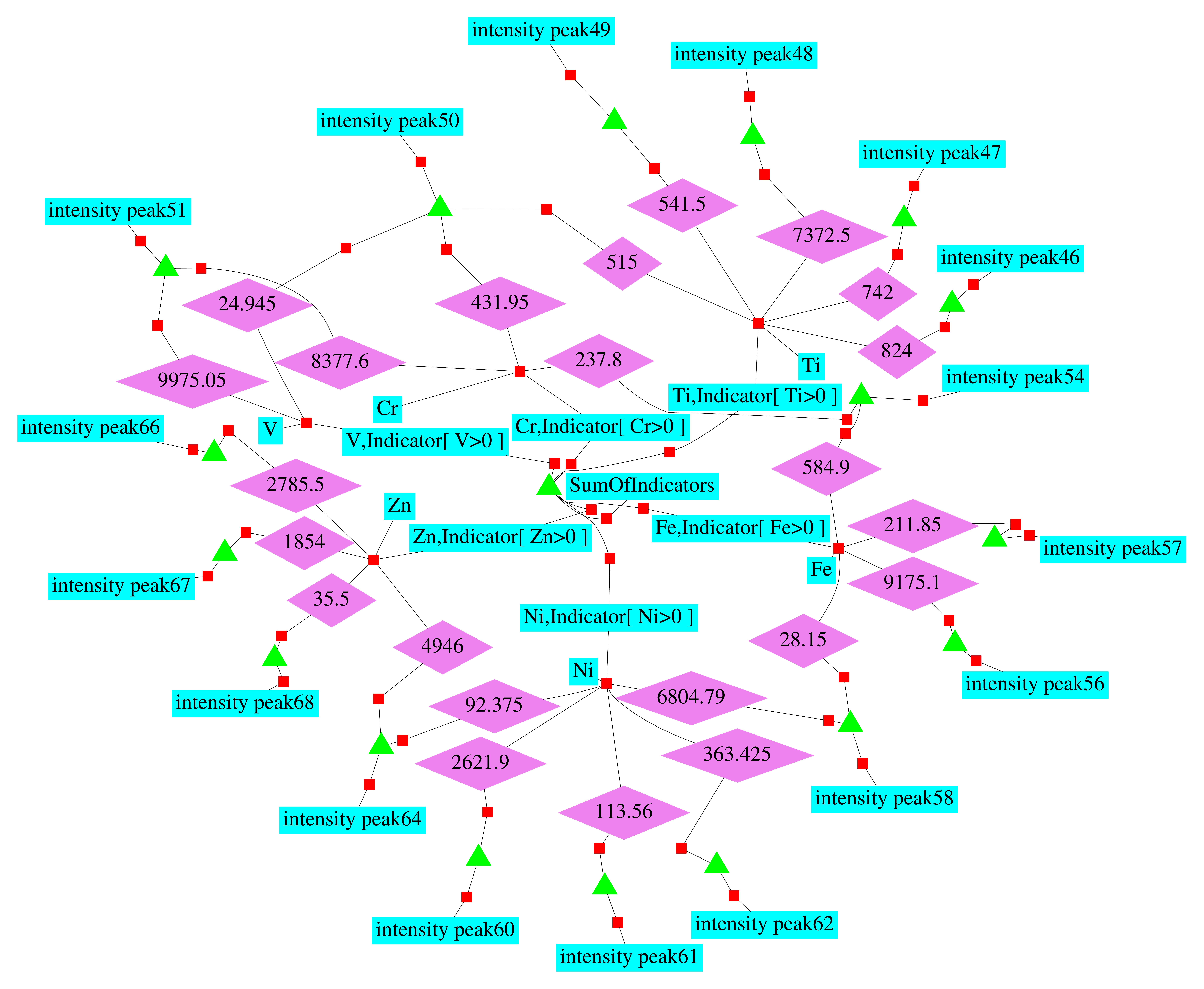}
\caption{{\bf Graphical model for elemental abundance with
    regularization.} A spectrum is generated from composition $Ni_2
  V_7 Zn_2 Fe_4 Ti_3$, and then that spectrum is used to construct the
  graph. Indicator variables are also inserted, and the sum those
  indicator variables (reflecting the total number of present
  elements) is constrained with a convolution tree. HUGIN nodes are
  drawn as cyan rectangles, constant multipliers are drawn as violet
  diamonds, hyperedges are drawn as red squares, and convolution trees
  are drawn as green triangles.
  \label{fig:isotope-graph-2-regularization}}
\end{figure}

The model elemental quantification model is exteneded to use indicator
variables and an extra additive dependency to enforce at most 5
elements are present. This is demonstrated using a spectra generated
from the formula $Ni_2 V_7 Zn_2 Fe_4 Ti_3$. The graph produced by this
spectrum without regularization is shown in
Figure~\ref{fig:isotope-graph-2} and the graph with regularization is
shown in Figure~\ref{fig:isotope-graph-2-regularization}.

\begin{figure}
\centering
\begin{tabular}{cc}
  \includegraphics[width=2.3in]{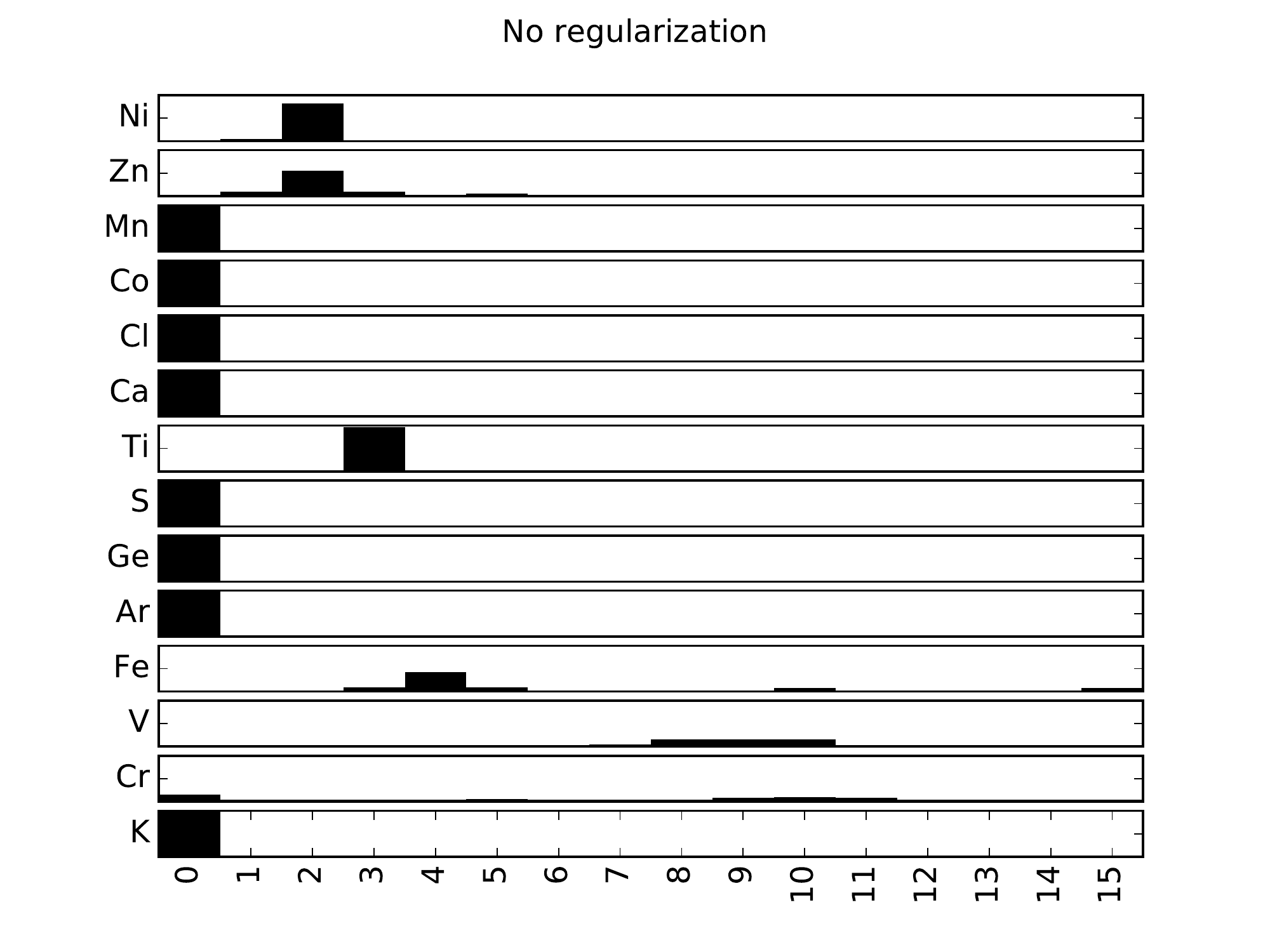} & \includegraphics[width=2.3in]{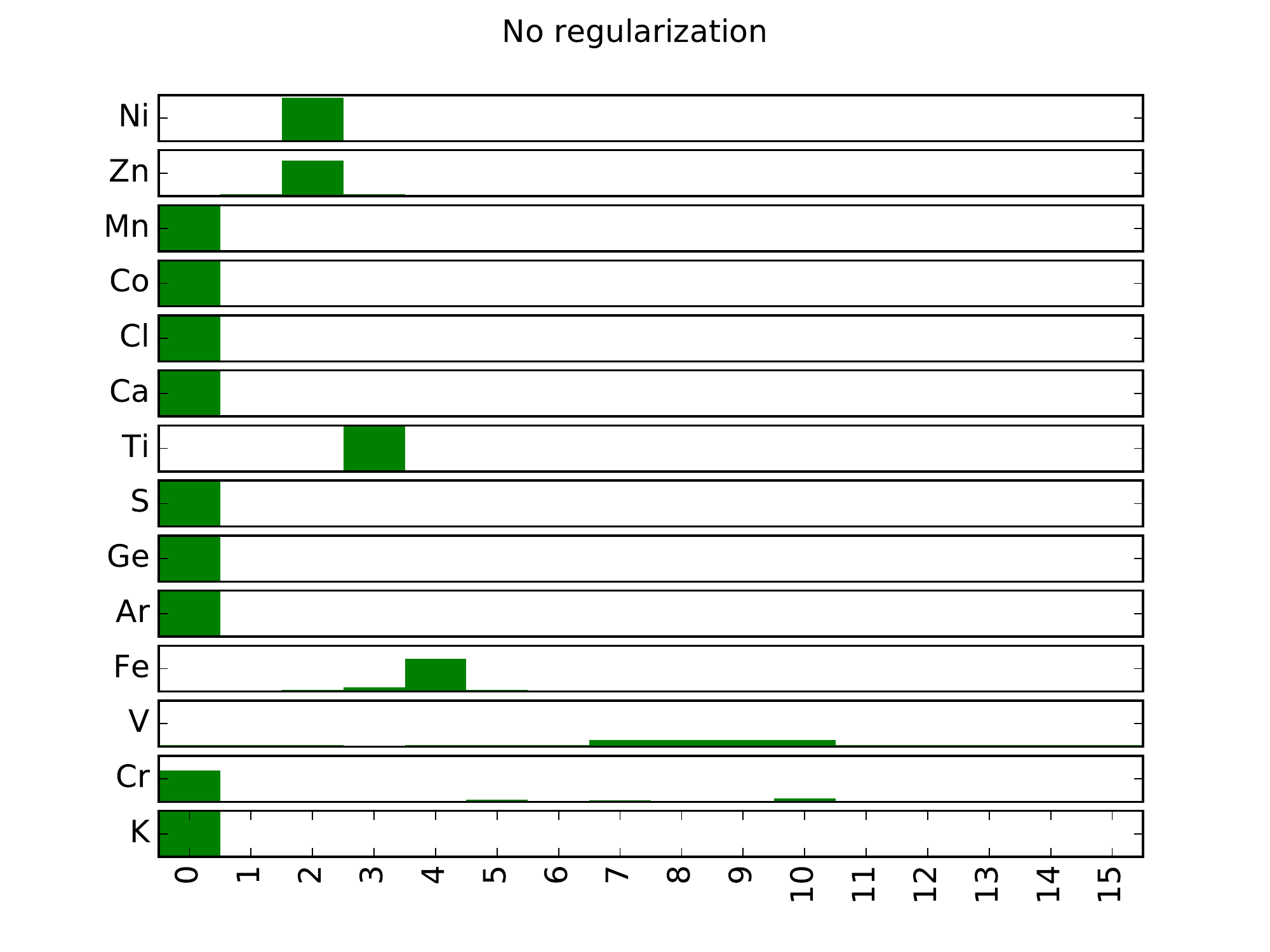}\\
  \includegraphics[width=2.3in]{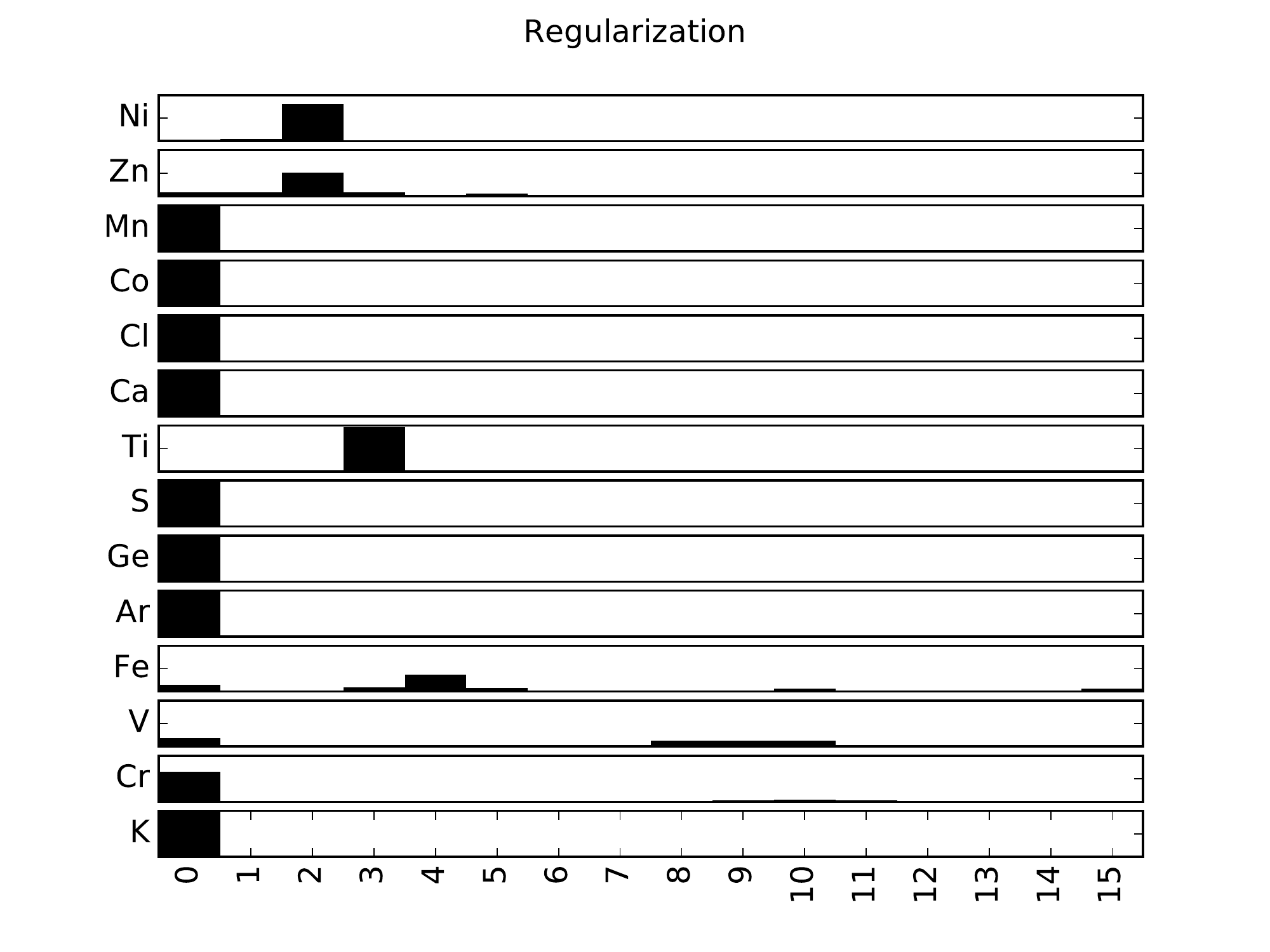} & \includegraphics[width=2.3in]{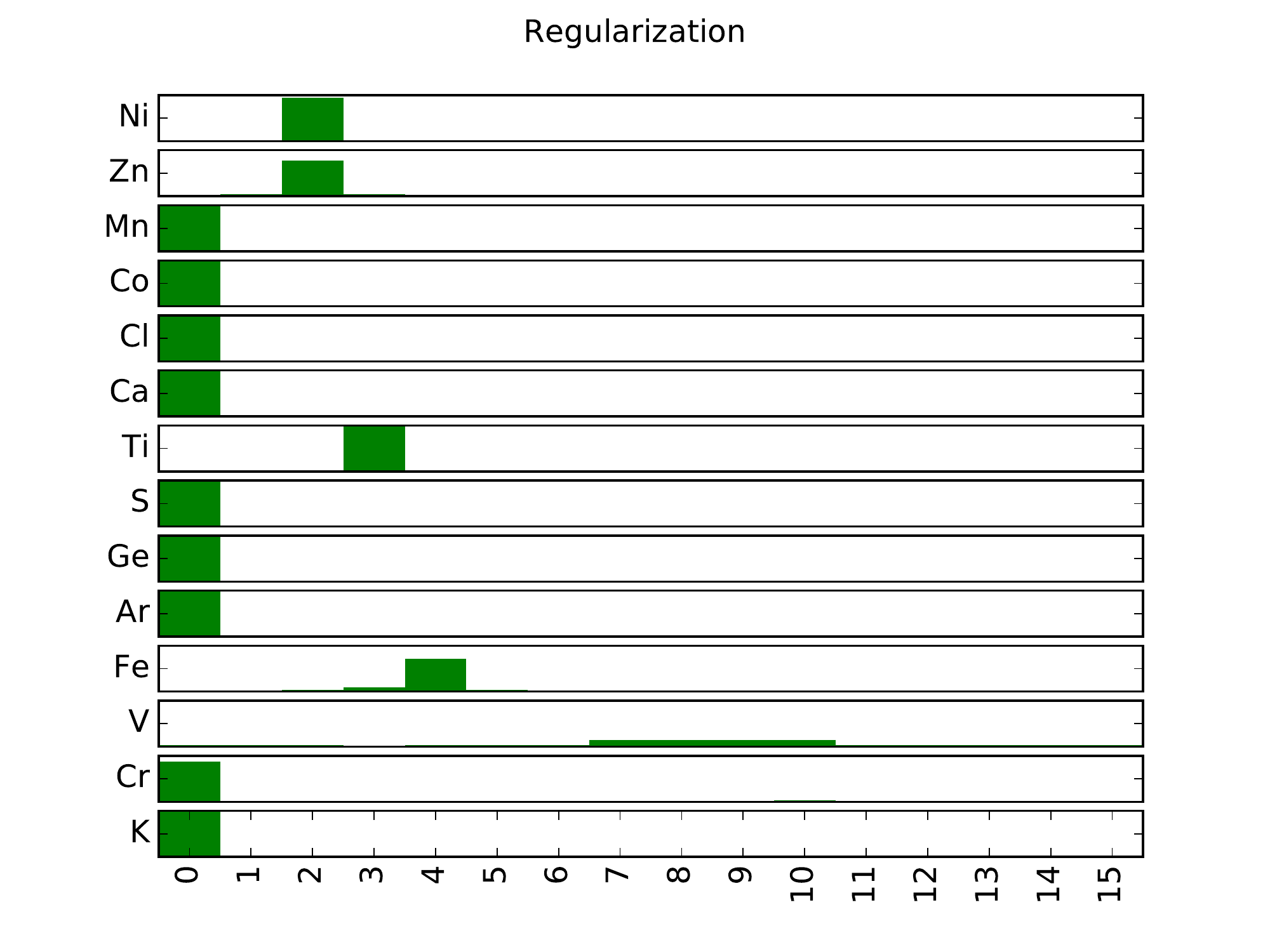}\\
\end{tabular}
\caption{{\bf Regularization for estimating elemental abundance.} The
  chemical formula $Ni_2 V_7 Zn_2 Fe_4 Ti_3$ is used to generate the
  spectrum. The correct abundance of chromium is $Cr=2$; the
  probability $\Pr(Cr=2)$ is improved by using $p=\infty$ (rather than
  $p=1$) and by including regularization information.
  \label{fig:isotope-posteriors-2}}
\end{figure}

Regularization adds additional information, and can thus be used to
improve the quality of the posteriors
(Figure~\ref{fig:isotope-posteriors-2}).

The low cost of solving additive models means that regularization is
not only inexpensive; it also means that these constraints can help
propagate sparsity of the solution space quickly through the
graph. This sparsity not only helps speed convergence, it also permits
greater trimming of convolution trees, which makes each message passed
faster.

\section*{Discussion}

The examples here only scratch the surface of the tools that could be
built with convolution forests. In the same way that linear
programming and quadratic programming are now ubiquitous for solving
combinatorial problems in applied settings, it should be possible for
the same to be said of probabilistic generalizations of the linear
diophantine equations.

One can imagine a future where combinatorial formulations of
probabilistic problems are delegated to a trusted engine for solving
or approximating them in the same black-box manner, enabling much more
widespread use of probabilistic graphical models on difficult,
combinatorial problems. Relaxations of applied problems into
elementary convex optimizations \cite{serang:conic} do not quantify the
uncertainty of the estimated solutions, and these kinds of projections
can be tricked by multimodality: for example, the expected value of a
bimodial distribution with two equal, symmetric modes will lie halfway
between the modes, and a quadratic projection of a distribution with
such a PMF will yield the point halfway between the modes, even though
it may have vanishing probability. Fully probabilistic models offer an
attractive approach around such problems.

{\tt EvergreenForest} is a first version of such an engine for
convolution forests. It is still in its infancy, but it has been
designed with an eye toward extensibility to supporting more complex
models and methods: For instance, it supports models with
heterogeneous $L_p$ spaces (\emph{i.e.}, mixing sum-product and
max-product inference). This can be quite useful for models where some
parameters are estimated via MAP inference (to avoid expected values
which themselves have low probability as point estimates as described
above), but where other random variables are estimated using
sum-product inference (thereby benefitting from the amount of
information in the aggregate of all possible paths rather than
focusing on the best).

An example of models where this is done can be found in polyploid
genotyping and mapping. These problems also feature additive
symmetries (\emph{e.g.}, each bin in a histogram of population
genotypes counts the sums of indicator variables of individuals
belonging to that particular genotype) \cite{serang:efficient2}.

Protein identification in mass spectrometry can also be phrased in
terms of additive models \cite{serang:probabilistic}. Using the
elemental quantification approach from this manuscript, it would be
possible to use both MS1 and MS2 information in the same graph,
unifying protein identification and quantification (identification
could simply be thought of as an indicator testing whether the protein
quantity is $>0$). Problems like bibliometrics (\emph{e.g.},
H-index \cite{hirsch:index}) could be rephrased in similar
probabilistic terms. For instance, a paper's citations are partitioned
as a sum of the authors' contributions (thus mitigating inflation from
large consortia, which effectively count the citations multiple times,
once for each author).

\begin{figure}
\centering
\includegraphics[width=2.3in]{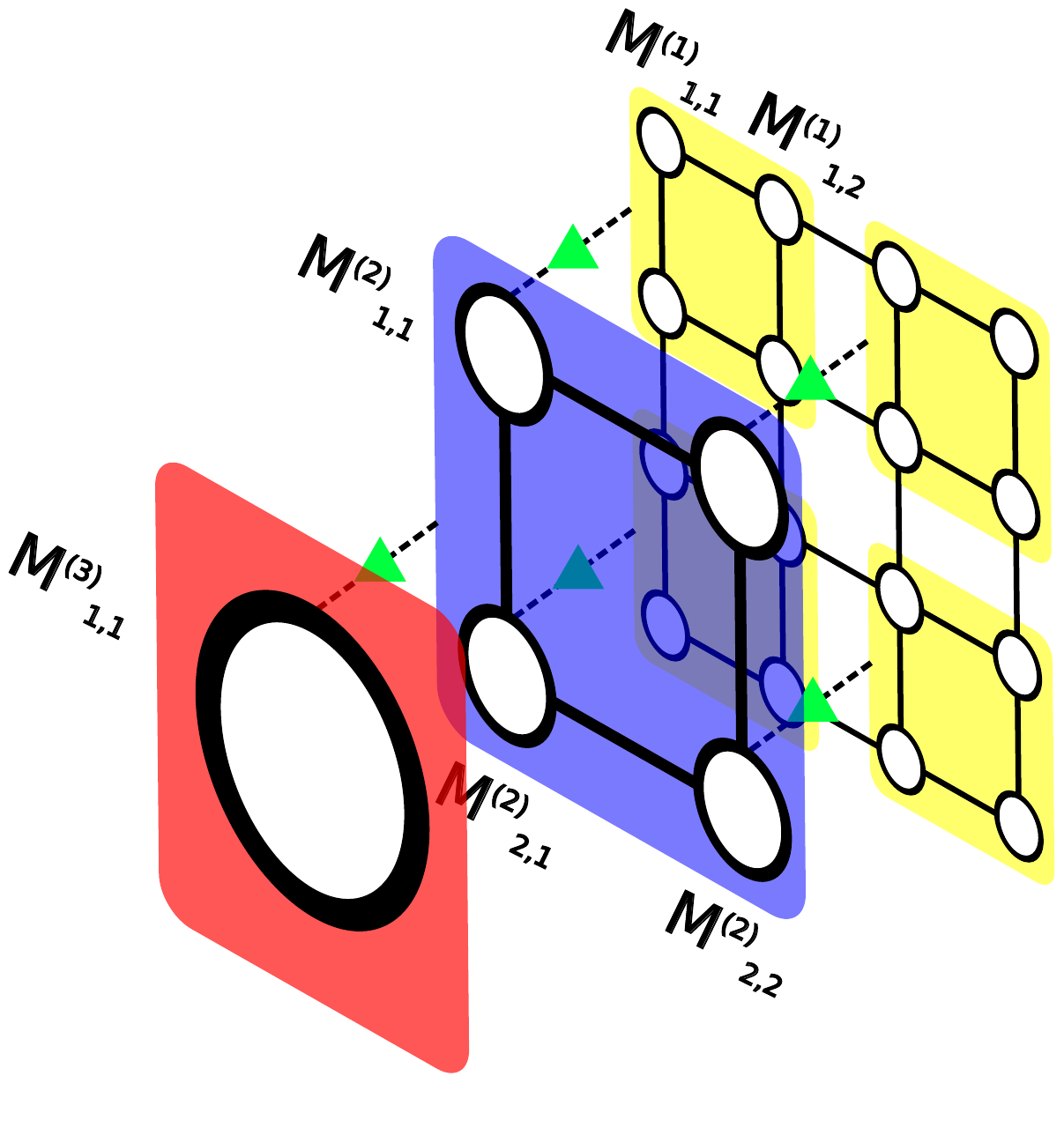}
\caption{{\bf Cascaded convolution trees for image analysis.} Three
  Ising layers, $M^{(1)}$, $M^{(2)}$, and $M^{(3)}$, are connected to
  one another by convolution trees. Convolution trees, drawn as green
  triangles, have their summand nodes marked with colored backgrounds
  and connected via a dashed line, and the node corresponding to the
  sum connected with a dashed line (drawn to distinguish them from the
  Ising connections in this 3D representation). For example,
  $M^{(2)}_{1,1}$, the upper-left corner of the middle layer, is equal
  to the sum of the four upper-left nodes in the rear layer (marked in
  yellow): $M^{(2)}_{1,1} =
  M^{(1)}_{1,1}+M^{(1)}_{1,2}+M^{(2)}_{2,1}+M^{(1)}_{2,2}$. By
  cascading the additive dependencies in this manner, all layers can
  be stacked to merge only four pixels at a time. By including an
  indicator variable for each layer $I_{L^{(i)}>0}$ (not shown), it
  would be possible to constrain the sum of the active layers with an
  extra convolution tree (which fixes the sum of the active layers
  $I_{L^{(1)}>0} + I_{L^{(2)}>0} \cdots = 1$). This will introduce
  loops into the graph, but the graph can nonetheless be solved
  adequately by loopy belief propagation.
  \label{fig:cascaded-ising}}
\end{figure}

Another example of a well-suited application is in image processing,
where it would be possible to easily use information about the total
brightness of an image, or even better, which cascade convolution
trees whose inputs are supersets of one another: A model including
information on the sum of all pixels (\emph{i.e.}, total brightness)
could potentially also include information on the sum of all pixels in
each quadrant of the image. Once the brightnesses of each quadrant
have been computed, the four of them can be summed once more to
compute the distribution on the total brightness of the image rather
than computing it from scratch. Taken to its logical conclusion, such
models would hierarchically merge four pixel chunks in an $n \times m$
image to produce an $\frac{n}{2} \times \frac{m}{2}$ image, then merge
the pixels on and on until only one pixel remains
(Figure~\ref{fig:cascaded-ising}). Each of these layers corresponds to
a level of resolution with which the image can be seen, where merging
pixels reduces error, but also reduces useful information. By pairing
indicator variables with the regularization described in this
manuscript, it would be possible to restrict the sum of the ``active''
layers to equal 1, and thus infer the most informative level of detail
with which an image should be analyzed. Such approaches could be
married with Ising models, which would enforce local dependencies
distributions within a particular layer.

Interestingly, this cascaded convolution tree design resembles the
topology of a ``convolutional'' neural network \cite{lecun:object,
  lecun:convolutional}; however, where a convolutional neural network
stores parameter point estimates (\emph{i.e.}, the model's weights),
and uses them to compute single point values at every node, this
cascaded convolution tree design stores all possible values at each
node (as a distribution). In this manner, max-convolution can be
thought of as somewhat analagous to max-pooling distributions. Because
of the topological similarity of the graphs, but difference in the
underlying types of analysis being performed, it could be interesting
to build convolution forests which also make use of neural network
methods like backpropagation. For instance, backpropagation could be
used to solve for hyperparameters of the cascaded convolution tree
model.

As mentioned previously, additive dependencies can be used to
represent DNA reads (or RNA or protein sequences), whose measured
abundances will be the sum of contributions from all candidate genomes
(or trasncriptomes or proteomes) \cite{serang:fast}. Convolution
forests extend this to arbitrary graphs on these reads, making
$p$-convolution trees applicable to metagenomics (or
metatranscriptomics or metaproteomics) problems, whose graphs are
dense and have many loops, and sometimes poor decompositions. It may
be interesting to hybridize the approach with disparate methods that
achieve high computational performance by employing deliberately
shortened reads. For example, \emph{de Bruijn} graphs shorten the
observed reads so that all possible reads (and a graph on them) is no
longer beyond computing; \emph{de Bruijn} graph have recently been
shown by Tang \emph{et al.} to be very promising results for
metaproteomics \cite{tang:graph}. As with the graph-theoretic analysis
used by Tang \emph{et al.}, deliberately shortening the observed reads
could be used to produce alternate convolution forests. An approach
qualitatively similar to the image analysis schema from
Figure~\ref{fig:cascaded-ising} could potentially permit several
\emph{de Bruijn} analyses, each with a different ``atomic'' substring
length, thereby simultaneously treating one data set as if it had
different read lengths.

Future work in convolution forests would benefit substantially from
the ability to selectively treat discrete distributions as either
sparse or dense, depending on the level of sparsity (thus enabling the
use of $O(n)$ sparse convolution, sparse max-convolution, and moste
generally, sparse $p$-convolution). Likewise, if an object oriented
approach were used to implement this dynamic sparsity (switching
between {\tt DensePMF} and {\tt SparsePMF} classes), then parametric
distributions could likewise inherit from the base {\tt PMF}
type. These parametric distributions could be related to discrete
distributions using the language of generating functions; for example,
a large uniform prior could be encoded by its generating function,
rather than actually initializing a vector with uniform values. That
generating function could be discretized into a discrete distribution
(\emph{i.e.}, the coefficients of a polynomial) once messages have
been passed and greater context is available.

The convolution tree algorithm is described in terms of discrete
distributions; however, these distributions need not be discrete as
long as the family of PMFs passed into a convolution tree is closed
under convolution. Families that are also closed under multiplication
of their PMFs or probability density functions (PDFs) can be iterated
through loopy belief propagation or through a collapsed Gibbs sampler
(\emph{i.e.}, using convolution trees to solve conditional problems
because Gibbs sampling can mix quite poorly on additive
constraints \cite{serang:probabilistic}).

Concerning linear diophantine equations of discrete distributions, a
better approach is possible by sending in the unscaled distributions
and their integer scales into a modified convolution tree, which could
factor out common denominators in scales of merged distribution pairs,
preventing unnecessary inflation of the distributions. This suggests a
more unified number-theoretic approach may make it possible to solve
dependencies based on sums of scaled distributions via the Chinese
remainder theorem and without ever scaling the distributions.

\section*{Supporting information}

The code for the engine, its modules, all demos presented here, and
utilities for visualizing graphs in Python are freely available under
an MIT software license and can be downloaded at
\url{https://bitbucket.org/orserang/evergreenforest}. The entire
library is implemented in a header-only fashion, so the essential
components of each module can be included via a single {\tt \#include}
statement.

\section*{Acknowledgments}
I am grateful to Karol W\k{e}grzycki for his great discussion on the
topic of min- and max-convolution and to Florian Heyl for his work on
early versions of peptide decomposition and elemental quantification
demos.

\end{document}